\documentclass[english,prd,superscriptaddress,nofootinbib,preprintnumbers]{revtex4}
\usepackage[latin1]{inputenc}
\usepackage{graphicx}
\usepackage{amssymb}
\usepackage{color}
\usepackage{float}
\usepackage{amsmath}
\usepackage{amsfonts}
\usepackage{dcolumn}
\usepackage{hyperref}
\usepackage{caption}
\usepackage{subcaption}
\usepackage{bigints}
\usepackage{mathtools}
\usepackage{soul,xcolor}
\hypersetup{
     breaklinks=true,
    pdfstartview={FitH},    
    colorlinks=true,       
    linkcolor=blue,          
    citecolor=red,        
    filecolor=magenta,      
    urlcolor=blue,           
    anchorcolor=green,      
    linktocpage=true
}

\newcommand{\be}{\begin{equation}}
\newcommand{\ee}{\end{equation}}
\newcommand{\bea}{\begin{eqnarray}}
\newcommand{\eea}{\end{eqnarray}}

\newcommand{\Mpl}{M_{\textrm{Pl}}}

\def\Tl{\Tilde}

\def\del{\delta}

\def\doi{http://doi.org}
\def\arxiv{http://arxiv.org/abs}

\allowdisplaybreaks

\begin{document}

\title{ What is needed of a scalar field if it is to unify inflation and late time acceleration?}

\author{Nur Jaman}
\email{nurjaman@ctp-jamia.res.in}
\affiliation{Centre for Cosmology and Science Popularization(CCSP), SGT University, Gurugram 12006, India}
\affiliation{Indian Institute of Science Education and Research Kolkata, Mohanpur, WB 741246, India}
\author{M. Sami}
\email{samijamia@gmail.com}
\affiliation{Centre for Cosmology and Science Popularization(CCSP), SGT University, Gurugram 12006, India}
\affiliation{ Center for Theoretical Physics, Eurasian National University,
Astana 010008, Kazakhstan.}

\begin{abstract}
ymmsQuintessential inflation refers to scenarios in which a single scalar field is used to describe inflation and late-time acceleration. This review is dedicated to the framework of quintessential inflation, with a focus on the building blocks of formalism. Consistent unification of inflation and late time acceleration using a single scalar field asks for a shallow field potential initially followed by steep behaviour thereafter and shallow again around the present epoch. The requirement of non-interference of scalar field with thermal history dictates steep nature of potential in the post-inflationary era, with a further restriction that late time physics be independent of initial conditions.
 We describe, in detail the scaling and asymptotic scaling solutions and the mechanism of exit from the scaling regime to late time acceleration. Review includes a fresh look at scaling solutions which are central to the theme of unification of inflation and late time acceleration.
 As for the exit mechanism, special attention is paid to the coupling of massive neutrino matter to the scalar field, which builds up dynamically and can give rise to late time acceleration. We present a detailed analytical treatment of scalar field dynamics in the presence of coupling. We briefly discuss the distinguished feature of quintessential inflation, namely, the blue spectrum of gravity waves produced during the transition from inflation to the kinetic regime.

\end{abstract}

\maketitle

\section{Introduction}
Accelerated expansion is generic to cosmological dynamics~\cite{Planck:2018vyg}, Universe has gone through inflation~\cite{Guth:1980zm,Sato:1980yn,Linde:1981mu,Albrecht:.1982wi,Starobinsky:1980te,Starobinsky:1982ee} (for review see, \cite{Liddle:1999mq,Tsujikawa:2003jp,Martin:2013tda,Vazquez:2018qdg}) at early epochs and is accelerating at present~\cite{SupernovaSearchTeam:1998fmf,SupernovaCosmologyProject:1998vns}. Inflation addresses the logical inconsistencies of hot big bang model and provides with a mechanism for primordial perturbations believed to have grown via gravitational instability into the structure we see today in the Universe~\cite{Starobinsky:1982ee,seed}. The age puzzle in hot big bang cries for late time acceleration~\cite{Krauss:1995yb,Turner:1997de} be  due to dark energy $-$ cosmological constant, quintessence or large scale modification of gravity. This phenomenon has been confirmed by direct as well as indirect observations~\cite{lateobs}, which, however, elude inflation. It is plausible to think that accelerated expansion is eternal, it had gone into hiding at an early epoch (end of inflation) allowing the thermal history to proceed as envisaged by the hot big bang
and it reappears today to account for the late time acceleration$-$re-incarnation of inflation. The term ``Quintessential Inflation" refers to a paradigm of unification of inflation and late-time acceleration that does not disturb the thermal history\cite{Peebles:1998qn,Sahni:2001qp, Peebles:1987ek,Huey:2001ae,Majumdar:2001mm,Dimopoulos:2000md,Sami:2003my,Dimopoulos:2002hm,Dias:2010rg,BasteroGil:2009eb,Chun:2009yu,Bento:2008yx,Matsuda:2007ax,Neupane:2007mu,Dimopoulos:2007bp,Rosenfeld:2006hs,Gardner:2007ib,BuenoSanchez:2006ah,Membiela:2006rj,Cardenas:2006py,Zhai:2005ub,Da,Rosenfeld:2005mt,Giovannini:2003jw,Dimopoulos:2001qu,Dimopoulos:2001ix,Yahiro:2001uh,Kaganovich:2000fc,Baccigalupi:1998mn,Lee:2014bwa,Capozziello:2005tf,Nojiri:2005pu,Elizalde:2008yf,Hossain:2014coa,Guendelman:2002js,WaliHossain:2014usl,Ahmad:2017itq,deHaro:2021swo,Dimopoulos:2020pas,Benisty:2020xqm,Karciauskas:2021fdu,Capozziello:2003tk,Sami:2004xk,Dimopoulos:2017zvq,Dimopoulos:2017tud,Bettoni:2021qfs,Wetterich:2022brb,Jaman:2018ucm,Rosati:2003yw,Salati:2002md,Akrami:2017cir,Akrami:2020zxw,Saba:2017xur}). 
In simple cases, inflation is driven by a scalar field,  which soon
after the end of inflation enters into an oscillatory phase and fast
decays in particle species giving rise to reheating/preheating of the
Universe \cite{Albrecht:1982mp,Dolgov:1982th,Abbott:1982hn,Dolgov:1989us,Ford:1986sy,Traschen:1990sw,Spokoiny:1993kt,Shtanov:1994ce,Kofman:1994rk,Kofman:1997yn,Garcia-Bellido:1997hex,Felder:1998vq,Lyth:2001nq,Feng:2002nb,delCampo:2009yc,Bassett:2005xm,Hardwick:2016whe,Campos:2002yk,Allahverdi:2010xz,Amin:2014eta,Garcia:2020eof,Tambalo:2016eqr,Dimopoulos:2017tud,Lopez:2021agu,Saha:2021kez,Pareek:2021lxz,Bhattacharya:2019ryo,Dimopoulos:2019gpz}). Attempts have been made to achieve the aforementioned goal by using a single scalar field as a source of both the phases of accelerated expansion in the framework of modified theories of gravity, particularly $f(R)$ theories~\cite{Capozziello:2003tk,Carroll:2003wy,DeFelice:2010aj,Sotiriou:2008rp,Gannouji:2012iy,Cosmai:2013nva} . For instance, putting the Starobinsky proposal for inflation and late time acceleration together using $f(R)$ could suffice, see Ref.\cite{Nojiri:2007uq} on the related theme. In these models, it is difficult to get rid of the strong dependence on initial conditions. In this review, we shall focus on the scalar field models where one has better control on this problem.

At the onset, it sounds fairly straightforward to carry the said unification in a single scalar field using a runaway type of potential which is shallow initially, suitable for slow roll, followed by steep behaviour for the entire history of the Universe and turning shallow again at late times to account for the observed acceleration.
Though the presence of early and late time phases of accelerated expansion is evident in the proposal, nonetheless, remarks related to thermal history and the independence of late time evolution from initial conditions are in order.
Since the potential is of the runaway type, the standard reheating mechanism is not operative in this framework. One needs to employ an alternative framework such as gravitational particle production~\cite{Ford:1986sy,Spokoiny:1993kt} (see also \cite{Chun:2009yu}), instant preheating~\cite{Felder:1998vq,Sami:2004xk,Campos:2002yk,Dimopoulos:2017tud}, curvaton preheating~\cite{Lyth:2001nq,Feng:2002nb,delCampo:2009yc,Hardwick:2016whe}, Ricci reheating~\cite{Dimopoulos:2018wfg} (see also, \cite{Bettoni:2021zhq,Opferkuch:2019zbd}) and so on. Radiation density produced
in these processes at the end of inflation is smaller than the field energy density by several orders of magnitude. Due to the steep nature of potential in the post-inflationary era, scalar field enters the kinetic regime soon after inflation ends  ($\rho_\phi\sim a^{-6}$)~\cite{Tashiro:2003qp} such that radiation domination takes place after a long kinetic regime and field energy density becomes sub-dominant. There are two requirements for the field dynamics before it emerges from sub-dominance to dominance at late times: (1) It must adhere to the nucleosynthesis constraint; and (2) Its emergence to late time acceleration must be unaffected by initial conditions. The latter puts a restriction on the class of field potentials that can be employed for quintessential inflation.

A comment on the duration of the kinetic regime is in order. The long kinetic regime might be problematic for nucleosynthesis due to relic gravity waves. Indeed, the energy density in relic gravity waves produced at the end of inflation increases in comparison to $\rho_\phi$ during the kinetic regime and might conflict with the nucleosynthesis constraint at the commencement of the radiative regime. The nucleosynthesis constraint puts a lower bound on the reheating temperature at the end of inflation and an upper bound on the duration of the kinetic regime. One of the reheating alternatives is provided by gravitational particle production which is a universal process that occurs due to a non-adiabatic change in space-time geometry when inflation ends; unlike other mechanisms, it does not require additional field(s); however, nucleosynthesis bound is challenged.  It should be noted that the presence of the kinetic epoch soon after the end of inflation induces a blue tilt in the gravitational waves spectrum. This feature enhances the prospects of detection of gravitational waves in the forthcoming observational missions.
 Further, a desirable feature of quintessential inflation should also be mentioned here, namely, field should track the background in the post inflationary era, and only at late times, it should exit to acceleration. Several exit mechanism have bee analysed in the literature. We shall, however, focus on a particular one due to 
 coupling of scalar field to massive neutrino matter.
Last but not least, one should ask whether the unification scheme is an academic exercise or has a distinguished observational signature. The presence of the kinetic regime, a generic feature of quintessential inflation, induces a novel feature in the spectrum of relic gravity waves, namely, the blue spectrum in the high-frequency regime. This is a generic model-independent prediction of the paradigm irrespective of its concrete realization, to be tested in future observations (see  Refs.~\cite{Chun:2007np,Kamada:2019ewe} on the related theme ). 

This is a pedagogical review written specifically for young researchers. Efforts have gone into highlighting the general requirements and the corresponding building blocks of the paradigm rather than discussing the technical details of concrete models of quintessential inflation. It goes without saying that the issues raised in the review are equally important for model building related to late time acceleration. We sincerely hope that the review will be helpful for beginners interested in related issues. The review should be read with footnotes, which include additional explanations and clarifications.
 The organisation of the review  is as follows. In Sec~\ref{Bblock}, we describe broad features of the paradigm: Suitable class of runaway potentials, alternative reheating models and exit mechanisms.
 In Sec~\ref{cosmodyn}, we discuss cosmological  dynamics of scalar field relevant to
  quintessential inflation. This includes field dynamics during inflationary era, some model independent feature of  inflation, nucleosynthesis constraint,   scaling and tracker behaviour in models of quintessential inflation. In Sec.~\ref{postdyn}, we describe  a  mechanism of exit from scaling regime to late time acceleration, which is realized by invoking coupling to massive neutrino matter. In this section prospects of detection of relic gravity waves are also discussed. Finally, we summarise the review in Sec.~\ref{postdyn}. We use the metric signature, $(-,+,+,+)$ and  the notation for the reduced Planck mass, $\Mpl^{-2} = 8\pi G$.

\newpage
\section{ Quintessential inflation: Building blocks }
\label{Bblock}
\begin{figure}[H]
    \centering
    \includegraphics[width=13cm, height=16cm]{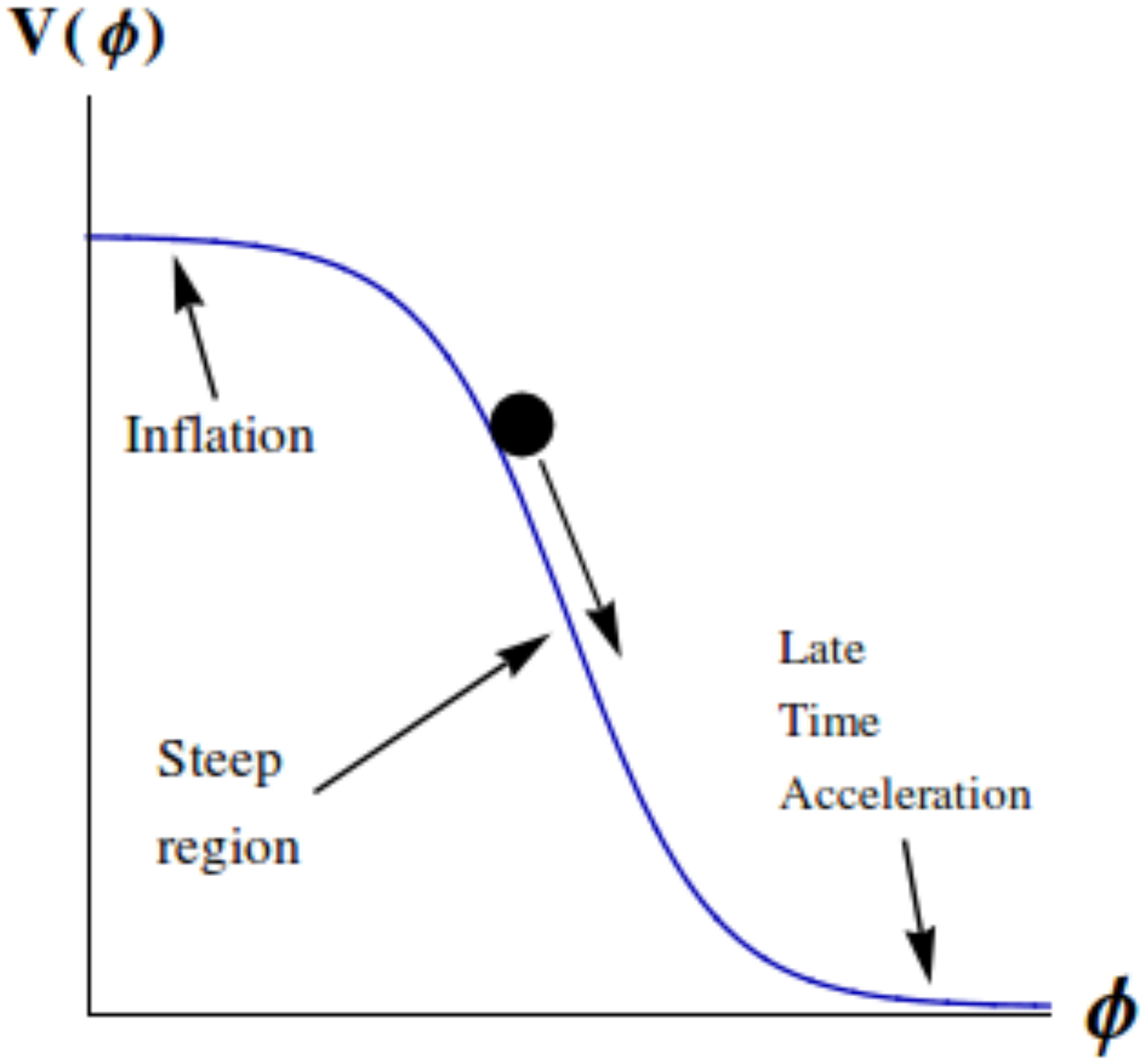}
    \caption{The figure displays a typical scalar field potential required to realise quintessential inflation: potential shallow initially, followed by steep behaviour thereafter, and shallow again around the present epoch.}
    \label{Fig:quintpot}
\end{figure}

In this section, we broadly indicate the basic features of model building for quintessential inflation and list the building blocks of the model. Fig.\ref{Fig:quintpot} shows the scalar field potential, flat at both the early and late time epochs and steep in between, joining the two ends. One immediate implication of the runaway type nature of the potential is the requirement for an alternative reheating mechanism. There are a few alternatives to be mentioned in the discussion to follow. It is important to note that in the context of warm inflation~\cite{Berera:1995wh,Berera:1995ie} a separate reheating mechanism may not be required as radiation is being produced during inflation itself due to the coupling of the inflaton field to the radiation bath. The idea of ``warm inflation" also works in the context of quintessential inflation as well~\cite{Lima:2019yyv,Dimopoulos:2019gpz,Basak:2021cgk,Levy:2020zfo,Gangopadhyay:2020bxn}
We shall also mention specific field tracks that could join the two phases of acceleration such that thermal history is left without interference and the late-time physics is insensitive to initial conditions.
\subsubsection{Reheating through gravitational particle production}
\label{sec:gravitionalreheating}
Gravitational particle production takes place at the end of inflation due to a non-adiabatic change of space time geometry \cite{Ford:1986sy}, (see also \cite{Spokoiny:1993kt}). It is a universal process and does not require the presence of additional field(s). Energy density of radiation produced in this process is given by, $\rho_r\simeq 0.01\times g_p H^4_{\rm end}$, where $g_p\sim 100$, the number of relativistic degrees produced in the process; it involves $H_{\rm end}$, the natural scale available at the end of inflation. In this case, typically, $\rho_\phi/\rho_r\simeq 10^{-10}$~\cite{Sahni:2001qp} and the process is inefficient and might challenge the nucleosynthesis constraint due to  over production of  relic gravity waves. Other alternatives include, instant preheating,  curvaton reheating mechanism and Ricci reheating.

\subsubsection{Instant Preheating} 
\label{inst}
Instant preheating~\cite{Felder:1998vq, Campos:2002yk} operates with an assumption that the inflaton field $\phi$ interacts with another scalar field $\chi$ (with coupling $g$) whose mass depends upon $\phi$ such that as $\phi$, after inflation, runs down its potential, $m_\chi$ increases and  decays into matter fields\footnote{In this framework, field $\phi$ is coupled to another scalar field $\chi$ which in turn couples to matter fields:
$\mathcal{L}_{int}=-(1/2) g^2\phi^2 \chi^2-h\chi\bar{\Psi}\Psi $. In this case,  $\chi$ does not bare mass, its effective mass depends upon $\phi$, $m_\chi=g\phi$. After inflation, as $\phi$ runs down its potential fast, $m_\chi$ changes in non-adiabatic fashion giving rise  to production of $\chi$ particles which might instantaneously decay into matter fields. Assuming that energy of produced particles is instantaneously thermalized (denoted by $\rho_r$), one finds at the end of inflation that
$(\rho_\phi/\rho_r)_{end}=(2 \pi)^3/g^2$.}   ($\chi$ couples to matter fields with coupling $h$). By appropriately choosing couplings ($g, h<1$), one can obtain instantaneous preheating at the end of inflation with the reheating temperature that satisfies the bound implied by the nucleosynthesis constraint.

\subsubsection{Curvaton Reheating} The Lagrangian is supplemented with an extra field $\sigma$ with a quadratic mass term of $m_ \sigma^2 \sigma^2$, such that $m_ \sigma\ll H_ {inf} $ and the $\sigma $ field ceases to play any role during inflation~\cite{Lyth:2001nq,Feng:2002nb,delCampo:2009yc,Hardwick:2016whe}. After inflation ends, the Hubble parameter decreases $m_\sigma \sim H$ and the  $\sigma$ field begins to oscillate and decay into matter fields, giving rise to reheating with desired temperature at the end of inflation.
\subsubsection{Ricci Reheating}
The problem associated with gravitational reheating, that is over-production of gravitational waves can be controlled by introducing another reheating mechanism called Ricci Reheating~\cite{Dimopoulos:2018wfg,Bettoni:2021zhq,Opferkuch:2019zbd}. In this mechanism a scalar field called reheaton is non-minimally coupled directly with the Ricci scalar. The Ricci scalar changes sign  after the end of inflation when the kinetic energy takes over the potential in the steep region of the potential. As a result the effective mass of the reheaton changes sign and becomes tachyonic. Due the tachyonic nature, reheaton filed grows as the filed rolls down towards new effective minima.  The rehating is realized by the transfer of energy from this reheaton to standard model particles. It is also possible to identify reheaton with Higgs field~\cite{Opferkuch:2019zbd}.

\subsubsection{Independence of late time dynamics from initial conditions}
The requirement of steep potential in the post inflationary era is dictated by
the need to achieve the commencement of radiation 
domination after which scalar field remains sub-dominant. However, a specific type of steep behaviour is asked for, which allows the scalar field to follow the background $-$ {\it scaling} solution which is an attractor of dynamical system.
The 
nucleosynthesis constraint, further, puts a  lower bound on the slope of these potentials. Scaling behaviour, for instance, is exhibited by a steep exponential potential, ($ V\sim Exp(-\alpha \phi/M_{Pl}; \alpha^2>3$)~\cite{Ferreira:1997hj,Copeland:1997et, Ferreira:1997au,SS}.
However, the scaling solution is non-accelerating as , in this case, $w_\phi=w_b=(0, 1/3)$ for cold dark matter and radiation, respectively. 
Hence, in this case, one needs to invoke an additional feature in the potential that could trigger an exit to late time acceleration.
Once the model parameters are set to exit to acceleration at the present epoch, changing the initial conditions within a reasonable range does not affect late-time physics $\hat{\rm a}$  {\it  la} a {\it tracker}~\cite{Steinhardt:1999nw,Chiba:2009gg}.

\subsubsection{Exit mechanisms}
There are several ways to realize the exit from scaling regime to accelerated expansion: (1) one might simply add a cosmological constant to scalar field Lagrangian or if one is willing to go beyond, one might add to scalar field potential, a term, $V_1 \sim Exp{(-\lambda_1 \phi/M_{Pl})}, \lambda_1^2<2$~\cite{Copeland:1997et,Barreiro:1999zs,Haro:2019peq}, such that it becomes operative only at late times giving rise to slow roll. (2) One might add to Lagrangian a Gauss-Bonnet term coupled to scalar field: $\xi(\phi) R^2_{GB}$ ( $
{R^2_{GB}}=R^2-4R_{\mu\nu}R^{\mu\nu}+R_{\mu\nu\alpha\beta}R^{\mu\nu\alpha\beta}
$ ) which can induce minimum in the run away potential and might give rise to late time acceleration by suitable choosing the coupling\cite{exitG}. However, it might be difficult to make this term invisible during inflation. (3a) It is rather straight forward to induce minimum in the scalar field potential by introducing coupling of scalar field to (cold) matter
which is proportional to trace of matter energy momentum tensor and vanishes identically during radiation era~\cite{Gumjudpai:2005ry}. In this case, scalar field can settle in the minimum of the effective potential giving rise to de Sitter like behaviour of interest to late time cosmology. However, it happens soon after matter domination is established where it is undesirable as matter phase should be left intact. (3b) This problem can be circumvented by invoking coupling of scalar field to massive neutrino matter which builds up dynamically only at late times as massive neutrinos turn non-relativistic. In this paper, we shall focus on the dynamics of this  process. 
\subsubsection{Suitable class of scalar field potentials }
It is difficult to
have tractable analytical expression of a potential shallow initially as required by inflation with a tracker like behaviour in the post inflationary era. We often encounter the following class of potentials in the literature.
(a) Potentials steep initially with a desired  behaviour thereafter, for instance, inverse power law potentials. The known way to have accelerated expansion in the initial phase in this case is provided by ``Brane Worlds"~\cite{Sami:2004xk,Sahni:2001qp,Bento:2008yx,Adhikari:2020xcg}, where high energy brane corrections can give rise to slow roll with a graceful exit to deceleration as the field runs down its potential and brane damping disappears. Unfortunately, this scenario gives rise to a large tensor to scalar ratio of perturbations, which is ruled out by observation~\cite{Copeland:2000hn}.
(b) Class of potentials, flat initially, followed by a steep behaviour. For instance, $V\sim Exp(-\lambda \phi^n/\Mpl^n), n>1$\cite{Geng:2015fla}.
Potentials with the said characteristic can also be constructed using  non-canonical form of scalar field Lagrangian.
In this case one requires to invoke an exit mechanism to late time acceleration. 
(c) If we do not adhere to tracker like behaviour in the post inflationary era, it is possible to 
find a class of potentials shallow at early and late times and steep in between  with a {\it thawing realization}~\cite{Dimopoulos:2017zvq}.

\section{Cosmological dynamics of Scalar field  in a nut shell}
\label{cosmodyn}
Scalar field dynamics has played an important role in modern cosmology, starting with inflation, which led to a paradigm shift resulting in the integration of cosmology with high energy physics. In what follows, we present the basic elements of scalar field dynamics with and without the presence of background matter. Our discussion would revolve around scaling solutions and slow roll attractors. Special attention would be paid to asymptotic scaling solutions, one of the important building blocks of quintessential inflation. However, since the scaling solutions are non-accelerating in nature, one requires mechanisms of exit to late time acceleration. We discuss in detail the exit caused by the coupling of the scalar field to massive neutrino matter. The generic coupling is proportional to the trace of the energy momentum tensor of massive neutrino matter, which vanishes as long as neutrinos are relativistic. Coupling acquires non-zero value at late stages of evolution when neutrinos become non-relativistic. This is a unique physical process that occurred in the Universe, and neutrino mass is the only available energy scale close to the mass scale associated with dark energy (see for review on dark energy,~\cite{Carroll:2003qq,Copeland:2006wr,Sahni:2006pa,Sahni:2004ai,Li:2012dt,Brax:2017idh,Mortonson:2013zfa,Sami:2013ssa}, see alos~\cite{Zhang:2021ygh}). The coupling of the scalar field to massive neutrino matter invokes an important feature in the scalar field potential responsible for late time acceleration. We will include details in the sub-section to follow.
\subsection{Field evolution in absence of background matter: Slow roll versus fast roll}
Scalar dynamics has been used extensively in the development of models for both inflation and late-time acceleration.In this section, we shall focus on dynamics of non-minimally coupled scalar field.
Let us consider the action of a minimally coupled scale field $\phi$ which populates the FLRW universe~\cite{Ratra:1987rm,Copeland:2006wr,samrev1,samrev2},
\begin{equation}
\mathcal{S}=\int{ d^4x\sqrt{-g}\left[ \frac{\Mpl^2}{2}R+\mathcal{L}(\phi) \right]} \equiv\int{ d^4x\sqrt{-g}\left[ \frac{\Mpl^2}{2}R-\frac{1}{2}\partial_\mu \phi \partial^\mu \phi-V(\phi)  \right]} 
\label{actionm}
\end{equation}
The energy momentum tensor for the scalar field reads,
\begin{equation}
T_{\mu\nu}(\phi)= \partial_\mu \phi \partial_\nu \phi+g_{\mu\nu}\mathcal{L}(\phi)  
\end{equation}
such that the field energy density and pressure are given by,
\begin{equation}
 T^0{_0}\equiv -\rho_\phi=\frac{1}{2}
 \dot{\phi}^2+V(\phi);~~T^i{_i}\equiv p_\phi=\frac{1}{2}
 \dot{\phi}^2-V(\phi)
\end{equation}

For FLRW background dominated by the scalar field energy density, we have,
\begin{eqnarray}
\label{FEphi}
&& \left(\frac{\dot{a}}{a}\right)^2\equiv H^2=\left(\frac{1}{3\Mpl^2} \right)\rho_\phi\\
&& -\frac{\dot{H}}{H^2}=\frac{3}{2}(1+w_\phi)=3 \frac{\dot{\phi}^2/{2}}{\rho_\phi}
\Longleftrightarrow  \frac{\ddot{a}}{a}=-\frac{1}{6 M^2_{Pl}}\rho_\phi\left(1+3w_\phi\right)
\label{FEphia}
\end{eqnarray}
where the equation of state parameter for the field is defined as,
\begin{equation}
w_{\phi}\equiv \frac {P_\phi}{\rho_\phi}=\frac{ \frac{1}{2}
 \dot{\phi}^2-V(\phi)}{\frac{1}{2}
 \dot{\phi}^2+V(\phi)},
\end{equation}
which interpolates between $1$ and $-1$. The lower (upper) limit  corresponds to the potential (kinetic) energy dominated situation. The field configuration with $w_\phi\simeq -1$ or $-\frac{\dot{H}}{H^2}\ll1$ is referred to (quasi) de Sitter for which, $H$ is approximately constant and $a(t)\sim e^{Ht}$~\footnote{Expansion has character of acceleration ($\ddot{a}/a>0$ for $w_\phi<-1/3$), see Eq.(\ref{FEphia}).}.
In this case, field rolls slowly,
\begin{equation}
\dot{\phi}^2/2\ll V(\phi)\Longrightarrow \epsilon\equiv 3\frac{\dot{\phi}^2/2}{\rho_\phi}  \simeq \frac{3}{2}\frac{\dot{\phi}^2}{V}\ll1  
\end{equation}
where $\epsilon$ is one of the slow roll parameters.
The (quasi) de Sitter configuration is admitted by the scalar field dynamics as a possible fixed point. Indeed, the scalar field equation that follows from (\ref{actionm}) in the FLRW background is,
\begin{equation}
 \ddot  {\phi}+3H\dot{\phi}+V_\phi=0;~~V_\phi\equiv \frac{d V}{d\phi}
 \label{Fieldeq}
\end{equation}
where the second term  with Hubble rate acts like a friction coefficient. Field equation (\ref{Fieldeq}) is equivalent to energy conservation,
\begin{equation}
\dot{\rho_\phi}+3 H \rho_\phi(1+w_\phi)=0 
\end{equation}
which formally integrates to,
\begin{equation}
\rho_\phi=\rho_{in} e^{-3\int{(1+w_\phi)\frac{da }{a} } } \, , 
\end{equation}
with $\rho_{in} $ as a integration constant. For potential energy dominated situation, $\dot{\phi}^2/2\ll V$ ($w_\phi\simeq -1$),  field energy density is approximately constant where as in the opposite case $(\dot{\phi}/2\gg V)$,  dubbed kinetic regime,  $\rho_\phi\sim 1/a^6$.  Kinetic regime is realized in case of a steep potential where field rolls down its potential fast making the potential energy redundant giving rise to $\rho_\phi\sim a^{-6}$.

As mentioned before, we are interested in realizing the quasi de Sitter configuration. To this effect, we notice that
when, friction term in Eq.(\ref{Fieldeq}) is large, 
\begin{equation}
\label{Hf}
\beta\equiv  -\frac{\ddot{\phi}}{3 H\dot{\phi}}\ll 1 \Longrightarrow~~\text{Slow roll regime} :~~
3 H\dot{\phi}+{V_\phi} \simeq 0;~~H^2\simeq \frac{1}{3 M^2_{Pl} }V(\phi) 
\end{equation}
where $\dot{\phi}$ is small and adheres to slow roll condition, namely, $\epsilon\ll1$, which  in turn implies that the potential is nearly flat,\footnote{The acceleration term, $\ddot{\phi}$ in (\ref{Fieldeq}) can no longer be be dropped in case of a steep potential.}
\begin{equation}
 \epsilon=\frac{M^2_{Pl} }{2}\left(\frac{V_\phi}{V}\right)^2   
\end{equation}
In this case, even if we kick off with a huge acceleration, $\ddot{\phi}$, it is taken care of by the friction term.
Thus slow roll dynamics confirms the existence of (quasi) de Sitter fixed point
also known as slow roll solution. We should , however, check the consistency of (\ref{Hf}) with the field equation (\ref{Fieldeq}). Taking the time derivative of slow roll equation (\ref{Hf}) and using Eq.(\ref{FEphia}), we have,
\begin{equation}
 \beta=-\frac{\ddot{\phi}}{3H\dot{\phi}}   =+\frac{V_{\phi\phi}}{9 H^2}-\frac{(1+w_\phi)}{2};~~V_{\phi\phi}=\frac{d^2 V}{d\phi^2}
\end{equation}
Keeping $\beta$ negligible
requires,
\begin{eqnarray}
 && \epsilon=\frac{M^2_{Pl} }{2}\left(\frac{V_\phi}{V}\right)^2\ll1\\
&&  \eta=M^2_{Pl} \frac{V_{\phi\phi}}{V}\ll1,
\end{eqnarray}
which express the self consistency of slow roll approximation with full scalar field dynamics. Thus, we are hereby led to two necessary conditions for slow roll.
The first condition, tells us that the slope of the potential small and potential is shallow
($M_{Pl} |V_\phi/V|\ll1$) where as the second condition is a statement about the smallness of the curvature which ensures that slow roll can be sustained for a long duration ( $\dot{\epsilon}/H\epsilon=\mathcal{O}(\eta, \epsilon$ ) needed to collect required number of e-foldings necessary to address the shortcomings of the hot big bang model. Smallness of $\eta$ is also dictated by the flatness of  perturbation spectrum. 
Thus, the smallness of slow roll parameters ensures the required number of e-foldings and the scale-independent spectrum of adiabatic density perturbations produced during inflation.
\subsection{ Inflation: Model Independent Features}
As mentioned before, we focus on generic features of the paradigm rather than the concrete models.
In what follows, we present the model independent estimates of  inflation to be used later.
We recall that the tensor-to-scalar ratio, $r$ is defined as the ratio of amplitudes of the tensor and scalar perturbations,
\begin{eqnarray}
r\equiv \frac{P_T}{P_S} \label{r_def}
\end{eqnarray}
which is expressed through $\epsilon$
in the case of canonical scalar field as $r=16 \epsilon$.
 Observations~\cite{Planck:2018jri} put a bound on the tensor to scalar ratio of perturbations, namely, $r \lesssim 0.032 $ \cite{Tristram:2021tvh}  which translates into a restriction on the slope of the potential, $\lambda_s \equiv -\Mpl \frac{V,\phi}{V}\lesssim 0.06$  which is a more tighter bound on the slope than the one imposed by slow roll condition.  
The scalar power spectrum,
\begin{eqnarray}
P_S= \left.{\frac{1}{8 \pi ^2}\frac{1}{\epsilon}\frac{H^2}{\Mpl^2}}\right|_{k=aH}= \frac{2}{\pi}\, . \frac{1}{r}\frac{H_{\rm inf}^2}{\Mpl^2}
\end{eqnarray}
is bound due to CMB observations and adhering to recent findings\cite {Planck:2018jri,Tristram:2021tvh}, we have $H_{\rm inf}\simeq 5.8 \times 10^{-5}\, r^{1/2} \Mpl=1.4\times 10^{14} r^{1/2}\rm GeV$ or equivalently, $V_{\rm inf}^{1/4}= 0.01 \times r^{1/2} \Mpl$. At the end of inflation, $\epsilon=1$ which implies that $\dot\phi_{\rm end}=V_{\rm end}^{1/2}$ and we have the Hubble parameter at the end of inflation,
\begin{eqnarray}
H_{\rm end }= \frac{1}{\sqrt{2}\Mpl}V_{\rm end}^{1/2}. \label{hend}
\end{eqnarray}

Since inflation is a quasi de-Sitter phase of expansion it is expected H does not change much from commencement to the end of inflation. The argument is supported by the observational upper-bound on tensor-to-scalar ratio $r \lesssim .032$ such that the field excursion during inflation is not much. This is supported by numerical calculation for generalized exponential potential which shows that not only $H_{inf} \& H_{end} $ but $H_{inf}, H_{end} \& H_{kin}$ are within the same order of magnitude~\cite{Ahmad:2017itq}.

Using Eq.~(\ref{hend}) we have
 the estimation for $V_{\rm end }$,
\begin{eqnarray}
V_{\rm end}\approx 2 H_{\rm inf}^2 \Mpl^2 \Rightarrow
\rho_{\phi, \rm end}= \frac{3}{2} V_{\rm end} \approx 3 H_{\rm inf}^2 \Mpl^2\label{rhophiend}\,
\end{eqnarray}

In the kinetic epoch followed by inflation, the energy densities of the filed and radiation evolve as
\begin{eqnarray}
\rho_{\phi} &=& \rho_{\phi, \rm end}\left(\frac{a_{\rm end}}{a}\right)^6\\
\rho_r  &=& \rho_{r, \rm end} \left(\frac{a_{\rm end}}{a}\right)^4 \label{rhoskineton}
\end{eqnarray}

Thus, the ratio of the scale factors at the end of inflation to the commencement of radiation epoch or that of the respective temperatures are obtained by using $\rho_\phi (a_r)= \rho_r (a_r)$ which gives\\
\begin{eqnarray}
\frac{a_r}{a_{\rm end}}= \frac{T_{\rm end}}{T_r}= \left[\frac{\rho_{\phi, \rm end}}{\rho_{r, \rm end}}\right]^{1/2}\, . \label{eq:arbyend}
\end{eqnarray}
The length of kinetic regime depends upon the ratio of field energy density to the radiation energy density at the end of inflation. Efficient reheating process implies smaller value of this ratio corresponding to shorter kinetic regime.
Owing to the steep nature of scalar field potential in the post-inflationary era, field runs down its potential fast  after inflation ends and in a short while, enters the kinetic.
In fact, it is found numerically that for the potential (\ref{potn})  ${H_{\rm inf}}~,{H_{\rm end}}~\& ~H_{\rm kin}$ are within  the same order of magnitude~\cite{Ahmad:2017itq}. We shall use, ${H_{\rm inf}}\simeq {H_{\rm end}}\simeq~H_{\rm kin}$ in our estimations for model independent predictions.

\subsection{Evolution in presence of background matter: Scaling solution}
Scalar field dynamics in presence of background matter (cold matter/radiation) exhibits distinguished features. Under specific conditions scalar field might mimic background matter  $\hat{\rm a}$  {\it  la} {\it scaling solution} which plays an important in model building for dark energy and quintessential inflation. In order to check for scaling behaviour of $\rho_\phi, $, let us consider evolution equations in the presence of background matter\cite{Copeland:2006wr,samrev1,Copeland:1997et,Liddle:1998xm,Skugoreva:2019blk}
\begin{eqnarray}
&& H^2=\frac{1}{3M^2_{Pl} }\left(\rho_b+\rho_\phi\right)  \\
&& \dot{H}=-\frac{1}{2M^2_{Pl}}\left((1+w_B)\rho_b+\dot{\phi}^2\right)\\
&& \dot{\rho}_b+3 H \rho_b(1+w_b)=0
\end{eqnarray}
where $\rho_b$ and $w_b$ designate energy density and equation of state parameter for background matter respectively. As for the scalar field equation, background dependence is reflected there through  Hubble parameter. It follows from continuity equation for background matter that $\rho_B\sim a^{-3(1+w_b)}$.
Scaling solution is defined requiring that scalar field evolves exactly as the background matter, 
\begin{equation}
\text{ Scaling solution}:~~w_\phi=w_b;~~\rho_\phi\sim   a^{-3(1+w_b)}
\end{equation}
Scaling solution is a specific dynamical arrangement which preserves the ratio of kinetic to potential energy of the field,
\begin{equation}
w_{\phi}\equiv \frac {P_\phi}{\rho_\phi}=\frac{ \frac{1}{2}
 \dot{\phi}^2-V(\phi)}{\frac{1}{2}
 \dot{\phi}^2+V(\phi)}=w_b\Rightarrow \dot{\phi}^2(1-w_b)=2V(1+w_b)
 \label{scalingdef}
\end{equation}
We recall from earlier discussion that the interplay between the friction term and the gradient of potential reflects the nature of the solution of the field equation; the scaling solution should impose certain restrictions on their ratio.
Indeed, differentiating, (\ref{scalingdef}) left right and using the field equation (\ref{Fieldeq}), we have,
\begin{equation}
(3H\dot{\phi} +V_\phi)\dot{\phi} (1-w_b)=-(1+w_b)V_\phi \dot{\phi}\Rightarrow
-3(1-w_b)H\dot{\phi}^2=2V_\phi \dot{\phi}
\label{friction}
\end{equation}
Eqs (\ref{scalingdef}) and (\ref{friction}) yield two important properties of scaling solutions, namely \cite{Skugoreva:2019blk},
\begin{eqnarray}
\label{X}
&&  X\equiv \frac{\dot{\phi}^2}{2V}=\frac{1+w_b}{1-w_b}\\
&& Y\equiv \frac{V_\phi}{\dot{\phi}H}=\frac{3}{2}(w-1)
\label{Y}
\end{eqnarray}
Let us also note that for scaling solution.
\begin{eqnarray}
&& -\frac{\dot{H}}{H^2}=\frac{3}{2}(1+w_b)\\
&& \frac{\ddot{\phi}}{\dot{\phi}H}=-\frac{3}{2}(1+w_b)\, . 
\end{eqnarray}
 These equation readily yield the behaviour of scalar field in the scaling regime~\cite{Skugoreva:2019blk,Haro:2019peq},
\begin{equation}
H(t)=\frac{1}{3(1+w_b) t};~~\dot{\phi}\propto t^{-1}\Rightarrow \phi(t)\propto \ln(t)    
\end{equation}
So far, we did not need any information on the field potential that would support the scaling behaviour.
It is interesting to compute the $\Omega_\phi$ for scaling solution that would tell us about the fraction of total energy, $\rho_\phi$ constitutes and also the nature of the potential.
Using Eqs.(\ref{X}) $\&$ and (\ref{Y}), we find,
\begin{equation}
 \Omega_\phi=\frac{\rho_\phi}{3 M^2_{Pl} H^2}=
 \frac{1}{3\Mpl^2}\left(\frac{XY^2}{1-w_b} \right)\left( \frac{V}{V_\phi}   \right)^2   =\frac{3(1+w_b)}{4 \lambda^2_s};~~\lambda_s\equiv -M_{Pl} \frac {V_\phi}{V}
\end{equation}
 where $\lambda_s$ is the slope of the potential. Since  $\rho_\phi$ scales as the energy density of background matter in the scaling regime, $\Omega_\phi=Const$, which in turn implies that the slope of the potential should be constant and consequently, the exponential potential is singled out,
\begin{equation}
\label{Exp}
 V(\phi)=V_0e^{-\alpha\phi/M_{Pl}} ~~~;~\lambda_s=\alpha  
\end{equation} 
 However, an important information is yet missing in our analysis. For instance, if the slope is small, namely, $\alpha^2<2$, slow roll condition applies and we have $w_\phi<-1/3\to \dot{\phi}^2/2<V/2$ and no scaling solution in this case. This brings us to the question of the existence and stability of the scaling solutions we are interested in, which is addressed by dynamical analysis using the autonomous form of evolution equations.
Using the following notations~\cite{Copeland:1997et}\footnote{It might look more natural to
use the variable $X \& Y$ as we know their values for scaling solution. However, in this case, one requires one more variable $A$ which is directly linked to the field $\phi$. But the equation for this variable decouples from the system and we can analyse equations for $X \& Y$ without bothering about the third equation for the variable $A$. These variables are useful in the analysis of asymptotic scaling solutions which occur  in case of steep potentials of variable slope where we have system of three coupled equations for $X, Y$ and $A$.},
\begin{equation}
x\equiv \frac{\dot{\phi}}{\sqrt{6}M_{Pl}H};~~~y\equiv  \frac{\sqrt{V}}{\sqrt{3}M_{Pl} H },
\end{equation}
one can cast evolution equations, with exponential potential, in the autonomous form,
\begin{equation}
x'=f(x,y);~~~ y'=g(x,y)  
\end{equation}
along with the Friedman constraint equation,
\begin{equation}
\frac{\rho_B}{3M^2_{Pl} H^2}+x^2+y^2=1  
\end{equation}
where prime denotes the derivative with respect to $N\equiv \ln(a)$; $f$ and $g$ are functions of $x$ $\&$ $y $ whose functional forms are not required here.
Fixed point of the dynamical system is the one for which, $x',y'=0$ whose stability is checked looking at the sign of eigen values of the perturbed matrix.
Using this analysis, one finds~\cite{Copeland:1997et,Liddle:1998xm,Haro:2019peq},
\begin{equation}
 \text{Scaling solution}:~~~~~~w_\phi=w_m;~~~
 \Omega_\phi=\frac{3(1+w_b)}{\alpha^2}~~~\lambda^2_s=\alpha^2>3   
\end{equation}
which is an attractor of the dynamical system and this completes the story.
In the scaling regime, $\rho_\phi$  constitutes a constant fraction of the total energy density. The fraction is smaller for the larger value of the slope of the potential. In fact, lower bound on the slope is fixed by the nucleosynthesis constraint.
Exponential potential with slope larger than $\sqrt{3}$ is termed as {\it steep} otherwise {\it shallow} which also serves as yardstick for an arbitrary potential. Let us note that field dominated solution ($\Omega_\phi =1$) is also an attractor for $\alpha^2<{3}$ that we are not interested in here. The demarcation is clear: For $\alpha^2>3$, system settles into a scaling regime, otherwise a field dominated solution, which is accelerating if $\alpha^2<2$.

 To summarize, scaling solution follows the background matter being sub-dominant. During radiation era, field energy density scales as radiation energy density ($\rho_r$), namely, $\rho_\phi\sim 
a^{-4}$ which adjusts itself to $\rho_m$ (energy density of cold matter) after matter domination is established and follows it for ever. Scaling regime is characterized by a fixed kinetic to potential  energy ratio, for instance, $\dot{\phi}^2/2=V$ in radiation era whereas, $\dot{\phi}^2/2=2V$ for matter domination. Last but not least, since scaling solutions are non-accelerating, one requires an additional  feature in the potential which would allow exit to late time acceleration.
In this framework, $\Gamma=1$ and $w_\phi=w_b$ for most of the history of Universe and only at late times, transition to acceleration would take place $\hat{\rm a}$  {\it  la}   a {\it perfect tracker }
In the forthcoming discussion, we shall mention exit mechanism discussed in the literature. 
\subsection{Nucleosynthesis Constraint on Extra Species}
\label{BBNs}
One of the major successes of hot big bang includes the  synthesis of light elements, Deuterium ($ ^2\rm {H} $), Tritium  ($^3\rm{H}$) and Helium ($^4\rm{He}$) in the early Universe. 
When temperature in the Universe was around $1~ \rm {TeV}$, all the standard model degrees of freedom were in equilibrium. All of them were relativistic and contributed to radiation energy density~\cite{Kolb:1990vq,Husdal:2016haj},

\begin{eqnarray}
\rho_r=\frac{\pi^2}{30}g_* (T) T^4
\end{eqnarray}
where  $g_*(T)$ denotes the effective number of relativistic degrees of freedom  in the universe at 
temperature T. Relativistic species including
both the bosons and the fermions contribute to $g_*(T)$,
\begin{equation}
 g_*(T)=\sum_{Bosons}{g_B {\left( \frac{T_i}{T} \right)^4 }}+\frac{7}{8}\sum_{Fermions}{g_F {\left( \frac{T_i}{T} \right)^4 }},
\end{equation}
where $T$ is the photon gas temperature and $g_B~ \&~ g_F$ denote bosonic and fermionic degrees of freedom with mass, $m_i \ll T$ ($m_i$ is the mass of a particular species).
The effective degrees of freedom  is defined relative to the photon gas. 
When temperature dropped to around $1$ $MeV$,
the relativistic species included $\gamma $, $e,e^{+}$ and three generations of neutrinos $\nu$ such that, $g_*=10.75$. Weak processes such as, $p+e^{-}\leftrightarrow
 n+\nu_e$ then froze out
leaving behind a definite ratio of $n/p$  densities. Freezing temperature can be estimated comparing the reaction rate $\Gamma_{int}=G_F^2 T^5_f$ with the Hubble parameter $H(T)$~\cite{Cyburt:2015mya},
\begin{equation}
    G_F^2 T_f^5=\frac{g_*^{1/2}}{\sqrt{3}\Mpl} T_f^2 \Rightarrow T_f=\left(\frac{g_*^{1/2}}{\sqrt{3}G^2_F \Mpl}\right)^{1/3}\simeq 1~ MeV
    \label{tf}
\end{equation}
where $G_F$ is Fermi constant and $T_f$ designates the freezing temperature. It should be noted that the $n/p$ density ratio is temperature sensitive, as $n/p \sim Exp(-Q/T_f$), where $Q$ denotes the neutron and proton mass difference. Since $T_f$ is less than the binding energy of Deuterium, its synthesis should naively begin. However, there is more than a billion photons per proton ($\eta=n_B/n_\gamma\simeq 6 \times 10^{-10}$, $n_B$ designated baryon number density) in the universe such that the number of photons whose energy exceeds  $T_f$ is large and Deuterium does not form due to photo-dissociation. This process is delayed and occurs at a lower temperature, $T_{BBN}\simeq 0.1 eV$, which then initiates the chain reaction for Helium formation.
 Helium abundance in Universe  has been measured accurately; it is sensitive to $T_f$, for obvious reasons, it also depends upon $\eta$.
 We should emphasize that $\eta $
is an important input here required for a successful nucleosynthesis.
It is further important to note that bringing in any extra relativistic degree of freedom like a scalar field would affect the numerical value of $T_f$
and consequently the Helium abundance. The latter should, therefore,impose a  constraint on any relativistic degree of freedom in the Universe over and above the standard model of particle physics $-$ {\it nucleosynthesis constraint} which, in particular, applies to often used quintessence as well as relic gravity waves. Indeed, for temperatures under consideration,
\begin{equation}
g_*=2+\frac{7}{8} \left(4+2 N_\nu \right)   
\end{equation}
where the first term is due to photon, first term in the brackets is attributed to electron and positron; $N_\nu=3$
in the standard model. Any extra radiation present in the Universe, can be parametrized through $N_\nu$\cite{Caprini:2018mtu}
\begin{equation}
\Delta \rho_r= \frac{\pi^2}{30}\frac{7}{4}  \Delta N_\nu  T^4
\end{equation}
where $\Delta N_\nu$ is anything over and above the standard model value of $N_\nu$ ($N_\nu=3$)
which should be constrained from the observed primordial  abundances and other data. Let us now assume an extra relativistic degree of freedom in the Universe dubbed "X" with energy density $\rho_X$ which should not exceed $\Delta \rho_r$ as a necessary requirement for not disturbing the Helium abundance~\cite{Caprini:2018mtu},
\begin{equation}
\left(\frac{\rho_X }{\rho_\gamma  } \right)_{1~MeV}\leq \frac{7}{8} \Delta N_\nu
\label{annT}
\end{equation}
where $\rho_\gamma=(\pi^2/15) T^4$
is the energy density for photon gas\footnote{Given that $\Delta N_\nu$ is constrained using primordial abundances and other data,  bound (\ref{annT}) is referred to as a "nucleosynthesis" constraint or "BBN" bound.}.
Let us note that Helium abundance depends upon two parameters, $T_f $ and $\eta$ which can be used to constrain $\Delta N_\nu$. The  constraint can be improved considerably by considering, in addition,  the Deuterium abundance and CMB data~\cite{Cyburt:2004yc}.
However, in this case, we should estimate the the ratio (\ref{annT}) below the $e^{+} e^{-}$ annihilation temperature, $T_{e^{+}e^{-}}$.
After electron positron annihilation, $T_\nu=(4/11)^{4/3} T$ such that,
\begin{equation}
 \left(\frac{\rho_X }{\rho_\gamma  } \right)_{T_{e^{+}e^{-}}}\leq \frac{7}{8} \left(\frac{4}{11}\right)^{4/3} \Delta N_\nu   
\end{equation}

Observational constraint on $\Delta N_\nu$  should put a restriction on the extra degree of freedom not the part of standard model.  
Remember that in scalar field models with tracking behavior, $\rho_\phi \propto a^4 $during the radiation era, and the fraction of total energy that $\rho_\phi$ comprises is determined by the slope of the steep exponential potential responsible for scaling behavior. Using combined data, gives $ N_\nu<3.2$ at $ 95 \% $ confidence level which implies a bound on the slope of the potential\cite{Caprini:2018mtu},
\begin{equation}
 \left(\frac{\rho_X }{\rho_\gamma  } \right)_{T_{e^{+}e^{-}}}\simeq \Omega_\phi|_{T_{e^{+}e^{-}}}=\frac{4}{\alpha^2}  \lesssim\frac{7}{8} \left(\frac{4}{11}\right)^{4/3}\times 0.2 \Rightarrow \alpha\gtrsim 10. 
\end{equation}
It may be noted that in case of thawing models,
 $\Omega_\phi\ll1$ and the $BBN$ constraint is trivially satisfied.
Similarly, for relic gravity waves, $\rho_{GW}$ should satisfy the following bound,
\begin{equation}
 \left(\frac{\rho_{GW }}{\rho_\gamma  } \right)_{T_{e^{+}e^{-}}}\leq \frac{7}{8} \left(\frac{4}{11}\right)^{4/3} \Delta N_\nu  
\end{equation}
which can be translated into a bound on $\Omega_{GW,0}$,
\begin{equation}
 \Omega_{GW,0}=   \frac{7}{8}\left(\frac{4}{11}\right)^{4/3} \Omega_{\gamma,0}\Delta N_\nu \simeq 5.6 \times 10^{-6} \Delta N_\nu
\end{equation}
where we used the fact that $\rho_{GW}\sim \rho_\gamma\sim a^{-4} $ and also used the observed value of fractional energy density of radiation at the present epoch, $\Omega_{\gamma,0}\simeq 2.47\times 10^{-5}$
Using the combined abundances of Helium and Deuterium with the CMB data $(\Delta N_\nu \lesssim0.2$), we finally have,
\begin{equation}
 \Omega_{GW,0} \lesssim 1.12 \times 10^{-6}   
 \label{omegagw0}
\end{equation}
which  puts  an upper bound on the kinetic regime or the lower bound on reheating temperature in models of quintessential inflation, see 
 section \ref{RG} for details.
\subsection{Dynamics with general class of potentials and emergence of scaling behaviour in the asymptotic regime} 
As discussed in the preceding sub-section, scaling solutions are specific to exponential potential, which are either suited to field dominated situations or scaling behaviour but not to both simultaneously. Thus, we might use a potential suitable for inflation that subsequently changes to exponential form soon after inflation ends or think of a non-exponential type of potential capable of addressing the inflationary requirements at an early time and dynamically mimicking the exponential behavior  in the post inflationary era. 

Exponential potential has a distinguished feature, its slope, $\lambda_s\equiv -M_{Pl} V_\phi/V$, is constant. In this case, we have two dynamical variable, $x$ and $y$ and  correspondingly two autonomous equations. For a potential  which not an exponential, slope becomes a variable quantity and one needs an additional dynamical equations for $\lambda_s$\cite{Copeland:2006wr,gammascale1,Nunes:2000yc,Ng:2001hs},
\begin{eqnarray}
&& x'=f(x,y,\lambda_s);~~~ y'=g(x,y,\lambda_s)  \\
\label{sleq}
&& \lambda'_s=-\sqrt{6}\lambda^2_s(\Gamma-1)x\\
&& \Gamma=\frac{V_{\phi\phi} V}{V^2_\phi}
\label{slopeeq}
\end{eqnarray}
where $\Gamma$ is an important field construct which controls the dynamics of the system with  a general type of potential.
 For exponential function, $\Gamma=1$, corresponding to constant slope of the potential making  Eq.(\ref{sleq}) redundant.  In this case, $\rho_\phi$ scales as background energy density $\rho_b$ giving rise to scaling solution which is an attractor of dynamics. Without the loss of generality, we may consider a class of potentials in which field $\phi$ rolls down the potential from smaller to larger values ($\phi\to \infty$) such that $x>0$.
 In case of potentials with $\Gamma>1$, for instance,
 \begin{equation}
  V=V_0 \frac{M^n_{Pl}}{\phi^n};~~\Gamma=1+1/n~(n > 1)   ,
 \end{equation}
  slope decreases with evolution, see Eq.(\ref{sleq}), eventually, giving rise to slow roll and forcing the field to evolve to quasi de Sitter ($\rho_\phi\simeq \textrm{const.}, w_\phi\simeq -1$) which is an attractor\cite{Nunes:2000yc}. While the field is rolling down the steep part of the potential ($\lambda_s$ is large), $\rho_\phi$ closely follows the background~\cite{Steinhardt:1999nw},
 \begin{equation}
 w_\phi=\frac{2 \Gamma}{2\Gamma-1}=w_b+\frac{2} {n}(1-w_b)   
 \end{equation} 
 It should be noted that for generic values of $n$, $w_\phi\simeq w_b$ and $\Gamma\simeq 1$\footnote{Notice that, $w_\phi=w_b$ for $n\to \infty$ which is not surprising as power law  corresponds to exponential in this limit. }.
 However, in this case ($\Gamma>1$), and slope  ($\lambda_s=n/\phi$ ) decreases with evolution such that the field enters the slow roll regime at late stages,
 \begin{equation}
  w_\phi=-1+\frac{\lambda^2_s}{3}   \to -1,~~\lambda_s\to 0
 \end{equation}
 and can account for late time acceleration. Scalar field solution, that mimics the background in the post inflationary era and only at late stages overtakes it to account
 for the observed late time acceleration, has been assigned a nomenclature as {\it tracker}, see Fig.\ref{fig:scaling_stainhert}.
 Obviously, the class of potentials with $\Gamma>1$ are of interest and have been widely investigated in the literature. 
 
 
On the other hand,  if $\Gamma<1$, slope of the potential increases with evolution such that field runs down its potential faster and faster making the potential energy redundant and naturally ends up in the kinetic regime with, $\rho_\phi\sim a^{-6}$.
 In this case, it follows from (\ref{slopeeq}) that slope increases,
 \begin{equation}
  \frac{1}{\lambda_s}=\int{x(\Gamma-1)}dN 
  \label{slopeeq1}
 \end{equation}
 and in process of evolution as $\lambda_s$ becomes large and $\Gamma \to 1$. In this case, system would eventually join the scaling track in the asymptotic regime. We shall demonstrate it in an example. On the other hand, during evolution, $\lambda_s\to 0$ if we begin from $\Gamma>1$ ( $\Gamma \neq Constant$), and in this case $\Gamma$ increases towards infinity. In a sense, "$\Gamma$" defines the type of steepness of the potential: {\it Slope tells us about the steepness of the potential, whereas $\Gamma$ expresses whether, with evolution, steepness is unchanged ($\Gamma=1 $), increases ($\Gamma<1 $), or decreases} ($\Gamma > 1 $). Potentials with $\Gamma<1$ ($\Gamma > 1)$ are sometimes referred to in the literature as steeper (shallower) than the exponential potential for brevity.
 It is interesting to think of slope $\lambda_s(\phi)$ as defined over the smooth set of functions $\{V(\phi\}$. Since,
 \begin{equation}
 \frac{ d\lambda_s}{d\phi}=-\lambda^2_s (\Gamma-1)=0\to  \Gamma=1,
 \end{equation}
  slope, $\lambda_s$, is extremum for an exponential function. The second derivative,
 \begin{equation}
 \frac{d^2\lambda _s}{d\phi^2}=-\frac{1}{V^4_\phi}\left( V^2_\phi(V_{\phi\phi\phi}V+V_{\phi\phi}V_\phi)-2V^2_{\phi\phi}V_\phi V \right)
 \end{equation}
 computed for exponential potentials also vanishes. Thus $\Gamma=1$ is saddle point in the functional space which explains the mentioned features of the dynamics.
 
 It is desirable that for most of the history of Universe, field follows the background   (scaling behaviour) and only at late stages,
  it  exits the scaling regime, takes over the background to become the dominant component and might mimics dark energy. In what follows, we discuss as how to incorporate this feature in the dynamics.
 \subsubsection{Making a perfect tracker from scaling solutions: $\Gamma=1~ \& ~\Gamma<1$ }
  In view of the aforesaid, one needs a steep exponential type of behaviour in the post inflationary era with a mechanism of late time exit to acceleration $\hat{\rm a}$  {\it  la} {\it tracker} not sensitive to initial conditions. In the framework of quintessential inflation, one could design a potential that changes to shallow exponential one suitable to inflation at early epochs and reduces to a steep exponential function in the post inflationary era. It is easier to accomplish this program using non-canonical scalar field action\cite{Geng:2015fla,Wetterich:2013jsa},
  \begin{eqnarray}
  && \mathcal{S}=\int{d^4x\sqrt{-g}\left[\frac{M^2_{Pl} } {2}R-k^2(\phi)\partial_\mu \phi\partial^\mu \phi-V(\phi)\right]}\\
  && k^2(\phi)=\left(\frac{\alpha^2-\tilde{\alpha}^2}{\tilde{\alpha}^2}\right)\frac{1}{1+\beta^2 e^{\alpha \phi/M_{Pl}}}+1\\
  && V=M^4_{Pl} e^{-\alpha \phi/M_{Pl}}   
  \label{Lnc}
  \end{eqnarray}
  where, $\alpha, \tilde{\alpha}$ and $\beta$ are constants to be chosen by 
  the requirement of model building. Let us use the following transformation,
  \begin{eqnarray}
  && \sigma=f(\phi)   \\
  && k^2=\frac{\partial f}{\partial \phi}
  \end{eqnarray}
  to change the action to canonical form,
  \begin{equation}
 \mathcal{S}=\int{d^4x\sqrt{-g}\left[\frac{M^2_{Pl} } {2}R-\partial_\mu \sigma \partial^\mu \sigma-V(f^{-1}(\sigma))\right]}     
  \end{equation}
  The effect of canonicalization is pushed into the potential which we need to to work out in the small and large field limits to demonstrate 
  the viability of the underlying construction to early and late time dynamics. To this effect, we need to express the canonical field $\sigma$ through field $\phi$ for $\tilde{\alpha}<1 $ and $\alpha\gg \tilde{\alpha}$. In the small field limit where we expect inflation, we have\footnote{We skip details here and refer the reader to \cite{Geng:2015fla} for the same.},
 \begin{eqnarray}
 &&    \phi \ll-2 M_{Pl}\ln\beta/\alpha \Rightarrow k^2(\phi)\simeq \frac{\alpha^2}{\tilde{\alpha}}\\
&& \sigma(\phi) \simeq\frac{\alpha}{\tilde{\alpha}}\phi\Rightarrow
 V\propto e^{-\tilde{\alpha} \sigma/M_{Pl}}
 \end{eqnarray} 
 which can give rise to inflation for $\tilde{\alpha}<1$.
 In the large field
 approximation, 
 \begin{equation}
  \phi \gg-2 M_{Pl}\ln\beta/\alpha \to k^2(\phi)\simeq 1\Rightarrow V\propto e^{-\alpha \sigma/M_{Pl}},  \end{equation}
 with a desired scaling behaviour
 in the post inflationary era. It is therefore not surprising that Lagrangian (\ref{Lnc}) can successfully describe inflation at early times and with scaling behaviour in the post-inflationary era.

In the canonical formulation,  an interesting possibility is provided by a class of non-exponential type of potentials flat initially followed by steep behaviour which is exponential like in the asymptotic regime, $\Gamma\to 1$ for large values of the field.
  To this effect, it is interesting to examine a class of generalized exponential potentials\cite{Geng:2015fla},
  \begin{equation}
  \label{potn}
 V(\phi)=V_0e^{-\alpha \phi^n/M_{Pl}^n};~~~n>1   
  \end{equation}
which can support slow roll ($\lambda_s=\alpha \phi^{n-1}$) at early times  when $\phi$ is small followed by a steep
behaviour after inflation ends. In this case, slope ($\lambda_s(\phi)$) is a dynamical variable whose evolution is dictated by $\Gamma$,
\begin{equation}
\Gamma(\phi)=1-\frac{n-1}{n\alpha}    \left(\frac{M_{Pl}}{\phi}\right)^n
\label{gamman}
\end{equation}
which approaches one from below for large values of $\phi/M_{Pl}$. Before the asymptotic region is reached, $\Gamma<1$ and (\ref{potn}) exhibits the properties of a function steeper than the exponential potential pushing the field into kinetic regime. However, for large values of the field (\ref{potn}) dynamically mimics the behaviour of an exponential potential ($\Gamma\to 1, \phi\to\infty$) allowing the field to finally catch up with the background. 
Does the field continue in kinetic regime when  $\Gamma<1$~?
 In what follows, we describe a phenomenon that occurs when the field $\phi$ runs down a steep potential in the presence of background matter density. Thereafter, we shall be able to narrate the complete story of field dynamics
in a steep potential.
\subsection{Slow roll in presence of background matter: Freezing regime}
The formalism of slow roll parameters can be applied to quintessence, keeping in mind that the Hubble rate is not solely defined by scalar field energy density, background matter density also contributes to it. 
\begin{figure}
     \centering
     \begin{subfigure}[b]{0.44\textwidth}
         \centering
         \includegraphics[width=\textwidth]{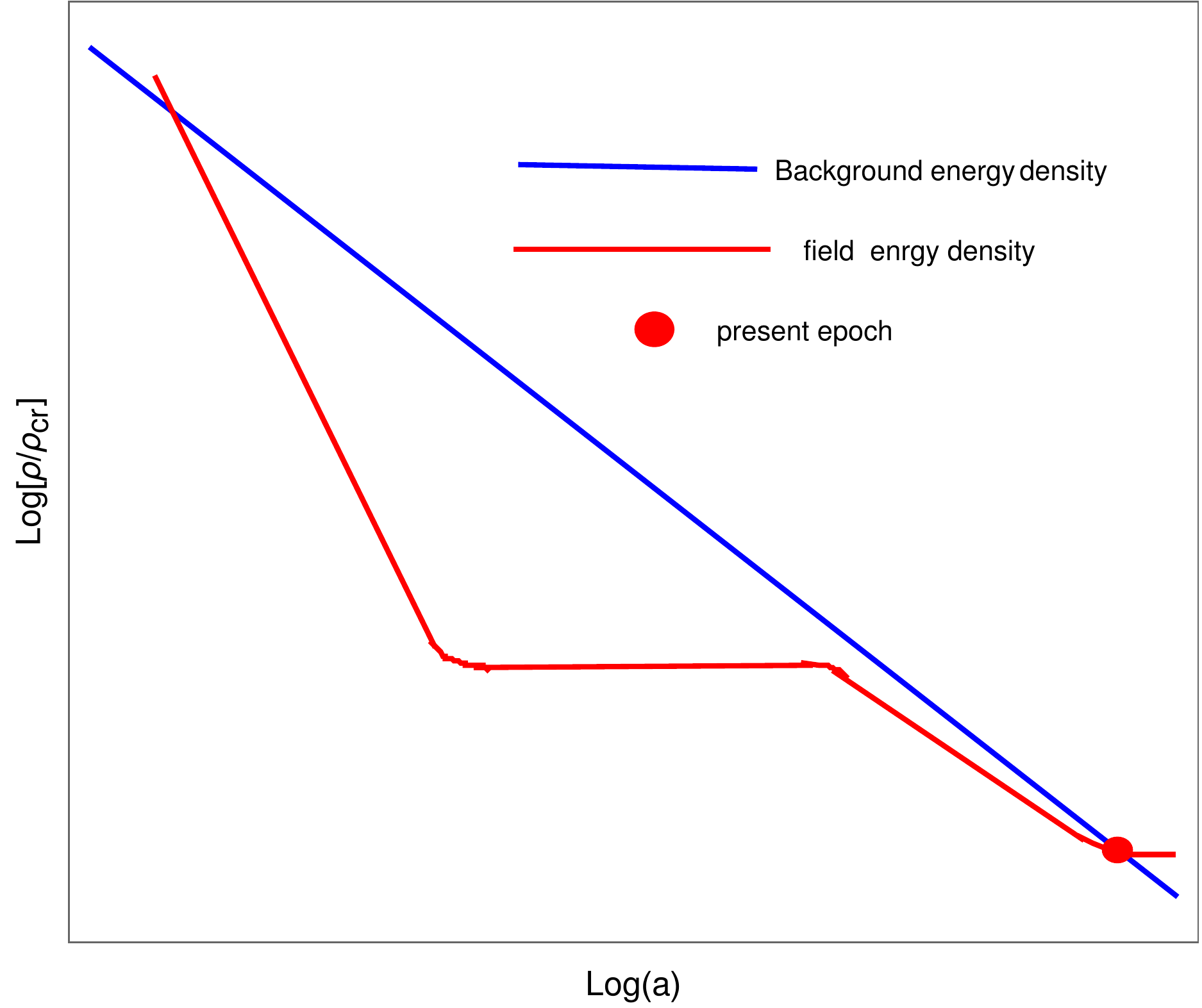}
         \caption{tracker}
         \label{fig:scaling_stainhert}
     \end{subfigure}
     \hfill
     \begin{subfigure}[b]{0.47\textwidth}
         \centering
         \includegraphics[width=\textwidth]{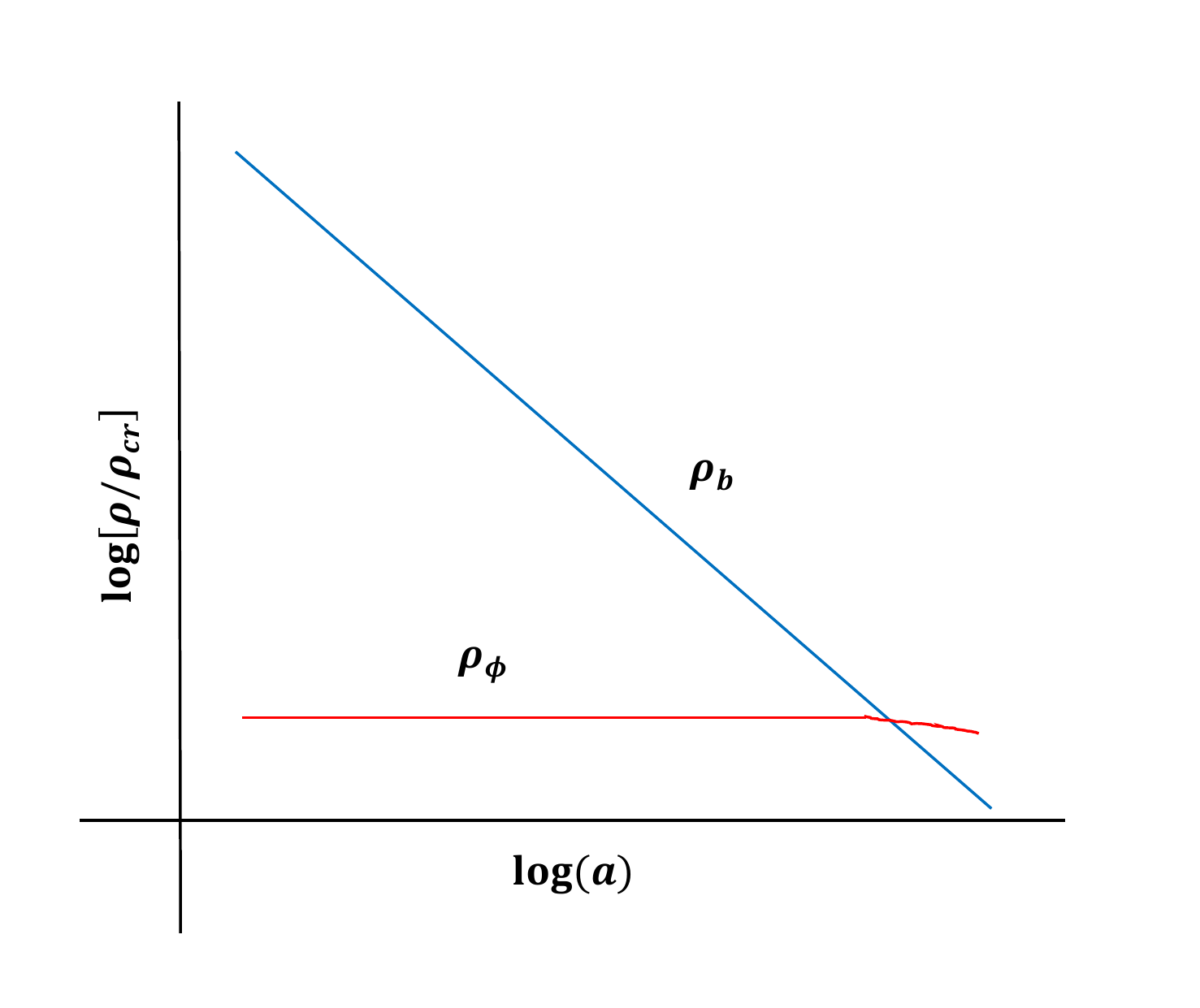}
         \caption{Thawing}
         \label{figthwaing}
     \end{subfigure}
    \caption{ Fig(a) shows evolution of field energy density $\rho_\phi$ and background matter density (radiation/cold matter) versus $\ln( a )$ for a potential  shallower than the exponential potential ($\Gamma>1$), for instance, inverse power law potentials, $V\sim \phi^{-n}, n>0$ with $\Gamma=1+1/n$. After the recovery from freezing regime, field yet on the steep part of potential evolves with equation of state parameter close to that of the background. In this case, slope of the potential gradually decreases and scalar field slowly moves towards the background and overtakes it and joins the slow roll with diminishing value of the slope $\hat{\rm a}$  {\it  la} a {\it tracker}. Fig.(b) shows schematic representation of thawing quintessence where the field is frozen on a flat potential due to Hubble damping. When $\rho_\phi$ approaches the background energy density at late times, the field resumes slow roll and accounts for late time acceleration.}
\end{figure}



\begin{figure}[H]
\begin{center}
$%
\begin{array}{c@{\hspace{.1in}}cc}
\includegraphics[width=2.8 in, height=2.8 in]{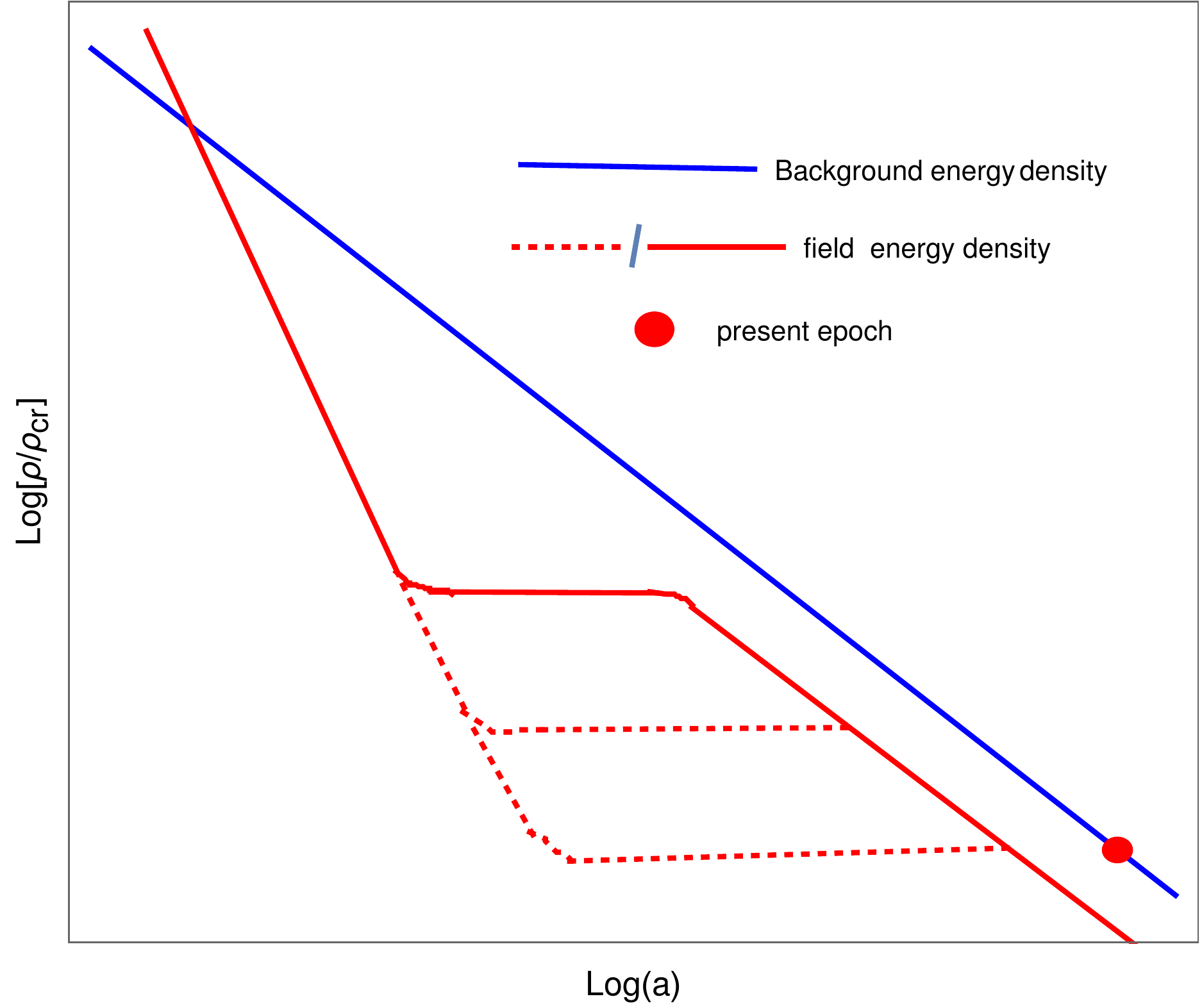} & %
\includegraphics[width=2.8 in, height=2.8 in]{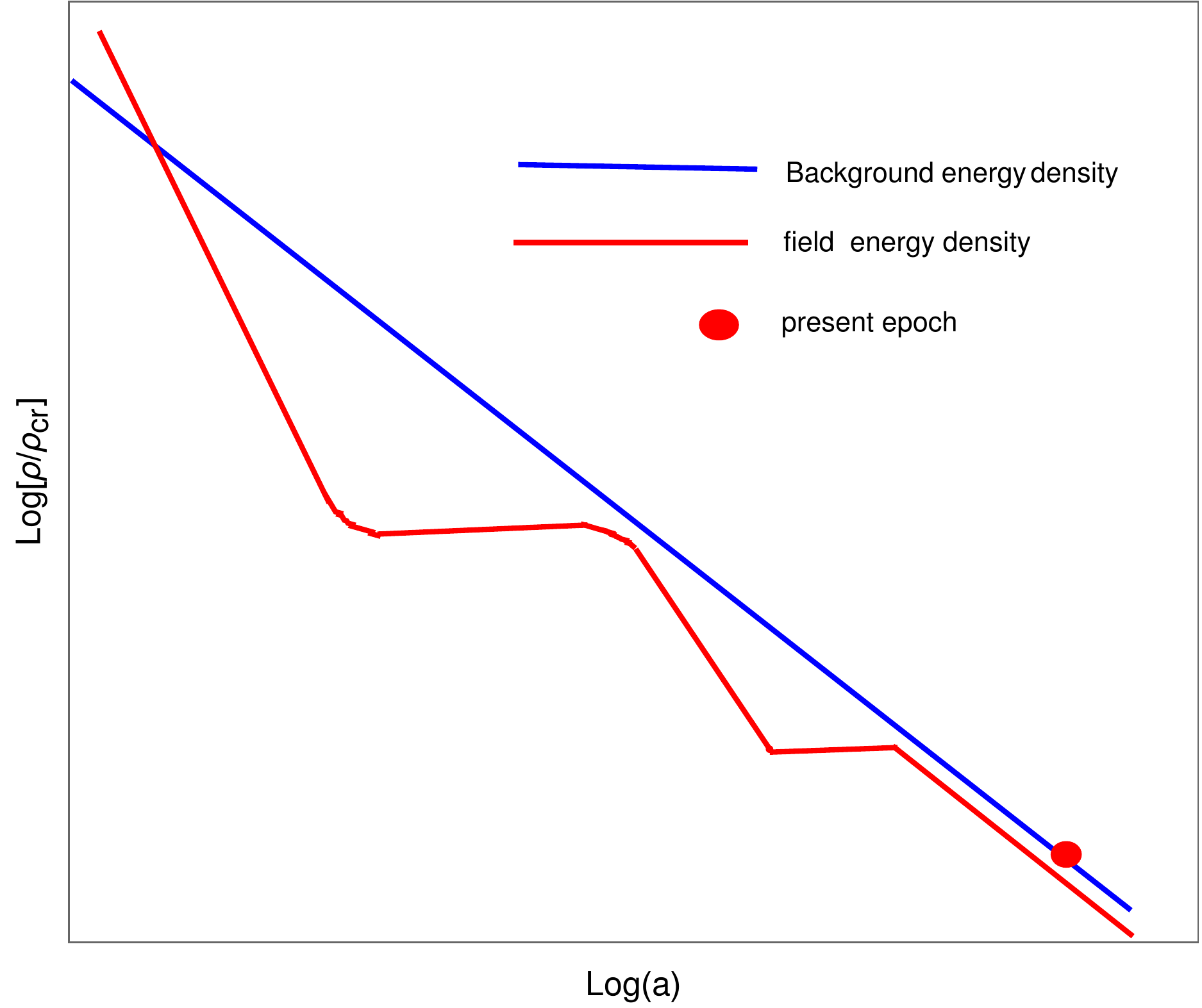} \\%
\mbox (a) & \mbox (b) &%
\end{array}%
$%
\end{center}

\begin{center}
$%
\begin{array}{c@{\hspace{.1in}}cc}
\includegraphics[width=2.8 in, height=2.8 in]{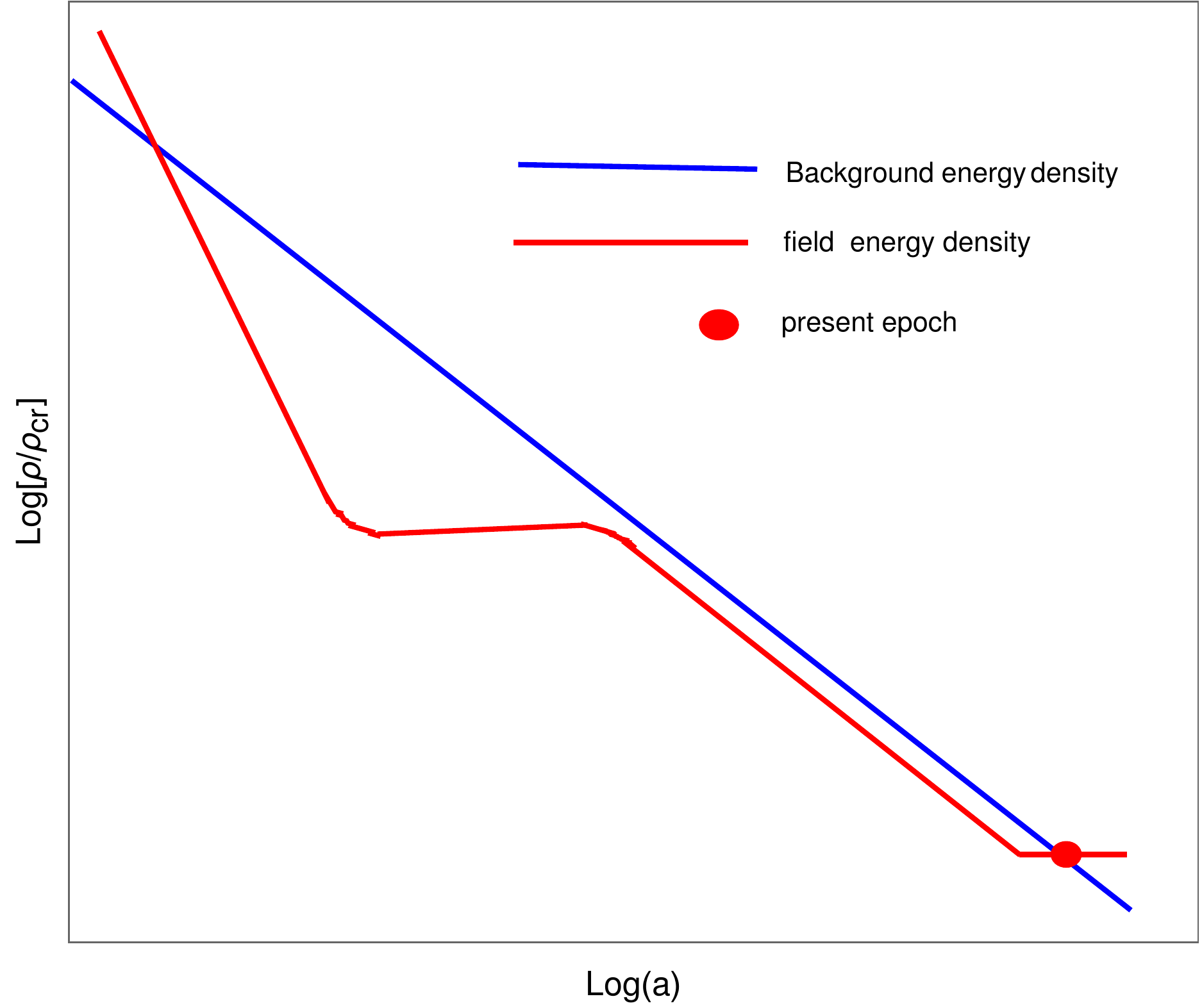} & %
\includegraphics[width=2.8 in, height=2.8 in]{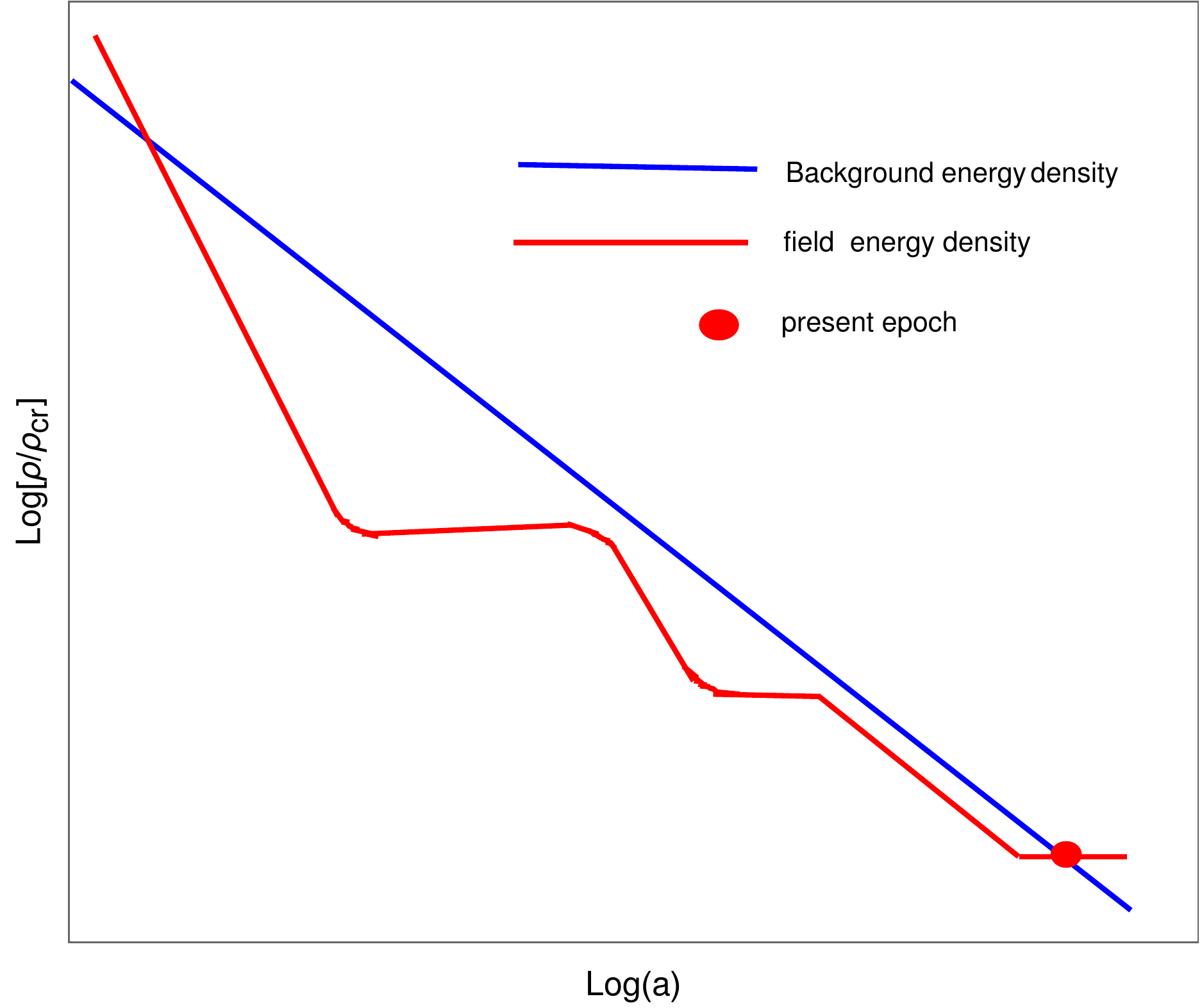} \\
\mbox (c) & \mbox (d)%
\end{array}%
$%
\end{center}

\caption{{\small  Figure shows the qualitative behaviour of $\rho_\phi$ along with the energy density of background matter (cold matter/radiation) versus $\ln a$  for a steep potential. As $\rho_\phi$  overshoot the background energy density ($\rho_\phi\ll\rho_b$  $\rho_b$ designates background matter energy density), field freezes on its potential due to Hubble damping. After the recovery from freezing, field evolution crucially depends upon the nature of steepness of potential. (a) In case of the exponential potential ($\Gamma=1$), field catches up with the background and tracks it for ever. Fig.(b)  exhibits the general feature of scalar field dynamics for a  field potential steeper than the exponential potential ($\Gamma<1$): after recovery from freezing, $\rho_\phi$ evolves in steps (down and right) and eventually catches up with the background\cite{Geng:2015fla}
$-$ asymptotic scaling solution.
Figs. (c) $\&$ (d) show exit from   scaling and asymptotic scaling regimes to late time acceleration at the present epoch $\hat{\rm a}$  {\it  la} a {\it perfect tracker}. Exit is triggered due a  late time feature in the runaway potential that makes it shallow at late stages of evolution. }}
\label{evolutiontest}
\end{figure}

To be fair, unlike in the case of inflation, slow roll parameters might be useful here only for a broad perspective. 
An interesting feature of field dynamics is noticed
for field rolling down a steep potential when its energy density is negligible compared to background energy density. In this case, field freezes on its potential and waits there for conducive situation to resume motion. Indeed, in case of slow roll\footnote{It should be noted that, in the present situation, $-\dot{H}/H^2=(1+w_b)/2$, is not related to slow roll parameter due to the presence of background matter. In case of slowly rolling quintessence, the friction term  need not to be large and $\beta$ may not be negligible. For tracker models, $|\beta|\ll1$, however, for thawing quintessence, $\beta=\mathcal{O}(1)$ and $\beta$ is nearly constant, $\dot{\beta}/\beta H\ll1$. Unlike the thawing case, the consistency of slow roll gives rise to, $\eta=3(1+w_b)/2=3/2$ in case of the trackers.}
\begin{equation}
 \dot{\phi}^2/2\ll V\Longrightarrow\epsilon=3\frac{\dot{\phi}^2/2}{\rho_\phi}   \simeq \frac{3}{2} \frac{\dot{\phi}^2}{V}=\frac{1}{6} \frac{V^2_\phi}{V H^2},
\end{equation}
where we used the slow roll value of $\dot{\phi}$.
In what follows, we shall be interested in the case when $\rho_\phi\ll\rho_b$
such that, $H^2\simeq \rho_b/3 M^2_{Pl}$,
\begin{eqnarray}
\label{slowmatter}
&&\epsilon=\frac{M^2_{Pl}}{2}\left(\frac{V_\phi}{V}\right)^2 \left(\frac{\rho_\phi}{\rho_B}\right)\equiv\epsilon_0 \left(\frac{\rho_\phi}{\rho_B}\right)\\
&& \epsilon=\frac{\lambda^2_s}{2}\left( \frac{\rho_\phi}{\rho_b}\right)
\label{slowmatter1}
\end{eqnarray}
Since, we are dealing with a steep potential, $\epsilon_0>1$, but interestingly, $\epsilon\ll1$ as $\rho_b\gg V(\phi)$ $-${\it Hubble damping}. Thus, unlike the case of inflation, slow roll becomes operative here due to Hubble damping. However, this is not useful for dark energy as $\rho_\phi\ll\rho_b$ in this case, but it is certainly useful for the understanding of scalar field evolution in a steep potential\footnote{Let us note that this feature is central to thawing models where field is frozen on a shallow potential such that field begins slow roll after it covers from Hubble damping and accounts for late time acceleration. Initial conditions are set specially or tuned allowing it to happen around the present epoch and model parameters are chosen to comply with observation}.
As a result of Hubble damping, field gets frozen on its potential and waits there 
till the ratio, $\rho_\phi/\rho_b$ acquires a specific numerical value
($\rho_b\sim a^{-3(1+w_B)}$,
 $\rho_\phi=Const$) given by Eq.(\ref{slowmatter1})
 when slow roll
is violated ($\epsilon=1$). Field evolution then commences but, hereafter, field dynamics, crucially depends upon the type of steepness of the potential. We shall come back to this point in the sub-section on post inflationary evolution.
\subsubsection{Recovery from freezing regime}
For the sake of illustration, let us focus on a steep exponential potential.
After recovering from the freezing regime, which happens due to slow roll violation, scalar field
joins the scaling track. But when it happens, crucially depends upon the numerical value of the slope of the potential. Indeed, slow roll (\ref{slowmatter}) is violated when,
\begin{equation}
 \epsilon\simeq 1\Rightarrow \left(\frac{\rho_\phi}{\rho_b}\right)\simeq \frac{2}{\alpha^2};~~\lambda^2_s=\alpha^2>3 , \end{equation}
 which tells us that  larger is the numerical value of $\alpha$, smaller is the ratio $\rho_\phi/\rho_b$, when field resumes evolution. Actually, when slow roll ends, scalar field begins running down the steep potential, its kinetic energy gradually increases to scaling value\footnote{$\dot{\phi}^2/2=2V, V$ for radiation  and matter domination respectively.}
 and it joins the scaling regime, it has no other option as scaling solution is an attractor in this case.
  Hence, as the field recovers from freezing and resumes evolution, it always finds itself on the entry to scaling regime irrespective of the numerical value  of the ratio, $\rho_\phi/\rho_b$.
 As a result, even if the field energy density constitutes a negligible fraction of the total energy,  scalar field joins the scaling track. In fact, this phenomenon has simple physical meaning, namely, a particle can move in a medium despite large friction if force acting  on it (gradient of the 
 force field potential) is large.
 This remark about the interplay between the slope of potential and ratio of field to background matter energy densities is specially important in relation to asymptotic scaling solutions that we shortly discuss.
 
 \subsubsection{Slowly rolling quintessence}
 Though slow roll parameters do not play a similar roll for quintessence, they are still helpful for a general perspective. The slow roll conditions
should be used with caution here because, unlike inflation, the Hubble parameter receives contribution from background matter in this case.
 Quintessence models can be classified as: Thawing models~\cite{Caldwell:2005tm,Linder:2006sv}(see also, \cite{Scherrer:2007pu,Linder:2015zxa})   and Tracker models~\cite{Steinhardt:1999nw,Urena-Lopez:2020npg}. In the first category, the
scalar field is initially frozen  on a shallow potential due to the Hubble
damping. Field recovers from freezing regime ($w_\phi=-1$) at late stages and rolls slowly along a shallow potential increasing
the equation of state parameter 
to the level consistent with observation, see Fig.\ref{figthwaing}. As for the trackers, after recovery from the freezing regime, the field (approximately) follows the background and only at late times does it quit to join the slow roll due to the shallow nature of the potential around the present epoch, see Fig.\ref{fig:scaling_stainhert}. For thawing models, $1/2<\beta<3/2$ and friction term is not operative here where as friction term is large, $\beta \ll1$, in case of trackers\cite{Caldwell:2005tm}. We shall assume that $\beta$ does not change much during slow roll in thawing case. In terms of $\beta$, scalar field equation (\ref{Fieldeq}) can be cast as (see \cite{Chiba:2009sj} for details),
\begin{equation}
 \dot{\phi}=-\frac{V_\phi}{3(1+\beta)H}   
 \label{Feqbeta}
\end{equation}
We than ask for the consistence of $\beta$ being $\mathcal{O}(1)$  (thawing models) or $\beta\simeq 0$ (trackers) with the field equation (\ref{Fieldeq}). Indeed, differentiating Eq.(\ref{Feqbeta})  with respect to time, we have,
\begin{equation}
 -\frac{\beta V_\phi}{1+\beta}=\ddot{\phi}\simeq -\frac{(1+w_b)}{2(1+\beta)}+\frac{V_{\phi\phi} V_\phi}{9 (1+\beta)^2H^2} \Rightarrow \beta\simeq \frac{(1+w_b)}{2}-\frac{V_{\phi\phi}}{9(1+\beta)H^2}
 \label{Cq}
\end{equation}
where we have ignored the term $\dot{\beta}/\beta H$ and used $-\dot{H}/H^2\simeq 3(1+w_b)/2 $. In thawing case, $\beta$ is assumed to be approximately constant and since the second  term on right hand side of (\ref{Cq}) is time dependent, it is necessary for the equality to hold that,
\begin{equation}
 |\eta|\ll1 \Rightarrow \beta=\frac{3}{2}(1+w_b)   
 \label{etat}
\end{equation}
where we used the fact that $\beta=\mathcal{O}(1)$. We find that our assumption that $\beta$ is constant is consistent with field equation.
On the other hand, $|\beta| \ll 1$ for trackers and we find the consistency condition,
\begin{equation}
 \eta=\frac{3}{2}(1+w_b)   
\end{equation}
which  gives an order of magnitude estimate for the mass of quintessence, namely, $m \simeq  H_0\sim 10^{-33}~\rm {eV}$.

\subsubsection{The asymptotic scaling solution}
We argued that the class of potentials with $\Gamma<1$ mimic
 scaling behaviour in the asymptotic regime. The  generalized exponential potential (\ref{potn}),
 shallow initially($\phi/M_{Pl}\ll1$) with asymptotic features of an exponential potential   in the post inflationary region, is a suitable candidate for quintessential inflation.
As $\Gamma\to 1$ for large values of $\phi/M_{Pl}$, we expect scaling regime in the asymptotic region. We need to set autonomous system with suitable choice of variables, the choice for $X \& Y$ is obvious, the third variable should explicitly include the information about asymptotic nature of scaling solution\cite{Skugoreva:2019blk},
\begin{eqnarray}
\label{XY}
&&  X\equiv \frac{\dot{\phi}^2}{2V};~~~ Y\equiv \frac{V_\phi}{\dot{\phi}H}\\
&& A\equiv\frac{1}{\phi/M_{Pl}+1}
\label{XYA}
\end{eqnarray}
Let us cast the equation of motion in autonomous form in case of potential (\ref{potn}) using variables defined in (\ref{XYA}) $\&$ (\ref{XY}),
\begin{eqnarray}
  && \frac{dx}{dN}=-2X(3+Y+XY)\\
&& \frac{dY}{dN}=2XY^2\left(1-\frac{(n-1)A^n}{\alpha n(1-A)^n}     \right)+Y(9/2+Y+3 w_b/2)-  \nonumber \\
&&~~~~~~~~~~~XY^3A^{2n-2}(X(w_b-1)+w_b+1)\left[\frac{A-1}{\alpha n(1-A)^n}   \right]^2\\
&& \frac{dA}{dN}=\frac{2XYA^{n+1}(1-A)}{\alpha n(1-A)^n}
\label{XYAA}
\end{eqnarray}
Let us also express other quantities of interest through autonomous variables,
\begin{eqnarray}
&&  \Gamma=1-\frac{(n-1)A^n}{\alpha n(1-A)^n}  \\
&& w_\phi=\frac{X-1}{X+1}\\
&&w_\phi=\frac{2}{3}(X+1)XY^2A^{2n-2}\frac{1}{\alpha^2 n^2(1-A)^{2n-2}}
\end{eqnarray}
Since $\Gamma=1$ for $A=0$ ($\phi\to \infty$), system should mimic the properties of exponential potential, thereby, we expect scaling behaviour in the asymptotic regime. Indeed, we find the critical point,
\begin{equation}
X=\frac{1+w_b}{1-w_b} ;~~Y=\frac{3}{2}(w_b-1):~~A=0  
\label{sasymp}
\end{equation}
as expected. It was demonstrated in reference \cite{Skugoreva:2019blk} that the asymptotic scaling solution (\ref{sasymp}) is an attractor. It should be noted that in case of $n=1$, the third Eq.(\ref{XYAA}) decouples from the other two equations which can be solved without the reference to (\ref{XYAA}) which encodes the information about asymptotic condition and we recover standard result
corresponding to exponential potential.
It is important  to note that unlike the case of exponential potential, here,
we can not infer time dependence of physical quantities in scaling regime by  the critical values in (\ref{sasymp}) as 
the fixed point is reached
in the limit, $t\to \infty; \phi\to \infty$.
Substituting the critical point in the expressions for $\Gamma,~ w_\phi~ \& ~\Omega_\phi$, we have
\begin{eqnarray}
\text{Asymptotic scaling solution}:~~~\Gamma\to 1;~~w_\phi \to w_m:~~\Omega_\phi\to 0
\end{eqnarray}

\begin{figure}
    \centering
    \includegraphics[height=7cm, width=9cm]{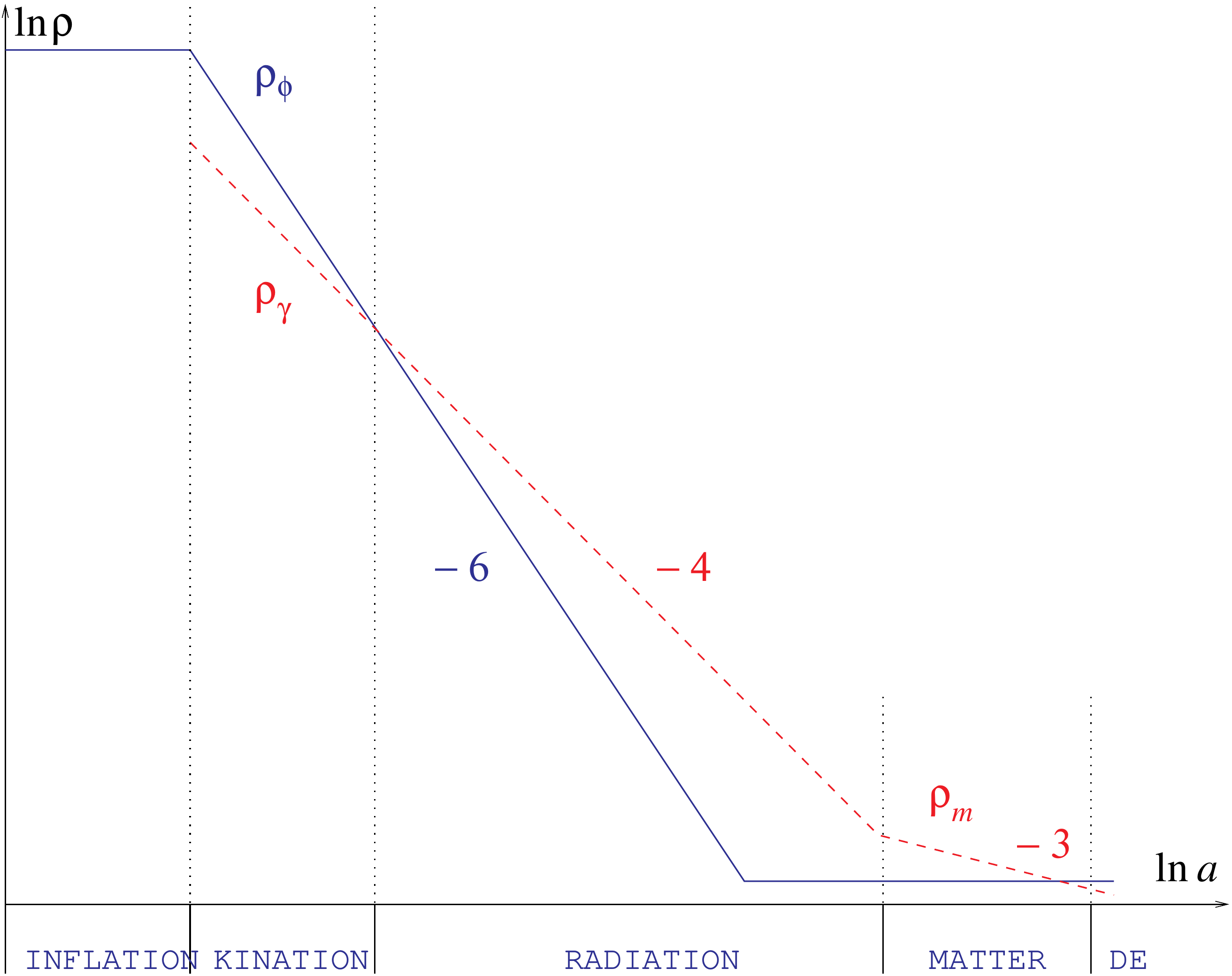}
    \caption{Qualitative behaviour of $\rho_\phi$ for the potential used in Ref.\cite{Dimopoulos:2017zvq} ($V\sim \exp\left[1-\tanh(\phi)\right]-1$) which is shallow initially and at late stages and steep in between. In this case, overshoot is deep enough that the field freezes at the shallow part of the potential. After recovering from Hubble damping, it rolls slowly and adheres to late time acceleration (figure is adapted from Ref.~\cite{Dimopoulos:2017zvq}).  }
    \label{plotdimoalphaattractor}
\end{figure}

In this case, passage to
fixed point can be studied numerically. To analytically understand the field dynamics, we need an anzats in the asymptotic regime whose consistency with the field equation should be checked. We have the following anzats in the asymptotic regime($t\to \infty$)\cite{staro} (see \cite{Scherrer:2022umm} for comparison),
\begin{equation}
\alpha\left(\frac{\phi(t)}{M_{Pl}}\right)^n=f_0 \ln\left(\frac{t}{t_1}\right)\ +f_1\ln\left[\ln \left(\frac{t}{t_1}\right)\right]+..  
\label{anzats}
\end{equation}
Substituting  (\ref{anzats}) in the field equation (\ref{Fieldeq} and keeping in mind the asymptotic behaviour, we find,
\begin{equation}
f_0=2;~~  f_1=\frac{2(n-1)}{2};~~t^2_1 =\frac{M^2_{Pl}(1-w_b)} {V_0 n^2(1+w_b) \alpha^{2/n}2^{(n-1)/n}},
\end{equation}
which allows us to compute important parameters in the asymptotic regime  showing passage to the critical point (\ref{sasymp}),
\begin{eqnarray}
&& w_\phi=w_b+(w^2_b-1)\frac{(n-1)^2}{n^2}  \frac{\ln\left[\ln \left(\frac{t}{t_1}\right)\right] } { \ln\left(\frac{t}{t_1}\right)}\to w_b,~t\to \infty\\
&& \Omega_\phi=\frac{3(1+w_b)}{4 n^2}\left(\frac{2}{\alpha}\right)^{2/n}\left[ \ln\left(\frac{t}{t_1}\right)\right]^{2(1-n)/n}\to 0,~~t\to \infty~~(n>1)
\end{eqnarray}
which obviously reduce to the standard results for $n=1$. We have, therefore demonstrated that the scalar field system in case of the generalized exponential potential (\ref{potn}) with $\Gamma<1$ 
eventually comes to the scaling track in the asymptotic regime. We also analytically demonstrated  the passage to the critical point.
We had argued that scalar field
with an arbitrary steep potential with $\Gamma<1$ should ultimately join the scaling track. In view of quintessential inflation, it is desirable that  $\Gamma\simeq 1$  for quite some time while $\phi$ is rolling on steep part of the potential\footnote{One should be little careful here with regard to $\Gamma$ which is one for exponential potential irrespective of its being steep or shallow.} and $\rho_\phi $ follows the background energy density before exiting to acceleration to mimic tracker like behaviour. Depending upon the underlying field potential,
approximate scaling regime  might be invisible if overshoot of background energy density is so deep that field freezes on its shallow part and after recovery from Hubble damping it enters the slow roll mimic dark energy $-$ {\it thawing realization}, see Fig.\ref{plotdimoalphaattractor}. Obviously, no exit mechanism is required in this model but late time physics is sensitive to initial conditions. Let us once again reiterate that
in case of the generalized exponential potential potential  (\ref{potn}), with an exit mechanism, we have the desired situation to realize a perfect tracker. 

Using the above given technical information on scalar field dynamics,
we would be in position to narrate scalar field dynamics in the presence of background in the post inflationary era. This would give us right perspective of unification of inflation with late time acceleration without interfering with the thermal history of Universe known to a good accuracy.
\section{Post inflationary dynamics: The exit mechanism via coupling with massive neutrino}
\label{postdyn}

In the preceding subsections, we described scalar field dynamics with and without the presence of background energy density $\rho_b$. We argued that the runaway potential generic to the underlying framework should be steep in the post-inflationary era. The energy density of radiation produced at the end of inflation $\rho^{end}_r$ in this framework is typically several orders of magnitude lower than $\rho^{end}_\phi$, and the radiative regime takes a long time to begin. After inflation ends, the field rolls down the steep potential fast, making the potential energy redundant. As a result field evolves in the kinetic regime ($\rho_\phi\sim a^{-6}$) and eventually undershoots the background such that
$\rho_\phi\ll\rho_b$. When this happens, field enters the slow roll due to Hubble damping  and freezes ($\rho_\phi=const$)  on its potential (see, Eq.(\ref{slowmatter})) and waits there till the background energy density ($\rho_b\sim a^{-m}$; m=4, 3 for radiation and matter respectively) becomes comparable to $\lambda^2_s \rho_\phi/2$, slow roll is then violated and evolution of field resumes. Hereafter, field dynamics crucially depends upon the nature of steepness of the potential which is dictated by $\Gamma$. For exponential potential $\Gamma=1$, scalar field would follow the background as scaling solution is an attractor in this case, see Fig.\ref{evolutiontest}(a). If  $\Gamma<1$, the slope of the potential increases giving rise to the increase in kinetic energy,
field energy density would then move away from the background and evolve in the kinetic regime ending up in the freezing regime when $\rho_\phi\ll\rho_b$ and so on and so forth. However, slope all the time increases during evolution and when it reaches large values, $\Gamma$ then approaches its scaling value, $\Gamma=1$, see Eq.(\ref{slopeeq1}). As a result, system eventually joins the scaling track, see Fig.\ref{evolutiontest}(b). In case of generalized exponential potential (\ref{potn}), we demonstrated that scaling solution is an  attractor in the asymptotic regime. In this case, $\lambda_s\sim (\ln(t))^{(1-1/n)}$, $\Omega_\phi\to 0$ and $w_\phi\to w_b$ for $t\to \infty$. Obviously, in this case, there is no issue related to nucleosynthesis. It should be noted that the nucleosynthesis constraint imposes a lower bound on $\alpha$ for exponential potentials corresponding to a small fraction of field energy density, which is automatically true for asymptotic scaling solutions as $\Omega_\phi\to 0$ in this case.

If $\Gamma > 1$ (inverse power law potentials), the slope of the potential $\lambda_s$ decreases when the field rolls along the steep part of the potential with a large value of $\lambda_s$, it evolves with an equation state $w_\phi$, which is close to the equation of state of background matter for generic values of $n $ $-$ {\it approximate scaling behavior}. However, $\lambda_s$ decreases with evolution, causing the field energy density to gradually move towards the background and eventually overtake it at late times; the field then enters a slow roll with a diminishing slope ($w_\phi\to-1$) that accounts for late time acceleration, see Fig.\ref{fig:scaling_stainhert}. Setting the model parameters appropriately, one makes the framework consistent with observation. The change of initial conditions related to scalar field dynamics, within a broad range changes the time for field entry into the (approximate) scaling regime, leaving the late time physics unchanged. A remark about "generic" initial conditions is in order. One can choose extraordinary field initial conditions, giving rise to a deep overshoot, allowing the field to recover from Hubble damping only at late times in the region of slow roll, such that the observational consistency is guaranteed. Late time physics, in this case, would become sensitive to initial conditions, necessitating a re-adjustment of the model parameters to ensure observational consistency$-$ {\it thawing realization}.

 In the case of the generalised exponential potential, an interesting dynamical feature is observed, where $\rho_\phi$ initially evolves in steps (down and forward) and eventually catches up with the background in the asymptotic region as $\Gamma\to 1$.
Let us once again reiterate that the scaling, or at least asymptotic scaling behaviour in the post-inflationary era, is required for viable cosmological dynamics. Secondly,  since the scaling solution is an attractor which is non-accelerating, one needs a late time feature in the potential  to mimic slow roll allowing  exit to late time acceleration dubbed a perfect tracker,
see Figs.\ref{evolutiontest}(c) $\&$ (d).

 In what follows, we shall analyze an exit mechanism based upon scalar field coupling to massive neutrino matter which modifies the scalar field potential. The coupling is effective only at late stages of evolution, where the effective potential has a minimum allowing the field to settle there at a late times and an exit to a quasi de Sitter phase is realized.
 
The late time exit for steep run away potential can be realized by non-minimal coupling of the quintessence field to  matter which modifies the runaway potential such that the effective potential has minimum (see Refs.~ \cite{Barreiro:1999zs,Bartolo:1999sq,Saridakis:2010mf} for other kind of exit mechanism). The conformal coupling of field to matter under consideration is proportional to the trace of energy momentum tensor of matter and vanishes identically during radiation. Coupling becomes operative after matter domination is established giving rise to effective potential that has minimum which is right feature in the  wrong time. In general, minimum in the potential where field can settle down leads to de Sitter like behaviour and is unwanted in this case, it can spoil the matter phase, it is more than desirable to leave the matter phase intact. Problem finds its solution in the coupling to massive neutrino matter which acquires non-vanishing value only at late times when massive neutrinos become non-relativistic~\cite{Wetterich:2013jsa,Wetterich:2013wza,Hossain:2014xha,Geng:2015fla}.


Let us consider the action which describes coupling between massive neutrino matter and  scalar field\cite{Sami:2021ufn,Gumjudpai:2005ry},\\
 \begin{eqnarray}
\mathcal{S}=\int d^4x \sqrt{-g}\left[ \frac{\Mpl^2}{2}R -\frac{1}{2} \partial_\mu \phi \partial^\mu \phi -V(\phi)\right]+ \mathcal{S}_m +\mathcal{S}_r +\mathcal{S}_\nu \left( A^2(\phi) g_{\alpha\beta}, \Psi\nu\right)\, ,
\label{nuaction}
\end{eqnarray}
where, $\mathcal{S}_m$ , $\mathcal{S}_r$ and  $\mathcal{S}_\nu $ are actions for cold matter, radiation  and neutrino matter respectively. $g_{\mu\nu}$ and  $\Tilde{g}_{\mu\nu}\equiv A^2(\phi)g_{\mu\nu}$
designate the Einstein and Jordan metrics.
It is important to note here that the metric for the neutrino action is different from the rest (including matter and gravity part). One should, therefore, be careful while deriving the relevant quantities from this action. In this section we shall explicitly derive the continuity equation using (\ref{nuaction}) for neutrino matter and scalar field in the Einstein frame.
 Let us note that in the Jordan frame neutrino matter is minimally coupled and its energy momentum tensor obeys the standard conservation law. 
In the Jordan frame, energy momentum-tensor for neutrino matter is given by,
\begin{eqnarray}
\Tl{T}_{\mu\nu}=\frac{-2}{\sqrt{-\Tl{g}}}\frac{\del{\mathcal{S}}}{\del{\Tl{g}_{\mu\nu}}}=\begin{pmatrix}
\Tl{\rho} & 0 & 0 & 0 \\
0   &    &    &   \\ \, .
0 &  &   \Tl{a}^2 \Tl{p} &\\
0 &  &    &    
\end{pmatrix}
\end{eqnarray}
Or equivalently,
\begin{eqnarray}
\Tl{T}^\mu~ _{\nu}=\rm {diag}\left(-\Tl{\rho}, \, \Tl{p}, \, \Tl{p},\,  \Tl{p}\right)\, ,
\end{eqnarray}
which satisfies the standard continuity equation, 
\begin{eqnarray}
&&\Tl{\nabla}_\mu \Tl{T}^{\mu\nu}=0\, \\
&& \dot{\Tl{\rho}}+3 \Tl{H}\left( \Tl{\rho}+\Tl{p}\right)=0\, .
\label{cont1}
\end{eqnarray}
Since the Einstein and Jordan metrics are related via the conformal factor $A^2(\phi)$, we have the following relations,
\begin{eqnarray}
g_{\mu\nu}&=& A^{-2} \Tl{g}_{\mu\nu}\nonumber\\
g^{\mu\nu} &=& A^2 \Tl{g}^{\mu\nu}\nonumber\\
\sqrt{-g} &= & A^{-4} \sqrt{\Tl{g}}\\\label{ident1}
a &=& A^{-1} \Tl{a}\nonumber\\
dt &=& A^{-1} \Tl{dt}\nonumber\, .
\end{eqnarray}
Let us note that the energy-momentum tensor in the Einstein frame with metric $g_{\mu\nu}$ is related to its counter part in the Jordan frame, \\ 
\begin{eqnarray}
T_{\mu\nu}=\frac{-2}{\sqrt{-g}}\frac{\delta \mathcal{S}}{\delta g^{\mu\nu}} = \frac{-2}{A^{-4}\sqrt{\Tl{g}}}\frac{\delta \mathcal{S}_\nu}{\delta \Tl{g}_{\mu\nu}}=A^2 \Tl{T}_{\mu\nu}\, \label{Tmunud} .
\end{eqnarray}
We can  compare the individual components  of stress-energy tensors of the two frames by the following manipulation,
\begin{eqnarray}
T^\mu ~_\nu = g^{\mu \alpha} T_{\alpha\nu} = A^2 \Tl{g}^{\mu\alpha}  A^2 \Tl{T}_{\alpha\nu} = A^4 \Tl{T}^{\mu}~_\nu \Rightarrow \Tl{\rho}= A^{-4}\rho, ~~ \Tl{p}= A^{-4} p\, .
\label{rhoprlnfrm}
\end{eqnarray}

The Hubble parameter for the two frames are related as,

\begin{eqnarray}
H= \frac{1}{a}\frac{d a}{d t}= \frac{A^{-1} \Tl{a}}{a}=\frac{1}{A^{-1} \Tl{a}}\left[ \Tl{a}\frac{d A^{-1}}{dt} +  \frac{d \Tl{a}}{d \Tl{t}}\right]=-\frac{\dot{A}}{A} + A \Tl{H}\, \label{hubrl},
\end{eqnarray}
where time deviate  is taken with respect to $t$, as we want to express all physical quantities in the Einstein frame without reference to Jordan frame. We also have the following important relations,

\begin{eqnarray}
\dot{\Tl{\rho}}=\frac{d \Tl{\rho}}{d \Tl{t}}= A^{-1} \frac{1}{dt}\left(A^{-4} \rho_\nu\right)=- 4 A^{-5} \frac{\dot A}{A} \rho_\nu + A^{-5} \dot{\rho}_\nu \label{iden2}
\end{eqnarray}
Using Eq.~(\ref{hubrl}) and (\ref{iden2}), the continuity equation (\ref{cont1}) gets transformed to the Einstein frame, \\
\begin{eqnarray}
&&\dot{\rho}_\nu + 3H \left(\rho_\nu +p_\nu\right)= \frac{\dot A}{A}\left(\rho_\nu - 3 p_\nu\right) \nonumber\\
&& \rm {or~equivalently }\nonumber\\
&& \dot{\rho}_\nu + 3H \rho_\nu\left(1 +w_\nu\right)= \frac{ A, \phi}{A}\dot\phi \rho_\nu \left(1- 3 w_\nu\right) \label{contphonuf}
\end{eqnarray}
\begin{figure}[H]
    \includegraphics[width=15 cm, height=10cm]{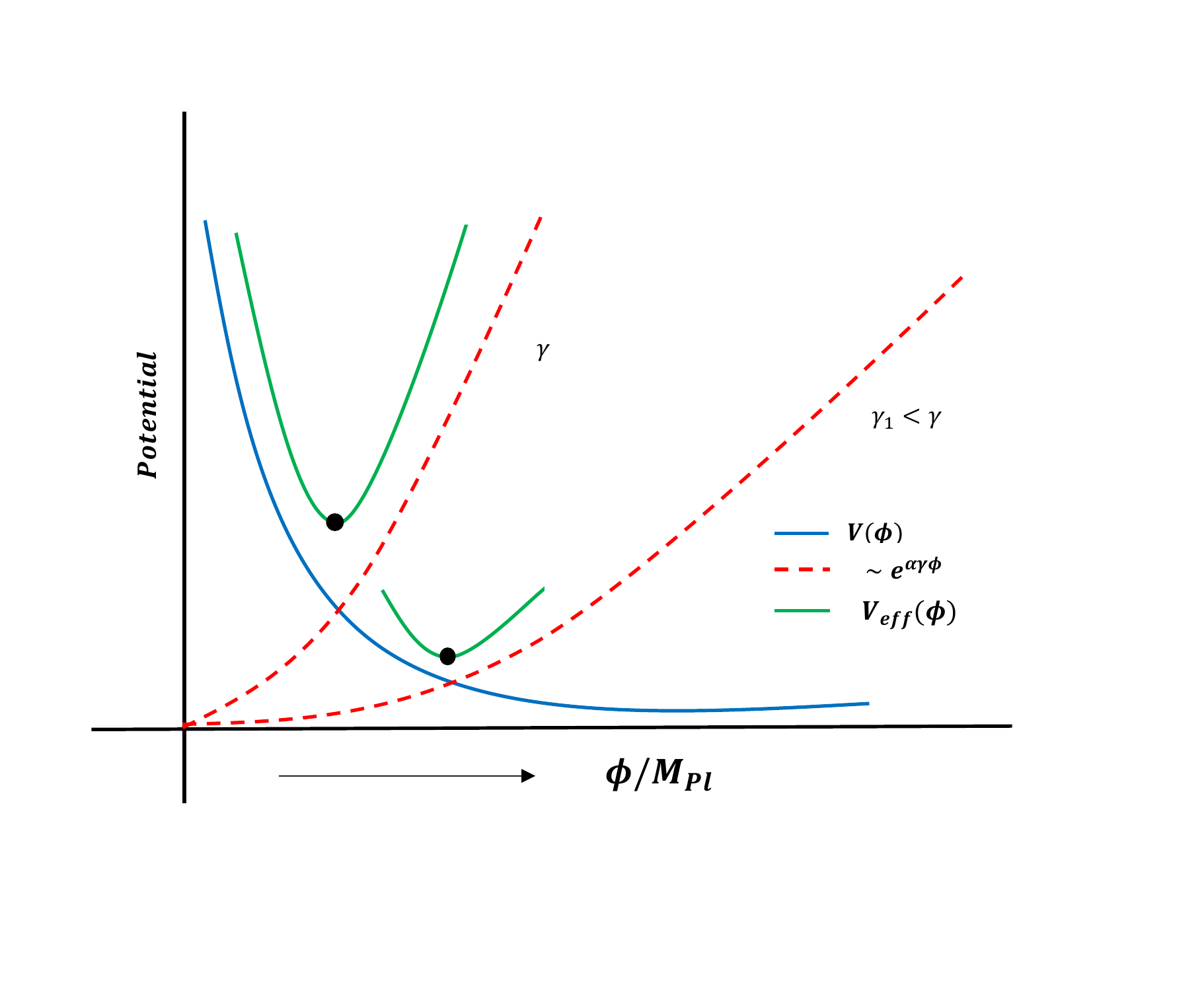}
    \caption{Schematic diagram showing the effective potential (green line); the original runaway type of potential is depicted by  blue line. Coupling of field $\phi$ to massive neutrino matter manifests itself in the effective potential (\ref{veff}) through $A(\phi)\hat{\rho}_\nu$ [$A(\phi)=Exp(\alpha\gamma\phi/\Mpl)$] shown by dashed lines.
     The graph shows that for smaller values of $\gamma$, the minimum of effective potential hits larger values of the field.}
    \label{effectivepotential}
\end{figure}
 In similar way, we derive the Klein-Gordon equation for scalar field,
 \begin{eqnarray}
 \ddot\phi+3H\dot\phi+V_{,\phi}=-\frac{A_{,\phi}}{A}(-\rho_\nu+3p_\nu) \equiv -\frac{A_{,\phi}}{A}\rho_\nu(1- 3 w_\nu),
 \label{kgnucp}
 \end{eqnarray}
 which, after multiplying on both sides by $\dot{\phi}$,
 can be cast in the following suggestive form,
 \begin{eqnarray}
 \dot{\rho_\phi}+3H\rho_\phi(1+w_\phi)=-\frac{A_{,\phi}}{A}\dot{\phi}\rho_\nu(1-3w_\nu).
 \label{contfield}
 \end{eqnarray}
 
 It is evident from Eq. ~(\ref{contphonuf}) and (\ref{contfield})  that conservation law does not hold    for neutrinos and scalar field individually but the  total energy density, $\rho_\phi+\rho_\nu$ adheres to standard conservation as it should be.
 
 Let us emphasize that neutrinos, due to their tiny masses ($m_\nu=\mathcal{O}(10^{-2})~ eV$), remain relativistic for most of the cosmic history with $w_\nu= \frac{1}{3}$ and their coupling to the scalar field vanishes. Only at the late times, when they turn non-relativistic, they start behaving like cold matter and the effect of the coupling comes into play which  has very interesting implications for late time dynamics. For a runaway type potential of the form (\ref{potn}), we argue that coupling modifies the potential such that the effective potential  has a minimum  around the present epoch. In case of $n=1$, we shall demonstrate analytically that the field settles in the minimum giving rise to a quasi de Sitter like behaviour which is exactly followed by
 massive neutrino matter. Numerical investigations, confirm that same conclusions hold for
 $n>1$. 
 
Let us first begin from general considerations.
 Owing to non-relativistic nature of neutrinos around the present epoch, we can approximate by $w_\nu \approx 0$. It is further convenient for analytical estimation that we choose, $\hat{\rho}_\nu\equiv A^{-1}\rho_\nu$, which is conserved in the Einstein frame,
 \begin{eqnarray}
 \hat{\rho}_\nu + 3 H  \hat{\rho}_\nu =0, 
 \label{rhonuhat}
 \end{eqnarray}
 and the field evolution equation in the approximation under consideration acquires the following form,
\begin{equation}
 \ddot\phi+3H\dot\phi=-V_{,\phi}-A_{,\phi}\hat\rho_\nu ,
 \label{frwfeqd}
\end{equation}
It is  indicative from Eq. (\ref{frwfeqd}), that we can write an effective potential, up to an additive constant as,
\begin{eqnarray}
V_{\rm eff}= V(\phi)  + A(\phi)\hat{\rho_\nu}\, . \label{veff}
\end{eqnarray}
Let us now use a specific form of the conformal factor,
\begin{equation}
\label{cc}
A(\phi)= e^{\alpha\gamma\frac{\phi}{\Mpl}}. 
\end{equation}
Although the original potential is runaway type with the chosen form of $A(\phi)$, the effective potential do posses a minimum, see  Fig.~(\ref{effectivepotential}). It should be noted here that Eqs.~(\ref{rhonuhat}) and (\ref{frwfeqd}) are exact for coupling to ordinary cold dark matter ($w=0$) whereas the equations hold only at late times when $w_\nu\approx 0$ in case of  coupling to massive  neutrino matter.
To capture the behaviour of massive neutrino matter which is relativistic  at early times and non relativistic near the present (late time), we assume  the following antsatz, 
\begin{eqnarray}
w_\nu= \frac{1}{6}\left\{ 1+ \tanh\left[ \frac{\ln(1+z)-z_{\rm eq}}{z_{\rm dur}}\right]\right\}
\label{wnu}, 
\end{eqnarray}
where $z$ is the red-shift and $z_{\rm dur} $ and $z_{\rm equal}$ are two  parameters to be fixed using fitting with observation. 
The aforesaid applies to any runaway type of potential. However,
in order to proceed further with analytical estimates, we shall  consider the exponential potential (\ref{Exp}) or (\ref{potn}) with $n=1$. In this case, we have the expression for the effective potential,
\begin{eqnarray}
V_{\rm eff}= V_0 e^{-\alpha\frac{\phi}{\Mpl}}+ \hat{\rho_\nu} e^{\alpha \gamma\frac{\phi}{\Mpl}}\, 
\label{veffn1}
\end{eqnarray}
 Minimising the potential w.r.t $\phi$, we find, 
\begin{eqnarray}
\phi_{\rm min}= \log\left( \frac{V_0}{\gamma \hat{\rho_\nu}}\right)^{\frac{\Mpl}{\alpha(1+\gamma)}} \implies V_{\rm eff, min}= V_0 \left(\frac{\gamma \hat{\rho_\nu}}{V_0}\right)^{\frac{1}{1+\gamma}}\, , 
\label{Veffmin}\, ,
\end{eqnarray}
where the potential (\ref{veffn1}) is defined up to an additive constant and the above relation is true for large values of $\gamma$ required by observation, (to be discussed below). The question thus arises whether the minimum given by (\ref{Veffmin}) corresponds to the present dark energy density of the Universe. We will demonstrate that it does. 
\begin{figure}[H]
    \centering
    \includegraphics[width=9cm, height=7cm]{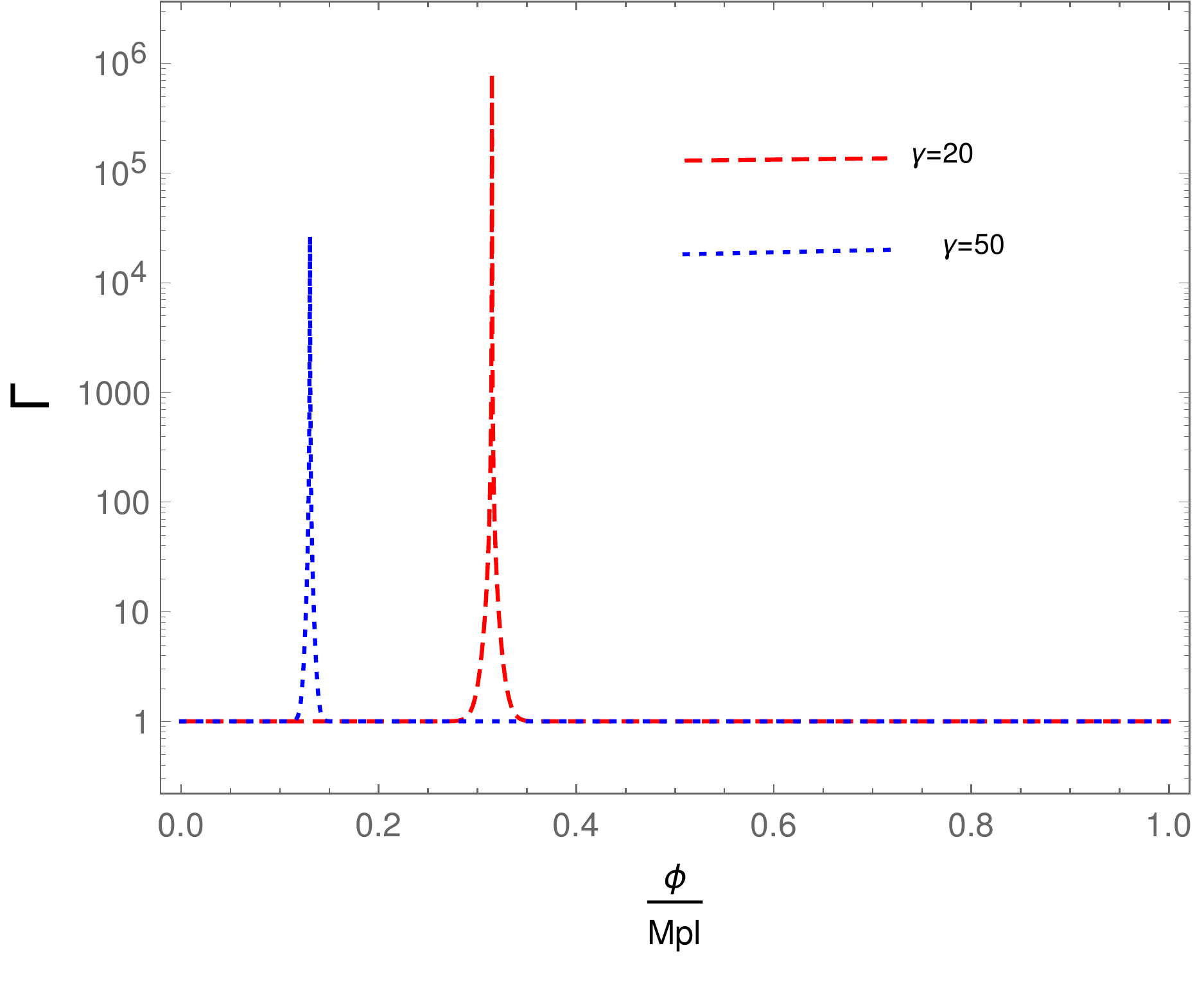}
    \caption{Figure shows the plot of $\Gamma$ for effective potential given by (\ref{veffn1}). $\Gamma$ increases fast during slow roll, which happens here around the minimum which shifts towards larger values of $\phi$ for smaller values of $\gamma$. The plot is for $\alpha=10$ and the potential is normalised by $V_0$. }
    \label{Fig:gammaeffective}
\end{figure}
It is instructive to rewrite Eqs.(\ref{contphonuf} ) and ( \ref{kgnucp}) for the conformal coupling (\ref{cc}),
\begin{eqnarray}
\label{contphonuf1}
&& \dot{\rho}_\nu + 3H \rho_\nu\left(1 +w_\nu\right)= \frac{\alpha \gamma}{\Mpl}\dot\phi \rho_\nu \left(1- 3 w_\nu\right) \\
&&\ddot\phi+3H\dot\phi+V_{,\phi}= -\frac{\alpha \gamma}{\Mpl}\rho_\nu(1- 3 w_\nu),
 \label{kgnucp1}
\end{eqnarray}

Using the effective potential picture, which is approximate in case of coupling to massive neutrino matter, we have put forward a general perspective to convince the reader that the coupling under consideration triggers transit to late time acceleration. In what follows, we present the exact analytical results   based upon dynamical analysis. Indeed, in this case, the autonomous system admits an attractor solution which corresponds to\cite{Hossain:2014xha},
\begin{eqnarray}
w_\phi &=& \frac{-3 + \alpha^2 (1+ \gamma)}{\alpha^2 (1+ \gamma)^2}\label{wphin1}\, , \\
\dot \phi &=& \frac{3 H \Mpl}{\alpha (1+ \gamma)} \label{phidotn1}\, .
\end{eqnarray}

 It is interesting to check for the late time behaviour of neutrino matter density. One might naively think that with the field approaching de Sitter: $\dot{\phi}\to 0$ and  $w_\nu \approx 0$ near the present epoch, neutrino matter should decouple from scalar field and, $\rho_\nu\sim a^{-3}$, see Eq.(\ref{contphonuf1}). 
 On the contrary, $\rho_\nu$ follows the scalar field and behave like dark energy which is counter intuitive. This behaviour can be  understood, recalling Eq.~(\ref{contphonuf}), and the fact  that $p_\nu \approx 0$ around the present epoch,. Indeed,
\begin{eqnarray}
\dot\rho_\nu + 3H \rho_\nu \left( 1- \frac{A, \phi}{A} \frac{\dot \phi}{3H} \right)=0\, .
\end{eqnarray}
Taking into account that $A_{, \phi}/{A}= \alpha \gamma /\Mpl$, we write the continuity equation in a suggestive form, 
\begin{eqnarray}
\dot \rho_\nu +3H \left( 1+ w_{\nu \rm}^{\rm eff}\right)\label{contrhonueff}=0\ ;~~w_\nu ^{\rm eff}\equiv -\frac{\alpha \gamma \dot{\phi}}{3 H \Mpl}
\label{wnuphidot}
\end{eqnarray}
 The behaviour of $w_\nu^{\rm eff}$ determines the late time evolution of neutrino matter density. Indeed, plugging $\dot{\phi}$   from Eq. (\ref{phidotn1}) in 
 Eq.(\ref{wnuphidot}), we get,
\begin{figure}
    \begin{center}
    \includegraphics[width=10cm, height=8cm]{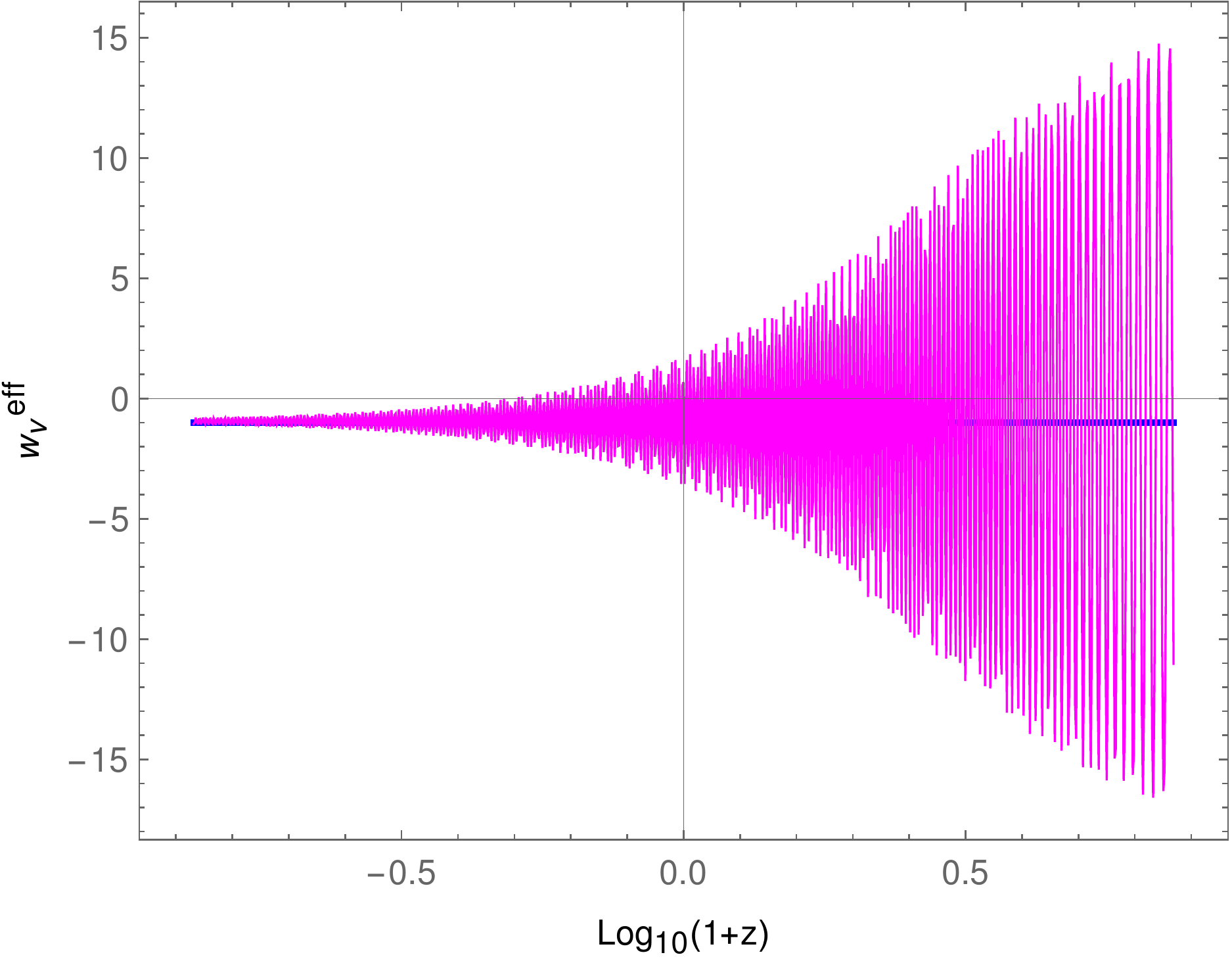}
        \end{center}
    \caption{Evolution of $w_\nu^{\rm eff}$ versus the red-shift in case of the generalized exponential potential (\ref{potn}) for $n=3$ and  $\alpha \gamma =1000$, compatible with observation \cite{Basak:2021cgk};  solid blue line corresponds $w^{\rm eff}_\nu=-1$. Figure shows that $w_\nu^{\rm eff}$ oscillates and eventually settles close to minus one.}
    \label{fig:wnueff}
\end{figure}
\begin{eqnarray}
w_\nu^{\rm eff}= - \frac{\alpha \gamma \dot \phi}{3 H \Mpl}= -\frac{\gamma}{1+\gamma} .
\end{eqnarray}
As mentioned before, {nucleosynthesis constraint forces, $\alpha\gtrsim 10$,  using Eq.(\ref{wphin1}) then we have  in the leading order in $\alpha$, 
\begin{equation}
w_\phi=  -\frac{\gamma}{1+\gamma}+\mathcal{O}(\alpha^{-2}) \end{equation}
which is close to $-1$ for large values of $\gamma$, consistent with observations on late time acceleration. } This in turn implies that $w_\nu^{\rm eff}\simeq -1$ for the fixed point and massive neutrino matter follows the scalar field at late times.
\begin{figure}
    \centering
    \includegraphics[height=9cm, width=10cm]{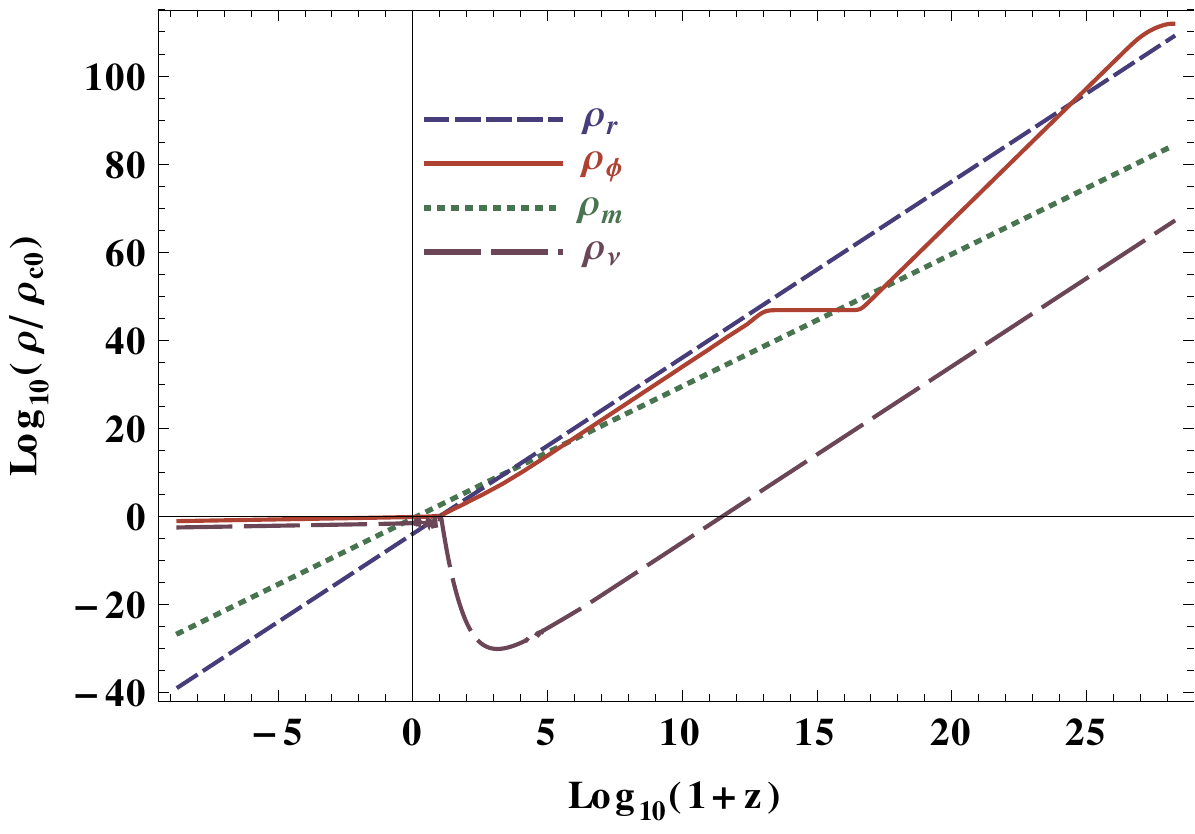}
    \caption{Plot of energy densities versus the red-shift on the log scale for generalized exponential potential (\ref{potn}). Evolution is shown from the end of inflation. Field is shown to track the background with exit   to de Sitter at late times followed by the massive neutrino matter density. Figure corresponds to numerical values of parameters:
     $\gamma = 800$ ,  $ \lambda = 10^{-8}$ , $ n = 6 $~  $z_{eq} = 2.55$~  $\&$~$ z_{dur }= 3$.}
    \label{fi:my_label}
\end{figure}
In fact, numerical results confirm that same features are shared by the model based upon the generalized exponential potential (\ref{potn});  see Figs.~\ref{fig:wnueff} $\&$ \ref{fi:my_label}, and Ref.~\cite{Basak:2021cgk}
for details.
In terms of the effective potential picture, it is clear that the minimum is time-dependent via $\hat{\rho}_\nu$, but the system essentially settles around its attractor and matches observational values of the dark energy paradigm at the present epoch by appropriately choosing $``\gamma."$.

A comment on coupling of scalar field to dark matter is in order.
In this case, the exit to acceleration might happen soon after mater domination is established. The only free parameter $\gamma$
 should then be tuned to delay the minimum to late stage which is accomplished by taking a $\gamma\ll 1$ (see  Figs. \ref{effectivepotential} $\&$ \ref{Fig:gammaeffective}) ) but this defies acceleration as
 $w_\phi \to 0, \gamma \to 0$ and the system is back to scaling track.
 We have demonstrated that this issue is  successfully addressed
 by invoking  coupling to massive neutrino matter.

Let us summarize our main findings in the subsection. Scalar field coupling to massive neutrino matter, due to tiny neutrino masses,  builds up dynamically only at late stages giving rise to exit from scaling regime to de Sitter like configuration around the present epoch.
In case of exponential potential, we supported our conclusions by analytical estimates.
Indeed, in this case,  as
$\gamma$ becomes larger and larger, $\dot{\phi}$ becomes smaller and smaller ($\dot{\phi}\to 0$)
 such that $\alpha \gamma \dot{\phi}\sim H_0 \Mpl\sim \rho^{1/2}_{cr}$ and the RHS of Eq.(\ref{contphonuf1}) is non-vanishing as de Sitter is approached. As a result neutrino matter is never decoupled from scalar field while de Sitter is approached. The latter forces the neutrino matter to follow the scalar field at late times. Numerical results confirms the same behaviour for $n>1$, see Fig.\ref{fi:my_label}.
\section{The distinguished features of quintessential inflation: Relic gravity waves}
\label{RG}
The gravity waves (GW) are described by transverse trace-less component of the metric perturbation over the background  space time, say specially flat FLRW space time, $ds^2= -dt^2 +a(t)^2 \left(h_{i j}.  +\delta_{ij}\right)dx^{i}dx^j$. The perturbation satisfies the transverse trace-less condition $h^{i}_{i}=0, \partial_ih^j=0, $. The Fourier decomposition for the the perturbation for a mode with wave number $\textbf{k}$ can be written as \begin{eqnarray}
h_{ij}(t,x)= \sum_{n=+,\times} \int \frac{d\textbf{k}}{(2\pi)^{2/3}} \varphi_{ij}^n(\textbf{k}) h_{\textbf{k}}^n (k,t) e^{\textbf{k}\cdot \textbf{x}}\, ,
\end{eqnarray}
where the polarization tensor $ \varphi_{ij}^n(\textbf{k})$ are symmetric and satisfies transverse traceless condition. The Fourier component of the tensor mode satisfies the wave equation~\cite{Sahni:1990tx} \\
\begin{eqnarray}
h_{k}^{n''}(\tau)+ 2 \frac{a'}{a} h_{k}^{n'}(\tau) + k^2 h_{k}^{n}=0 \, ,
\end{eqnarray}
where the $'$ denotes derivative with respect to conformal time, $d\tau\equiv \frac{dt}{a}$. Energy spectrum for the GWs with wave number $k$ are given by 
\begin{eqnarray}
\Omega_{\rm GW}(k, \tau)\equiv \frac{1}{\rho_{\rm cr}}\frac{d \rho_{\rm GW}}{d\ln k}\, .
\end{eqnarray}
The GW energy density is given by\\
\begin{eqnarray}
\rho_{\rm GW}= -T^0_{~0}=\frac{\Mpl^2}{4}\large \int \frac{d^3 k}{(2\pi)^3}\frac{k^2}{a^2}2 \sum_{n} |h_{k}^n|^2\, .
\end{eqnarray}
The GW energy spectrum for the present time  can be written as 
\begin{eqnarray}
\Omega_{\rm GW, 0}= \frac{1}{12}\left(\frac{k^2}{a_0^2 H_0^2}\right)P_T(k) T^2(k)\, \label{gwspec}\, ,
\end{eqnarray}
where, $P_T (k)= \left. \frac{2}{\pi^2} \frac{H_{\rm inf}^2}{\Mpl^2}\right |_{k=aH}$ is the primordial tensor power spectrum, defined at the horizon crossing and  $T^2(k)$ is the transfer function which
describes time evolution of
each mode after the end of inflation. The transfer function today is given by\cite{Boyle:2005se,Watanabe:2006qe,Kuroyanagi:2008ye}. 
\begin{figure}
    \centering
    \includegraphics{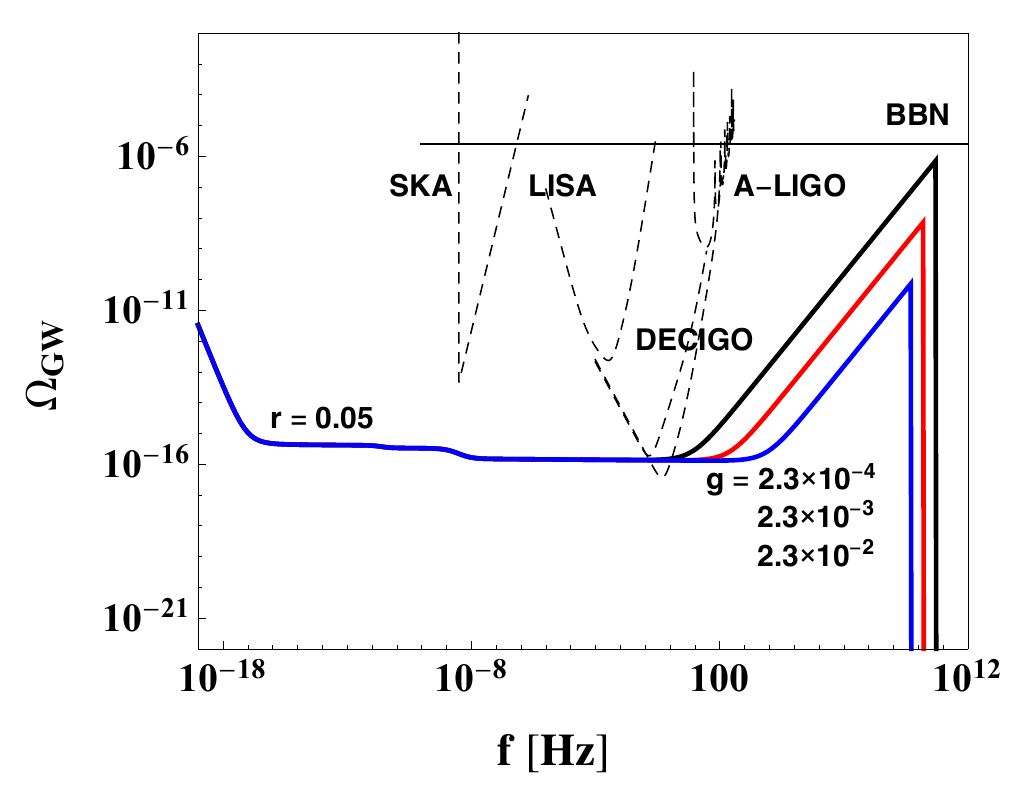}
    \caption{The GW spectrum versus the frequency in Hz for the paradigm of quintessential inflation. The blue tilt in the spectrum is visible on the high frequency side due to the presence of long kinetic epoch. Instant preheating is assumed as reheating mechanism with coupling constant $g$(consistent with nucleosynthesis constraint), see subsection \ref{inst} and  Ref.\cite{Ahmad:2019jbm}. }
    \label{fig:GWtilt}
\end{figure}
\begin{eqnarray}
T^2(k)=\frac{1}{2} \frac{a_{\rm hc}^2}{a_0^2}\, , \label{tranferfun}
\end{eqnarray}
where the subscript ``$0$'' represents value at present time and $\rm hc$ for horizon crossing. Here it should be mentioned that in the super-horizon limit $k\ll a H$, $h_k$ stays constant and vary as $h_k \propto \frac{1}{a}$ in the sub-horizon limit, $k\gg aH$.
Different mode of GW  enter the horizon at different  epochs and to evaluate the spectrum today, we need to calculate the Hubble parameter precisely, 
\begin{eqnarray}
H\approx H_0 \sqrt{\Omega_\phi (a) + \Omega_{r0}  \left(\frac{g_*}{g_{*0}}\right) \left(\frac{g_{*s}}{g_{*s0}}\right)^{-4/3}\left(\frac{a}{a_0}\right)^{-4} +\Omega_{m0}\left(\frac{a}{a_0}\right)^{-3} }\label{HubbleGW}, 
\end{eqnarray}
where $g_*(s)$ denotes the effective number of relativistic degrees of freedom at temperature $T$ contributing to entropy.
It should mentioned that the major contribution of scalar field to Hubble comes in two phases, namely,  kinetic regime which follows inflation and the dark energy era around the present epoch. Let us put it in the following form,\\
\begin{eqnarray}
\Omega_\phi (a)= \Omega_{\phi K}+ \Omega_{\Lambda 0}\, ,
\end{eqnarray}
where the latter is identified with the present value of dark energy. The kinetic part, can be evaluated as,\\
\begin{eqnarray}
\Omega_{\phi, K}\equiv \frac{\rho_{\phi, K}}{3 H_0^2 \Mpl^2}= \frac{\rho_{\phi, \rm end}}{3 H_0\Mpl^2}\left(\frac{a}{a_{\rm end}}\right)^{-6}= \frac{3 H_{\rm end}^2\Mpl^2}{3 H_0^2\Mpl^2}\left(\frac{a}{a_{\rm end}}\right)^{-6}=  \frac{ H_{\rm end}^2}{ H_0^2}\left(\frac{a}{a_{\rm end}}\right)^{-6}\, .
\end{eqnarray}
Making use of the relation, $k= a_{\rm hc}H_{\rm hc}$ and Eq.~(\ref{HubbleGW}), the scale factor $a_{\rm hc} $, in different regimes, can be expressed as,
\begin{eqnarray}
a_{\rm hc, \rm MD}& =& \frac{H_0^2 a_0^3}{k^2}\Omega_{\rm m0}\label{amatter}\\
a_{\rm hc, \rm RD}&=& \frac{a_0^2 H_0}{k}\left(\frac{g_*}{g_{*0}} \right)^{1/2}\left( \frac{g_{*s}}{g_{*s0}}\right)^{-2/3}\sqrt{\Omega_{r0}}\label{arad}
\end{eqnarray}
The scale factor for those mode entering during kinetic epoch can be found by using the fact that during kinetic epoch up-to the radiation commencement the energy density goes as $\rho_{\rm tot}\approx \rho_{\phi}\propto a^{-6}$ (or equivalently $H\propto a^{-3}$) and $k_r= H_r a_r$
\begin{eqnarray}
k &= & a_{\rm hc, \rm KD} H_{\rm hc}= \frac{a_{\rm hc\rm KD}}{a_r}\frac{H_{\rm hc}}{H_r}~~ a_r H_r =  \frac{a_{\rm hc\rm KD}}{a_r} \frac{H_{\rm end} \left(\frac{a_{\rm end}}{a_{\rm hc, \rm KD}}\right)^3}{H_{\rm end}\left(\frac{a_{\rm end}}{a_r}\right)^3}k_r\nonumber\\
&=& \frac{a_r^2}{a_{\rm hc, \rm KD}^2}\implies 
\frac{a_{\rm hc, \rm KD}}{a_r}= \left(\frac{k_r}{k}\right)^{1/2}\,. \label{akin}
\end{eqnarray}


Substituting the values of the scale factors (\ref{amatter}, \ref{arad}) and (\ref{akin}), to the transfer function (\ref{tranferfun}), we obtain, 
\begin{eqnarray}
\Omega_{\rm GW, 0}^{(\rm MD)} &=&\frac{1}{6 \pi^2} \Omega_{\rm m0}^2 \frac{H_{\rm inf} ^2}{\Mpl^2}~ \frac{a_0^2H_0^2}{k^2} ~ ~ ~ ~ ~ ~ ~ ~{\rm for}~~~(k_0 < k < k_{\rm eq})\\
\Omega_{\rm GW,0}^{\rm RD} &=& \frac{1}{6 \pi^2} \Omega_{\rm r0} \frac{H_{\rm inf} ^2}{\Mpl^2}\left(\frac{g_{*}}{g_{*0}}\right) \left(\frac{g_{*,s}}{g_{*,s0}}\right)^{-4/3}~ ~ {\rm for}~~~(k_{\rm eq}<k<k_{\rm r})\\
\label{omega0rd}
\Omega_{\rm GW, 0}^{\rm (KD)} &= & \Omega_{\rm GW, 0}^{\rm (RD)} \left(\frac{k}{k_r}\right)\label{omega0kd}~~~~ {\rm for } ~~~~~({k_{\rm end}<k < k_r })\, 
\end{eqnarray}
where  $k_0$, $k_{\rm eq}$, $k_{\rm end}$, $k_{\rm r}$  represent values of $k$ at present, matter-radiation equality, end of inflation and commencement of radiation era respectively. The corresponding frequencies can easily be evaluated using the relation, $f=\frac{a H}{2 \pi}$\\

\begin{eqnarray}
f_0 &=& \frac{a_0 H_0}{2 \pi } \sim 3 \times 10^{-19 } {\rm Hz}\\
f_{\rm eq} & =& \frac{a_{\rm eq} H_{\rm eq}}{2\pi}\sim 10^{-17} \rm Hz\\
f_{\rm end}&=& \frac{a_{\rm end} H_{\rm end}}{2 \pi}= \frac{T_0}{T_{\rm end}}\left(\frac{43}{11 g_{*s}}\right)^{1/3}\frac{H_{\rm end}}{2 \pi}\gtrsim 10^8 \rm Hz
\end{eqnarray}
In the last step we have given an lower bound on frequency at end of inflation in a model independent way by considering instantaneous energy transfer at reheating.

\textbf{Nucleosynthesis  constraint on reheating temperature: }
As discussed in section \ref{BBNs}, the presence of any relativistic degree of freedom in the Universe, over and above the standard model of particle physics, is subject to the nucleosynthesis constraint. In particular, we have have a bound on relic
gravity waves given by (\ref{omegagw0}).
And since the  major contribution to GW spectrum comes from the blue-tilted spectrum due to kinetic regime ($k_{\rm end}< k < k_r$), we have
\cite{Cyburt:2015mya,Figueroa:2018twl},\\
\begin{eqnarray}
\Omega_{\rm GW, 0}\simeq \Omega^{\rm KD}_{\rm GW, 0}\equiv \Omega^{(\rm max)}_{\rm GW, 0}<1.12\times10^{-6}
\label{maxbound}
\end{eqnarray}
  Substituting the value of $\Omega_{\rm GW, 0}^{\rm (KD)}$ from Eq.~(\ref{omega0rd}),  and using the fact that for highest frequency at kinetic epoch,  $a=a_{\rm kin}\approx a_{\rm end}$, we get, 
\begin{eqnarray}
\Omega_{\rm GW, 0}^{\rm max}= \frac{1}{6 \pi^2} \Omega_{\rm r0}~ \frac{H_{\rm inf}^2}{\Mpl^2}~ \frac{g_*}{g_{*0}}~\left(\frac{g_{*s}}{g_{*s0}}\right)^{\frac{-4}{3}}\left( \frac{a_r}{a_{\rm end}}\right)^2 \label{omegamax1}
\end{eqnarray}
Putting the values, $\Omega_{\rm r0}=9.24 \times 10^{-5} $, $ g_* ( a_{\rm end})= g_{*s} (a_{\rm end})=106.75$, $g_{*0}=3.36$, $g_{*s0}=3.91$; $H_{\rm inf}= 1.4 \times 10^{14} r^{\frac{1}{2}} ~\rm GeV$ and using the upper bound on tensor to scalar ratio of perturbations, $r\approx .05$, we obtain the upper bound on the GW spectrum, \\
\begin{eqnarray}
&& \Omega_{\rm GW, 0}^{\rm max}=3.8 \times 10^{-16}\left(\frac{a_r}{a_{\rm end}}\right)^2=3.8 \times 10^{-16} \left(\frac{T_{\rm end}}{T_r}\right)^2\label{omegagwmaxrel}\\
&& \Omega_{\rm GW, 0}^{\rm max}< 1.12\times 10^{-6} \implies \left(\frac{T_{\rm end}}{T_r}\right)< 5.5 \times 10^{4} , .
\end{eqnarray}
where we have used the relation, ${a_r}/{a_{\rm end}}={T_{\rm end}}/{T_r}$. This bound enables us to constrain the duration of kinetic regime: from the end of inflation to the commencement of radiation era.  Higher is the radiation energy density produced at the end of inflation,  shorter would be the kinetic regime and the easier it would be to comply with the
BBN bound.
 Let us consider a natural mechanisms dubbed gravitational particle production and check for its consistency with nucleosynthesis.
In gravitational reheating (see, \ref{sec:gravitionalreheating}), the energy density produced after the end of inflation is given by 
\begin{eqnarray}
\rho_{r, \rm end} = 10^{-2} g_p H_{\rm end}^4 \approx 1.15 \times 10^{-17} \Mpl^4 r\, ,
\end{eqnarray}
where in the last step we used, $g_p=100$. 
From Eq.~(\ref{rhophiend}) we have $\rho_{\phi, \rm end}\approx 1 \times 10^{-8} \Mpl^4 r^2$. Thus,  Eq.~(\ref{eq:arbyend}), yields,\\
\begin{eqnarray}
\left(\frac{a_r}{a_{\rm end}}\right)^2= \left[ \frac{\rho_{\phi, \rm end}}{\rho_{\rm end}}\right]\approx 1.8 \times 10^{10}\,.
\end{eqnarray}
Finally, using Eq.~(\ref{omegagwmaxrel}), we arrive at the following estimate,
\begin{eqnarray}
\Omega_{\rm GW, 0}^{\rm max}\approx6 \times 10^{-6}\, ,
 \end{eqnarray}
which  challenges the nucleosynthesis constraint (\ref{omegagw0}). This is attributed to inefficiency (smaller value of $\rho_{r,\rm end}$ and correspondingly larger value of $(\rho_\phi/\rho_r)_{end}$) of reheating  process based upon gravitational particle production\cite{Ahmad:2019jbm}. In fact, other reheating mechanisms can give rise to larger values of $\rho_{r,\rm end}$ circumventing the problem due to relic gravity waves.  For example in instant preheating scenario, we have $(\rho_\phi/\rho_r)_{end}=(2 \pi)^3/g^2$, with $g$ being a coupling constant in the scenario. Constraining,  $g$, one can easily comply with the nucleosynthesis bound, see Ref.\cite{Ahmad:2019jbm} for details.\\


\section{Summary}
\label{sumary}
This review is a pedagogical exposition of scalar field dynamics in the FLRW background applied to quintessential inflation. In this framework, two specific roles are assigned to a scalar field, namely, a consistent description of inflation and late time acceleration, with the understanding that the field is invisible starting from the commencement of the radiative regime till the beginning of late time acceleration. The other important demand from scalar field dynamics is related to the insensitivity of late-time physics concerning the initial (field) conditions. These requirements broadly characterise the class of scalar field potentials suited to the framework under consideration: potentials flat initially, followed by steep behaviour in the post inflationary era, and again flat around the present epoch. Without the loss of generality, one assumes that the runaway potential is such that the field rolls away from the origin towards plus infinity. After inflation ends, the field energy density $\rho_\phi$ is larger by several orders of magnitude than the energy density of radiation assumed to be produced due to some alternative mechanism. Consequently, the field evolves in the kinetic regime for a long time before the commencement of radiation domination. Following that, $\rho_\phi$ overshoots the background matter (radiation)  density such that $\rho_\phi\ll\rho_r$. Field then freezes on its potential due to Hubble damping and waits till the radiation density redshifts to the level that $\rho_r/\rho_\phi$ becomes comparable to $\lambda_s^2/2$ ($\lambda_s$ is the slope of steep potential). Hereafter, field dynamics are sensitive to the type of steepness of the potential, which is quantified by the parameter $\Gamma$.  If $\Gamma=1$ (exponential potential), the field exactly follows the background as the scaling solution is an attractor in this case. In case, $\Gamma>1$, for instance, inverse power law potentials ($V\sim \phi^{-n}, \Gamma=1+1/n$), the slope decreases such that the field nearly follows the background while evolving along the steep part of the potential and at late times enters the slow-roll regime. This class of potentials is suitable for tracker behaviour. Slope, on the other hand, increases with evolution if $\Gamma<1$ and in this case, $\Gamma$ eventually approaches its scaling value in the asymptotic region. We demonstrated that the generalised exponential potential that falls in this class gives rise to a scaling solution in the asymptotic regime, which is an attractor. This class of potentials, shallow initially, is suitable for inflation and, in the post-inflationary era, retains the characteristics of a steep exponential potential. We further demonstrated that a perfect tracker can be designed from the classes of potentials with $\Gamma\leq 1$ by invoking a late time feature in the potentials that could facilitate exit to acceleration. In case, $\Gamma=1$,  we also need to invoke an early time feature in the potential which, in particular, can be achieved using a non-canonical kinetic term in the scalar field Lagrangian. 

As for the exit to late time acceleration, we mentioned different ways to realise it and presented detailed investigations of a mechanism based upon the coupling of the field to massive neutrino matter. In this framework, the distinguished physical process in the late Universe, namely, the turning of massive neutrinos from relativistic to non-relativistic, can trigger the desired exit.

The focus in this review was on the building blocks of quintessential inflation rather than the concrete models. We emphasised the model-independent and generic features of the paradigm. For instance, a long kinetic regime after the end of inflation and before the commencement of a radiative regime is an essential component of the framework, which distinguishes it from the standard inflationary scenario. The latter induces a novel feature in the spectrum of relic gravity waves, namely a blue tilted spectrum in the high-frequency regime. To this effect, we have included a brief discussion on relic gravity waves and presented the model-independent estimates to highlight the distinguishing features of the paradigm.

Last but not least, our discussion, especially on scalar field dynamics in the post-inflationary era, is pedagogical, and we hope the review will be helpful for young researchers interested in topics related to quintessential inflation and dark energy.
\section{ Acknowledgements}
We thank V. Sahni, R. Gannouji, Anzhong Wang, Shibesh Kumar, S. Capozziello, Azam Mofazzal and Sang Pyo Kim for useful comments.
We are indebted to Konstantinos Dimopoulos for his invitation to write this review and for making valuable comments that allowed us to improve the draft.
MS is partially supported by the Ministry of Education and Science of the Republic of Kazakhstan, Grant
No. 0118RK00935. NJ is supported by the National Postdoctoral Fellowship of the Science and
Engineering Research Board (SERB), Department of Science and Technology (DST), Government of India, file No. PDF/2021/004114. NJ thanks  the Centre for Cosmology and Science Popularisation (CCSP), SGT University for hospitality where the work was initiated.


\begin{thebibliography}{}
\bibitem{Planck:2018vyg}
N.~Aghanim \textit{et al.} [Planck],
\href{\doi/10.1051/0004-6361/201833910}
{Astron. Astrophys. \textbf{641}, A6 (2020)}
[erratum: Astron. Astrophys. \textbf{652}, C4 (2021)]
[\href{\arxiv/1807.06209}{arXiv:1807.06209} [astro-ph.CO]].
\bibitem{Guth:1980zm}
A.~H.~Guth,
\href{\doi/10.1103/PhysRevD.23.347}{Phys. Rev. D \textbf{23}, 347-356 (1981)}.
\bibitem{Sato:1980yn}
K.~Sato,
Mon. Not. Roy. Astron. Soc. \textbf{195}, 467-479 (1981)
NORDITA-80-29.Phys. Rev. D \textbf{23}, 347-356 (1981).


\bibitem{Linde:1981mu}
A.~D.~Linde,
\href{\doi/10.1016/0370-2693(82)91219-9}{
Phys. Lett. B \textbf{108}, 389-393 (1982)}.
\bibitem{Albrecht:.1982wi}
A.~Albrecht and P.~J.~Steinhardt,
\href{\doi/10.1103/PhysRevLett.48.1220}{Phys. Rev. Lett. \textbf{48}, 1220-1223 (1982)}.

\bibitem{Starobinsky:1980te}
  A.~A.~Starobinsky,
  ``A New Type of Isotropic Cosmological Models Without Singularity,''
  \href{\doi/10.1016/0370-2693(80)90670-X}{Phys.\ Lett.\ B {\bf 91}, 99 (1980)}.

\bibitem{Starobinsky:1982ee}
  A.~A.~Starobinsky,
  ``Dynamics of Phase Transition in the New Inflationary Universe Scenario and Generation 
of 
Perturbations,''
  \href{\doi/10.1016/0370-2693(82)90541-X}{Phys.\ Lett.\ B {\bf 117}, 175 (1982)}.

\bibitem{Liddle:1999mq}
A.~R.~Liddle,
[\href{\arxiv/astro-ph/9901124}{ arxXiv:astro-ph/9901124}[astro-ph]].
\bibitem{Tsujikawa:2003jp}
S.~Tsujikawa,
[\href{\arxiv/hep-ph/0304257}{arXiv:hep-ph/0304257} [hep-ph]]

\bibitem{Martin:2013tda}
J.~Martin, C.~Ringeval and V.~Vennin,
\href{\doi/10.1016/j.dark.2014.01.003}{Phys. Dark Univ. \textbf{5-6}, 75-235 (2014)}.

[\href{\arxiv/1303.3787}{arXiv:1303.3787} [astro-ph.CO]].
\bibitem{Vazquez:2018qdg}
J.~A.~V\'azquez, L.~E.~Padilla and T.~Matos,
\href{\doi/10.31349/RevMexFisE.17.73}{publication}
[\href{\arxiv/1810.09934}{arXiv:1810.09934} [astro-ph.CO]].


\bibitem{SupernovaSearchTeam:1998fmf}
A.~G.~Riess \textit{et al.} [Supernova Search Team],
\href{\doi/10.1086/300499}{Astron. J. \textbf{116}, 1009-1038 (1998)}
[\href{\arxiv/astro-ph/9805201}{arXiv:astro-ph/9805201} [astro-ph]].
\bibitem{SupernovaCosmologyProject:1998vns}
S.~Perlmutter \textit{et al.} [Supernova Cosmology Project],
Astrophys. J. \textbf{517}, 565-586 (1999)
[\href{\arxiv/astro-ph/9812133}{arXiv:astro-ph/9812133} [astro-ph]].
\bibitem{seed}
S.~W.~Hawking,
\href{\doi/10.1016/0370-2693(82)90373-2}{Phys. Lett. B \textbf{115}, 295 (1982)}; J.~M.~Bardeen, P.~J.~Steinhardt and M.~S.~Turner,
\href{\doi/10.1103/PhysRevD.28.679}{Phys. Rev. D \textbf{28}, 679 (1983)}.


\bibitem{Krauss:1995yb}
L.~M.~Krauss and M.~S.~Turner,
\href{\doi/10.1007/BF02108229}{Gen. Rel. Grav. \textbf{27}, 1137-1144 (1995)}
[\href{\arxiv/astro-ph/9504003}{arXiv:astro-ph/9504003} [astro-ph]].

\bibitem{Turner:1997de}
M.~S.~Turner,
[\href{\arxiv/astro-ph/9703161}{arXiv:astro-ph/9703161} [astro-ph]].

\bibitem{lateobs}
C.~B.~Netterfield \textit{et al.} [Boomerang],
\href{\doi:10.1086/340118}{Astrophys. J. \textbf{571}, 604-614 (2002)}
[\href{\arxiv/astro-ph/0104460}{arXiv:astro-ph/0104460} [astro-ph]];
N.~W.~Halverson, \textit{et al.}
\href{\doi/10.1086/338879}{Astrophys. J. \textbf{568}, 38-45 (2002)}
[\href{\arxiv/astro-ph/0104489}{arXiv:astro-ph/0104489} [astro-ph]].

\bibitem{Peebles:1998qn}
P.~J.~E.~Peebles and A.~Vilenkin,
\href{\doi/10.1103/PhysRevD.59.063505}{Phys. Rev. D \textbf{59}, 063505 (1999)}
[\href{\arxiv/astro-ph/9810509}{arXiv:astro-ph/9810509} [astro-ph]].
\bibitem{Peebles:1987ek}
P.~J.~E.~Peebles and B.~Ratra,
\href{\doi/10.1086/185100}{Astrophys. J. Lett. \textbf{325}, L17 (1988)}.
\bibitem{Sahni:2001qp}
V.~Sahni, M.~Sami and T.~Souradeep,
\href{\doi/10.1103/PhysRevD.65.023518}{Phys. Rev. D \textbf{65}, 023518 (2002)}
[arXiv:gr-qc/0105121 [gr-qc]].
\bibitem{Huey:2001ae}
  G.~Huey and J.~E.~Lidsey,
  ``Inflation, brane worlds and quintessence,''
  \href{\doi/10.1016/S0370-2693(01)00808-5}{Phys.\ Lett.\ B }
  \href{\doi/10.1016/S0370-2693(01)00808-5}{{\bf 514}, 217 (2001)}
  [\href{\arxiv/astro-ph/0104006}{astro-ph/0104006}].

\bibitem{Majumdar:2001mm}
  A.~S.~Majumdar,
  ``From brane assisted inflation to quintessence through a single scalar field,''
  \href{\doi/10.1103/PhysRevD.64.083503}{Phys.\ Rev.\ D {\bf 64}, 083503 (2001)}
  [\href{\arxiv/astro-ph/0105518}{astro-ph/0105518}].

\bibitem{Dimopoulos:2000md}
  K.~Dimopoulos,
  ``Towards a model of quintessential inflation,''
  \href{\doi/10.1016/S0920-5632(01)01058-1}{Nucl.\ Phys.\ Proc.}\
  \href{\doi/10.1016/S0920-5632(01)01058-1}{Suppl.\  {\bf 95}, 70 (2001)}
  [\href{\arxiv/astro-ph/0012298}{astro-ph/0012298}].

\bibitem{Sami:2003my}
  M.~Sami, N.~Dadhich and T.~Shiromizu,
  ``Steep inflation followed by Born-Infeld reheating,''
  \href{\doi/10.1016/j.physletb.2003.07.001}{Phys.\ Lett.\ B}
  \href{\doi/10.1016/j.physletb.2003.07.001}{{\bf 568}, 118 (2003)}
  [\href{\arxiv/hep-th/0304187}{hep-th/0304187}].

\bibitem{Dimopoulos:2002hm}
  K.~Dimopoulos,
  ``The Curvaton hypothesis and the eta-problem of quintessential inflation, with and 
without 
branes,''
  \href{\doi/10.1103/PhysRevD.68.123506}{Phys.\ Rev.\ D {\bf 68}, 123506 (2003)}
  [\href{\arxiv/astro-ph/0212264}{astro-ph/0212264}].

\bibitem{Dias:2010rg}
  M.~Dias and A.~R.~Liddle,
  ``On the possibility of braneworld quintessential inflation,''
  \href{\doi/10.1103/PhysRevD.81.083515}{Phys.\ Rev.\ D {\bf 81}, 083515}
  \href{\doi/10.1103/PhysRevD.81.083515}{(2010)}
  [\href{\arxiv/1002.3703}{1002.3703} [astro-ph.CO]].

\bibitem{BasteroGil:2009eb}
  M.~Bastero-Gil, A.~Berera, B.~M.~Jackson and A.~Taylor,
  ``Hybrid Quintessential Inflation,''
  \href{\doi/10.1016/j.physletb.2009.06.025}{Phys.\ Lett.\ B {\bf 678}, 157 (2009)}
  [\href{\arxiv/0905.2937}{0905.2937} [hep-ph]].

\bibitem{Chun:2009yu}
  E.~J.~Chun, S.~Scopel and I.~Zaballa,
  ``Gravitational reheating in quintessential inflation,''
  \href{\doi/10.1088/1475-7516/2009/07/022}{JCAP {\bf 0907}, 022}
  \href{\doi/10.1088/1475-7516/2009/07/022}{(2009)}
  [\href{\arxiv/0904.0675}{0904.0675} [hep-ph]].

\bibitem{Bento:2008yx}
  M.~C.~Bento, R.~G.~Felipe and N.~M.~C.~Santos,
  ``Brane assisted quintessential inflation with transient acceleration,''
  \href{\doi/10.1103/PhysRevD.77.123512}{Phys.}\
  \href{\doi/10.1103/PhysRevD.77.123512}{Rev.\ D {\bf 77}, 123512 (2008)}
  [\href{\arxiv/0801.3450}{0801.3450} [astro-ph]].

\bibitem{Matsuda:2007ax}
  T.~Matsuda,
  ``NO Curvatons or Hybrid Quintessential Inflation,''
  \href{\doi/10.1088/1475-7516/2007/08/003}{JCAP {\bf 0708}, 003}
  \href{\doi/10.1088/1475-7516/2007/08/003}{(2007)}
  [\href{\arxiv/0707.1948}{0707.1948} [hep-ph]].


\bibitem{Neupane:2007mu}
  I.~P.~Neupane,
  ``Reconstructing a model of quintessential inflation,''
  \href{\doi/10.1088/0264-9381/25/12/125013}{Class.\ Quant.}\
  \href{\doi/10.1088/0264-9381/25/12/125013}{Grav.\  {\bf 25}, 125013 (2008)}
  [\href{\arxiv/0706.2654}{0706.2654} [hep-th]].

\bibitem{Dimopoulos:2007bp}
  K.~Dimopoulos,
  ``Trapped Quintessential Inflation from Flux Compactifications,''
  \href{\arxiv/hep-ph/0702018}{hep-ph/0702018} [HEP-PH].

\bibitem{Gardner:2007ib}
  C.~L.~Gardner,
  ``Braneworld quintessential inflation and sum of exponentials potentials,''
  \href{\arxiv/hep-ph/0701036}{hep-ph/0701036}.
\bibitem{Da} M. Sami, N. Dadhich, TSPU Vestnik 44N7 (2004) 25-36
[\href{\arxiv/hep-th/0405016}{arXiv:hep-th/0405016}].
\bibitem{Rosenfeld:2006hs}
  R.~Rosenfeld and J.~A.~Frieman,
  ``Cosmic microwave background and large-scale structure constraints on a simple 
quintessential 
inflation model,''
  \href{\doi/10.1103/PhysRevD.75.043513}{Phys.\ Rev.\ D {\bf 75},}
  \href{\doi/10.1103/PhysRevD.75.043513}{043513 (2007)}
  [\href{\arxiv/astro-ph/0611241}{astro-ph/0611241}].

\bibitem{BuenoSanchez:2006ah}
  J.~C.~Bueno Sanchez and K.~Dimopoulos,
  ``Trapped quintessential inflation in the context of flux compactifications,''
  \href{\doi/10.1088/1475-7516/2007/10/002}{JCAP {\bf 0710},}
  \href{\doi/10.1088/1475-7516/2007/10/002}{002 (2007)}
  [\href{\arxiv/hep-th/0606223}{hep-th/0606223}].

\bibitem{Membiela:2006rj}
  A.~Membiela and M.~Bellini,
  ``Quintessential inflation from a variable cosmological constant in a 5D vacuum,''
  \href{\doi/10.1016/j.physletb.2006.08.043}{Phys.\ Lett.\ B {\bf 641}, 125}
  \href{\doi/10.1016/j.physletb.2006.08.043}{(2006)}
  [\href{\arxiv/gr-qc/0606119}{gr-qc/0606119}].

\bibitem{Cardenas:2006py}
  V.~H.~Cardenas,
  ``Tachyonic quintessential inflation,''
  \href{\doi/10.1103/PhysRevD.73.103512}{Phys.\ Rev.\ D {\bf 73}, 103512 (2006)}
  [\href{\arxiv/gr-qc/0603013}{gr-qc/0603013}].

\bibitem{Zhai:2005ub}
  X.~-h.~Zhai and Y.~-b.~Zhao,
  ``Dynamics of quintessential inflation,''
  \href{\doi/10.1088/1009-1963/15/10/046}{Chin.\ Phys.\  {\bf 15}, }
  \href{\doi/10.1088/1009-1963/15/10/046}{2465 (2006)}
  [\href{\arxiv/astro-ph/0511512}{astro-ph/0511512}].

\bibitem{Rosenfeld:2005mt}
  R.~Rosenfeld and J.~A.~Frieman,
  ``A Simple model for quintessential inflation,''
  \href{\doi/10.1088/1475-7516/2005/09/003}{JCAP }
  \href{\doi/10.1088/1475-7516/2005/09/003}{{\bf 0509}, 003 (2005)}
  [\href{\arxiv/astro-ph/0504191}{astro-ph/0504191}].

\bibitem{Giovannini:2003jw}
  M.~Giovannini,
  ``Low scale quintessential inflation,''
  \href{\doi/10.1103/PhysRevD.67.123512}{Phys.\ Rev.\ D {\bf 67}, 123512 (2003)}
  [\href{\arxiv/hep-ph/0301264}{hep-ph/0301264}].





\bibitem{Dimopoulos:2001qu}
  K.~Dimopoulos,
  ``Models of quintessential inflation,''
    [\href{\arxiv/astro-ph/0111500}{astro-ph/0111500}].


\bibitem{Dimopoulos:2001ix}
  K.~Dimopoulos and J.~W.~F.~Valle,
  ``Modeling quintessential inflation,''
  \href{\doi/10.1016/S0927-6505(02)00115-9}{Astropart.}\
  \href{\doi/10.1016/S0927-6505(02)00115-9}{Phys.\ {\bf 18}, 287 (2002)}
  [\href{\arxiv/astro-ph/0111417}{astro-ph/0111417}].

\bibitem{Yahiro:2001uh}
  M.~Yahiro, G.~J.~Mathews, K.~Ichiki, T.~Kajino and M.~Orito,
  ``Constraints on cosmic quintessence and quintessential inflation,''
  \href{\doi/10.1103/PhysRevD.65.063502}{Phys.\ Rev.\ D {\bf 65}, 063502 (2002)}
  [\href{\arxiv/astro-ph/0106349}{as}
  \href{\arxiv/astro-ph/0106349}{tro-ph/0106349}].

\bibitem{Kaganovich:2000fc}
  A.~B.~Kaganovich,
  ``Field theory model giving rise to 'quintessential inflation' without the cosmological 
constant 
and other fine tuning problems,''
  \href{\doi/10.1103/PhysRevD.63.025022}{Phys.\ Rev.\ D {\bf 63}, 025022 }
  \href{\doi/10.1103/PhysRevD.63.025022}{(2000)}
  [\href{\arxiv/hep-th/0007144}{hep-th/0007144}].



\bibitem{Baccigalupi:1998mn}
  C.~Baccigalupi and F.~Perrotta,
  ``Perturbations in quintessential inflation,''
  \href{\arxiv/astro-ph/9811385}{as}
  \href{\arxiv/astro-ph/9811385}{tro-ph/9811385}.



\bibitem{Lee:2014bwa}
  J.~Lee, T.~H.~Lee, P.~Oh and J.~Overduin,
  ``Cosmological Coincidence without Fine Tuning,''
  \href{\arxiv/1405.7681}{1405.7681} [hep-th].

\bibitem{Capozziello:2005tf}
  S.~Capozziello, S.~Nojiri and S.~D.~Odintsov,
  ``Unified phantom cosmology: Inflation, dark energy and dark matter under the same 
standard,''
  \href{\doi/10.1016/j.physletb.2005.11.012}{Phys.\ Lett.\ B {\bf 632}, 597 (2006)}
  [\href{\arxiv/hep-th/0507182}{hep-th/0507182}].

\bibitem{Nojiri:2005pu}
  S.~Nojiri and S.~D.~Odintsov,
  ``Unifying phantom inflation with late-time acceleration: Scalar phantom-non-phantom 
transition 
model and generalized holographic dark energy,''
  \href{\doi/10.1007/s10714-006-0301-6}{Gen.\ Rel.\ Grav.\  {\bf 38}, 1285 (2006)}
  [\href{\arxiv/hep-th/0506212}{hep-th/0506212}].

\bibitem{Elizalde:2008yf}
  E.~Elizalde, S.~Nojiri, S.~D.~Odintsov, D.~Saez-Gomez and V.~Faraoni,
  ``Reconstructing the universe history, from inflation to acceleration, with phantom and 
canonical 
scalar fields,''
  \href{\doi/10.1103/PhysRevD.77.106005}{Phys.\ Rev.\ D {\bf 77}, 106005 (2008)}
  [\href{\arxiv/arXiv:0803.1311}{arXiv:0803.1311} [hep-th]].
  \bibitem{Hossain:2014coa}
  M.~W.~Hossain, R.~Myrzakulov, M.~Sami and E.~N.~Saridakis,
  ``Class of quintessential inflation models with parameter space consistent with 
BICEP2,''
  \href{\doi/10.1103/PhysRevD.89.123513}{Phys.\ Rev.\ D {\bf 89},}
  \href{\doi/10.1103/PhysRevD.89.123513}{123513 (2014)}
  [\href{\arxiv/1404.1445}{arXiv:1404.1445} [gr-qc]].
\bibitem{Guendelman:2002js} 
  E.~I.~Guendelman and O.~Katz,
  ``Inflation and transition to a slowly accelerating phase from SSB of scale invariance,''
  \href{\doi/10.1088/0264-9381/20/9/309}{Class.\ Quant.\ Grav.\  {\bf 20}, 1715 (2003)}
  [\href{\arxiv/gr-qc/0211095}{gr-qc/0211095}].
 
\bibitem{WaliHossain:2014usl}
M.~Wali Hossain, R.~Myrzakulov, M.~Sami and E.~N.~Saridakis,
Int. J. Mod. Phys. D \textbf{24}, no.05, 1530014 (2015)
\href{\doi/10.1142/S0218271815300141}{Int. J. Mod. Phys. D \textbf{24}, no.05, 1530014 (2015)}
[\href{\arxiv/1410.6100}{arXiv:1410.6100} [gr-qc]].
\bibitem{Ahmad:2017itq}
S.~Ahmad, R.~Myrzakulov and M.~Sami,
\href{\doi/10.1103/PhysRevD.96.063515}{Phys. Rev. D \textbf{96}, no.6, 063515 (2017)}
[\href{\arxiv/1705.02133}{arXiv:1705.02133} [gr-qc]].
\bibitem{deHaro:2021swo}
J.~de Haro and L.~A.~Sal\'o,
\href{\doi/10.3390/galaxies9040073}{Galaxies \textbf{9}, no.4, 73 (2021)}
[\href{\arxiv/2108.11144}{arXiv:2108.11144} [gr-qc]].
%
 

\bibitem{Dimopoulos:2020pas}
K.~Dimopoulos and S.~S\'anchez L\'opez,
\href{\doi/10.1103/PhysRevD.103.043533}{Phys. Rev. D \textbf{103}, no.4, 043533 (2021)}
[\href{\arxiv/2012.06831}{arXiv:2012.06831} [gr-qc]]

\bibitem{Benisty:2020xqm}
D.~Benisty and E.~I.~Guendelman,
\href{\doi/10.1142/S021827182042002X}{Int. J. Mod. Phys. D \textbf{29}, no.14, 2042002 (2020)}

[\href{\arxiv/2004.00339}{arXiv:2004.00339} [astro-ph.CO]].
\bibitem{Karciauskas:2021fdu}
M.~Kar\v{c}iauskas, S.~Rusak and A.~Saez,
[\href{\arxiv/2112.11536}{arXiv:2112.11536}[astro-ph.CO]].
\bibitem{Capozziello:2003tk}
S.~Capozziello, S.~Carloni and A.~Troisi,
Recent Res. Dev. Astron. Astrophys. \textbf{1}, 625 (2003)
[\href{\arxiv/astro-ph/0303041}{arXiv:astro-ph/0303041} [astro-ph]].

\bibitem{Sami:2004xk}
M.~Sami and V.~Sahni,
\href{\doi/10.1103/PhysRevD.70.083513}{Phys. Rev. D \textbf{70}, 083513 (2004)}
[\href{\arxiv/hep-th/0402086}{arXiv:hep-th/0402086} [hep-th]].
\bibitem{Dimopoulos:2017zvq}
K.~Dimopoulos and C.~Owen,
\href{\doi/10.1088/1475-7516/2017/06/027}{JCAP \textbf{06}, 027 (2017)}
[\href{\arxiv/1703.00305}{arXiv:1703.00305} [gr-qc]].
\bibitem{Bettoni:2021qfs}
D.~Bettoni and J.~Rubio,
[\href{\arxiv/2112.11948}{arXiv:2112.11948} [astro-ph.CO]].


\bibitem{Dimopoulos:2017tud}
K.~Dimopoulos, L.~Donaldson Wood and C.~Owen,
\href{\doi/10.1103/PhysRevD.97.063525}{Phys. Rev. D \textbf{97}, no.6, 063525 (2018)}
[\href{\arxiv/1712.01760}{arXiv:1712.01760} [astro-ph.CO]].

\bibitem{Wetterich:2022brb}
C.~Wetterich,
[\href{\arxiv/2201.12213}{arXiv:2201.12213} [astro-ph.CO]].
\bibitem{Jaman:2018ucm}
N.~Jaman and K.~Myrzakulov,
\href{\doi/10.1103/PhysRevD.99.103523}{Phys. Rev. D \textbf{99}, no.10, 103523 (2019)}
[\href{\arxiv/1807.07443}{arXiv:1807.07443} [gr-qc]].
\bibitem{Rosati:2003yw}
F.~Rosati,
Phys. Lett. B \textbf{570}, 5-10 (2003)
\href{\doi/10.1016/j.physletb.2003.07.048}{Phys. Lett. B \textbf{570}, 5-10 (2003)
}
[\href{\arxiv/hep-ph/0302159}{arXiv:hep-ph/0302159} [hep-ph]].
\bibitem{Salati:2002md}
P.~Salati,
\href{\doi/10.1016/j.physletb.2003.07.073}{Phys. Lett. B \textbf{571}, 121-131 (2003)
}
[\href{\arxiv/astro-ph/0207396}{arXiv:astro-ph/0207396} [astro-ph]].

\bibitem{Akrami:2017cir}
Y.~Akrami, R.~Kallosh, A.~Linde and V.~Vardanyan,
\href{doi:10.1088/1475-7516/2018/06/041}{JCAP \textbf{06}, 041 (2018)}
[\href{\arxiv/1712.09693}{arXiv:1712.09693} [hep-th]].

\bibitem{Akrami:2020zxw}
Y.~Akrami, S.~Casas, S.~Deng and V.~Vardanyan,
\href{\doi/10.1088/1475-7516/2021/04/006}{JCAP \textbf{04}, 006 (2021)}
[\href{\arxiv/2010.15822}{arXiv:2010.15822} [astro-ph.CO]].
\bibitem{Saba:2017xur}
N.~Saba and M.~Farhoudi,
\href{\doi/10.1142/S0218271818500414}{Int. J. Mod. Phys. D \textbf{27}, no.04, 1850041 (2017)}
[\href{\arxiv/1711.09682}{arXiv:1711.09682} [gr-qc]].

\bibitem{Albrecht:1982mp}
A.~Albrecht, P.~J.~Steinhardt, M.~S.~Turner and F.~Wilczek,
\href{\doi/10.1103/PhysRevLett.48.1437}{Phys. Rev. Lett. \textbf{48}, 1437 (1982)}.
\bibitem{Dolgov:1982th}
A.~D.~Dolgov and A.~D.~Linde,
\href{\doi/10.1016/0370-2693(82)90292-1}{Phys. Lett. B \textbf{116}, 329 (1982)}.
\bibitem{Abbott:1982hn}
L.~F.~Abbott, E.~Farhi and M.~B.~Wise,
\href{\doi/10.1016/0370-2693(82)90867-X}{Phys. Lett. B \textbf{117}, 29 (1982)}.
\bibitem{Ford:1986sy}
L.~H.~Ford,
\href{\doi/10.1103/PhysRevD.35.2955}{Phys. Rev. D \textbf{35}, 2955 (1987)}.

\bibitem{Dolgov:1989us}
A.~D.~Dolgov and D.~P.~Kirilova,
Sov. J. Nucl. Phys. \textbf{51}, 172-177 (1990)
JINR-E2-89-321.
\bibitem{Traschen:1990sw}
J.~H.~Traschen and R.~H.~Brandenberger,
\href{\doi/10.1103/PhysRevD.42.2491}{Phys. Rev. D \textbf{42}, 2491-2504 (1990)}.
\bibitem{Spokoiny:1993kt}
B.~Spokoiny,
\href{\doi/10.1016/0370-2693(93)90155-B}{Phys. Lett. B \textbf{315}, 40-45 (1993)}
[\href{\arxiv/gr-qc/9306008}{arXiv:gr-qc/9306008} [gr-qc]].
\bibitem{Shtanov:1994ce}
Y.~Shtanov, J.~H.~Traschen and R.~H.~Brandenberger,
\href{\doi/10.1103/PhysRevD.51.5438}{Phys. Rev. D \textbf{51}, 5438-5455 (1995)}
[\href{\arxiv/hep-ph/9407247}{arXiv:hep-ph/9407247} [hep-ph]].
\bibitem{Kofman:1994rk}
L.~Kofman, A.~D.~Linde and A.~A.~Starobinsky,
\href{\doi/10.1103/PhysRevLett.73.3195}{Phys. Rev. Lett. \textbf{73}, 3195-3198 (1994)}
[\href{\arxiv/hep-th/9405187}{arXiv:hep-th/9405187} [hep-th]].
\bibitem{Kofman:1997yn}
L.~Kofman, A.~D.~Linde and A.~A.~Starobinsky,
\href{\doi/10.1103/PhysRevD.56.3258}{Phys. Rev. D \textbf{56}, 3258-3295 (1997)}
[\href{\arxiv/hep-ph/9704452}{arXiv:hep-ph/9704452} [hep-ph]].
\bibitem{Garcia-Bellido:1997hex}
J.~Garcia-Bellido and A.~D.~Linde,
\href{\doi/10.1103/PhysRevD.57.6075}{Phys. Rev. D \textbf{57}, 6075-6088 (1998)}
[\href{\arxiv:hep-ph/9711360}{arXiv:hep-ph/9711360} [hep-ph]].


\bibitem{Felder:1998vq}
G.~N.~Felder, L.~Kofman and A.~D.~Linde,
Phys. Rev. D \textbf{59}, 123523 (1999)
[\href{\arxiv/hep-ph/9812289}{arXiv:hep-ph/9812289} [hep-ph]].
\bibitem{Lyth:2001nq}
D.~H.~Lyth and D.~Wands,
\href{\doi/10.1016/S0370-2693(01)01366-1}{Phys. Lett. B \textbf{524}, 5-14 (2002)}
[\href{\arxiv/hep-ph/0110002 [hep-ph]}{arXiv:hep-ph/0110002 [hep-ph]}].


\bibitem{Feng:2002nb}
B.~Feng and M.~z.~Li,
\href{\doi/10.1016/S0370-2693(03)00589-6}{Phys. Lett. B \textbf{564}, 169-174 (2003)}
[\href{\arxiv/hep-ph/0212213}{arXiv:hep-ph/0212213} [hep-ph]].

\bibitem{delCampo:2009yc}
S.~del Campo, R.~Herrera, J.~Saavedra, C.~Campuzano and E.~Rojas,
\href{\doi/10.1103/PhysRevD.80.123531}{Phys. Rev. D \textbf{80}, 123531 (2009)}
[\href{\arxiv/0912.4721}{arXiv:0912.4721} [astro-ph.CO]].

\bibitem{Bassett:2005xm}
B.~A.~Bassett, S.~Tsujikawa and D.~Wands,
\href{\doi/10.1103/RevModPhys.78.537}{Rev. Mod. Phys. \textbf{78}, 537-589 (2006)}
[\href{\arxiv/astro-ph/0507632}{arXiv:astro-ph/0507632} [astro-ph]].
\bibitem{Hardwick:2016whe}
R.~J.~Hardwick, V.~Vennin, K.~Koyama and D.~Wands,
\href{\doi/10.1088/1475-7516/2016/08/042}{JCAP \textbf{08}, 042 (2016)}
[\href{\arxiv/1606.01223}{arXiv:1606.01223} [astro-ph.CO]].
\bibitem{Campos:2002yk}
A.~H.~Campos, H.~C.~Reis and R.~Rosenfeld,
\href{\doi/10.1016/j.physletb.2003.09.064}{Phys. Lett. B \textbf{575}, 151-156 (2003)}
[\href{\arxiv/hep-ph/0210152}{arXiv:hep-ph/0210152} [hep-ph]].
\bibitem{Allahverdi:2010xz}
R.~Allahverdi, R.~Brandenberger, F.~Y.~Cyr-Racine and A.~Mazumdar,
\href{\doi/10.1146/annurev.nucl.012809.104511}{Ann. Rev. Nucl. Part. Sci. \textbf{60}, 27-51 (2010)}
[\href{\arxiv/1001.2600}{arXiv:1001.2600} [hep-th]].

\bibitem{Amin:2014eta}
M.~A.~Amin, M.~P.~Hertzberg, D.~I.~Kaiser and J.~Karouby,
\href{\doi/10.1146/annurev.nucl.012809.104511}{Int. J. Mod. Phys. D \textbf{24}, 1530003 (2014)}
[\href{\arxiv/1410.3808 }{arXiv:1410.3808}[hep-ph]].
\bibitem{Garcia:2020eof}
M.~A.~G.~Garcia, K.~Kaneta, Y.~Mambrini and K.~A.~Olive,
\href{\doi/10.1103/PhysRevD.101.123507}{Phys. Rev. D \textbf{101}, no.12, 123507 (2020)}
[\href{\arxiv/2004.08404}{arXiv:2004.08404} [hep-ph]].
\bibitem{Tambalo:2016eqr}
G.~Tambalo and M.~Rinaldi,
\href{\doi/10.1007/s10714-017-2217-8}{Gen. Rel. Grav. \textbf{49}, no.4, 52 (2017)}
[\href{\arxiv/1610.06478}{arXiv:1610.06478} [gr-qc]].

\bibitem{Lopez:2021agu}
M.~L\'opez, G.~Otalora and N.~Videla,
\href{\doi/10.1088/1475-7516/2021/10/021}{JCAP \textbf{10}, 021 (2021)}
[\href{\arxiv/2107.07679 }{arXiv:2107.07679 }[gr-qc]].
\bibitem{Saha:2021kez}
P.~Saha,
[\href{\arxiv/2108.06612}{arXiv:2108.06612} [astro-ph.CO]].

\bibitem{Pareek:2021lxz}
P.~Pareek and A.~Nautiyal,
\href{\doi/10.1103/PhysRevD.104.083526}{Phys. Rev. D \textbf{104}, no.8, 083526 (2021)}
[\href{\arxiv/2103.01797}{arXiv:2103.01797} [astro-ph.CO]].
\bibitem{Bhattacharya:2019ryo}
S.~Bhattacharya, K.~Das and M.~R.~Gangopadhyay,
{\doi/10.1088/1361-6382/abbb64}{Class. Quant. Grav. \textbf{37}, no.21, 215009 (2020)}
[\href{arXiv:1908.02542}{arXiv:1908.02542} [astro-ph.CO]].
\bibitem{Dimopoulos:2019gpz}
K.~Dimopoulos and L.~Donaldson-Wood,
\href{\doi/10.1016/j.physletb.2019.07.017}{Phys. Lett. B \textbf{796}, 26-31 (2019)}
[\href{arXiv:1906.09648}{arXiv:1906.09648} [gr-qc]].
\bibitem{Carroll:2003wy}
S.~M.~Carroll, V.~Duvvuri, M.~Trodden and M.~S.~Turner,
\href{\doi/10.1103/PhysRevD.70.043528}{Phys. Rev. D \textbf{70}, 043528 (2004)}
[\href{\arxiv/astro-ph/0306438}{arXiv:astro-ph/0306438} [astro-ph]].
\bibitem{DeFelice:2010aj}
A.~De Felice and S.~Tsujikawa,
\href{\doi/10.12942/lrr-2010-3}{Living Rev. Rel. \textbf{13}, 3 (2010)}
[\href{\arxiv/1002.4928 }{arXiv:1002.4928}[gr-qc]].
\bibitem{Sotiriou:2008rp}
T.~P.~Sotiriou and V.~Faraoni,
\href{\doi/10.1103/RevModPhys.82.451}{Rev. Mod. Phys. \textbf{82}, 451-497 (2010)}
[\href{\arxiv/0805.1726}{arXiv:0805.1726} [gr-qc]].
\bibitem{Gannouji:2012iy}
R.~Gannouji, M.~Sami and I.~Thongkool,
\href{\doi/10.1016/j.physletb.2012.08.015}{Phys. Lett. B \textbf{716}, 255-259 (2012)}
[\href{\arxiv/1206.3395}{arXiv:1206.3395} [hep-th]].

\bibitem{Cosmai:2013nva}
L.~Cosmai, G.~Fanizza and L.~Tedesco,
\href{\doi/10.1007/s10773-015-2713-0}{Int. J. Theor. Phys. \textbf{55}, no.2, 754-765 (2016)}
[\href{\arxiv/1311.7281}{arXiv:1311.7281} [astro-ph.CO]].

\bibitem{Nojiri:2007uq}
S.~Nojiri and S.~D.~Odintsov,
\href{\doi/10.1016/j.physletb.2007.12.001}{Phys. Lett. B \textbf{659}, 821-826 (2008)}
[\href{\arxiv/0708.0924}{arXiv:0708.0924} [hep-th]].






\bibitem{Dimopoulos:2018wfg}
K.~Dimopoulos and T.~Markkanen,
\href{\doi/10.1088/1475-7516/2018/06/021}{JCAP \textbf{06}, 021 (2018)}
[\href{\arxiv/1803.07399}{arXiv:1803.07399} [gr-qc]].

\bibitem{Bettoni:2021zhq}
D.~Bettoni, A.~Lopez-Eiguren and J.~Rubio,
\href{\doi/10.1088/1475-7516/2022/01/002}{JCAP \textbf{01}, no.01, 002 (2022)}
[\href{\arxiv/2107.09671}{arXiv:2107.09671} [hep-ph]].

\bibitem{Opferkuch:2019zbd}
T.~Opferkuch, P.~Schwaller and B.~A.~Stefanek,
\href{\doi/10.1088/1475-7516/2019/07/016}{JCAP \textbf{07}, 016 (2019)}
[\href{\arxiv/1905.06823}{arXiv:1905.06823} [gr-qc]].

\bibitem{Tashiro:2003qp}
H.~Tashiro, T.~Chiba and M.~Sasaki,
\href{\doi/10.1088/0264-9381/21/7/004}{Class. Quant. Grav. \textbf{21}, 1761-1772 (2004)}
[\href{\arxiv/gr-qc/0307068}{arXiv:gr-qc/0307068} [gr-qc]].
\bibitem{Chun:2007np}
E.~J.~Chun and S.~Scopel,
\href{\doi/10.1088/1475-7516/2007/10/011}{JCAP \textbf{10}, 011 (2007)}
[\href{\arxiv/0707.1544}{arXiv:0707.1544} [astro-ph]].
\bibitem{Kamada:2019ewe}
K.~Kamada, J.~Kume, Y.~Yamada and J.~Yokoyama,
\href{\doi:/10.1088/1475-7516/2020/01/016}{JCAP \textbf{01}, 016 (2020)}
[\href{\arxiv/1911.02657}{arXiv:1911.02657} [hep-ph]].
\bibitem{Berera:1995wh}
A.~Berera and L.~Z.~Fang,
\href{\doi:10.1103/PhysRevLett.74.1912}{Phys. Rev. Lett. \textbf{74}, 1912-1915 (1995)}
[\href{\arxiv/astro-ph/9501024}{arXiv:astro-ph/9501024} [astro-ph]].
\bibitem{Berera:1995ie}
A.~Berera,
\href{\doi/10.1103/PhysRevLett.75.3218}{Phys. Rev. Lett. \textbf{75}, 3218-3221 (1995)}
[\href{\arxiv/astro-ph/9509049}{arXiv:astro-ph/9509049} [astro-ph]].
\bibitem{Lima:2019yyv}
G.~B.~F.~Lima and R.~O.~Ramos,
\href{\doi/10.1103/PhysRevD.100.123529}{Phys. Rev. D \textbf{100}, 123529 (2019)}
[\href{arXiv:1910.05185}{arXiv:1910.05185} [astro-ph.CO]].
\bibitem{Basak:2021cgk}
S.~Basak, S.~Bhattacharya, M.~R.~Gangopadhyay, N.~Jaman, R.~Rangarajan and M.~Sami,
[\href{\arxiv/2110.00607}{arXiv:2110.00607} [astro-ph.CO]].
\bibitem{Levy:2020zfo}
M.~Levy, J.~G.~Rosa and L.~B.~Ventura,
\href{\doi/10.1007/JHEP12(2021)176}{JHEP \textbf{12}, 176 (2021)}
[\href{\arxiv/2012.03988}{arXiv:2012.03988} [hep-ph]].
\bibitem{Gangopadhyay:2020bxn}
M.~R.~Gangopadhyay, S.~Myrzakul, M.~Sami and M.~K.~Sharma,
\href{\doi/10.1103/PhysRevD.103.043505}{Phys. Rev. D \textbf{103}, no.4, 043505 (2021)}
[\href{\arxiv/2011.09155}{arXiv:2011.09155} [astro-ph.CO]].
\bibitem{Ferreira:1997au}
P.~G.~Ferreira and M.~Joyce,
\href{\doi/10.1103/PhysRevLett.79.4740}{Phys. Rev. Lett. \textbf{79}, 4740-4743 (1997)}
[\href{\arxiv/astro-ph/9707286}{arXiv:astro-ph/9707286} [astro-ph]].
\bibitem{Ferreira:1997hj}
P.~G.~Ferreira and M.~Joyce,
\href{\doi/10.1103/PhysRevD.58.023503}{Phys. Rev. D \textbf{58}, 023503 (1998)}
[\href{\arxiv/astro-ph/9711102}{arXiv:astro-ph/9711102} [astro-ph]].
\bibitem{Copeland:1997et}
E.~J.~Copeland, A.~R.~Liddle and D.~Wands,
\href{\doi/10.1103/PhysRevD.57.4686}{Phys. Rev. D \textbf{57}, 4686-4690 (1998)}
[\href{\arxiv/gr-qc/9711068}{arXiv:gr-qc} [gr-qc]].
\bibitem{SS} S. Tsujikawa, M. Sami,
\href{\doi/10.1016/j.physletb.2004.10.023}{Phys. Lett. B \textbf{603}, 113-123 (2004)}
[\href{\arxiv/hep-th/0409212}{arXiv:hep-th/0409212} [hep-th]].
\bibitem{Steinhardt:1999nw}
P.~J.~Steinhardt, L.~M.~Wang and I.~Zlatev,
\href{\doi:10.1103/PhysRevD.59.123504}{Phys. Rev. D \textbf{59}, 123504 (1999)}
[\href{\arxiv/astro-ph/9812313}{arXiv:astro-ph/9812313} [astro-ph]].
\bibitem{Chiba:2009gg}
T.~Chiba,
\href{\doi/10.1103/PhysRevD.81.023515}{Phys. Rev. D \textbf{81}, 023515 (2010)}
[\href{\arxiv/0909.4365}{arXiv:0909.4365} [astro-ph.CO]].
\bibitem{Barreiro:1999zs}
T.~Barreiro, E.~J.~Copeland and N.~J.~Nunes,
\href{doi:10.1103/PhysRevD.61.127301}{Phys. Rev. D \textbf{61}, 127301 (2000)}
[\href{\arxiv/astro-ph/9910214}{arXiv:astro-ph/9910214} [astro-ph]].
\bibitem{Haro:2019peq}
J.~Haro, J.~Amor\'os and S.~Pan,
\href{doi:10.1140/epjc/s10052-020-7950-6}{Eur. Phys. J. C \textbf{80}, no.5, 404 (2020)}
[\href{\arxiv/1908.01516}{arXiv:1908.01516} [gr-qc]].
\bibitem{exitG} S. Tsujikawa, M. Sami,
\href{\doi/10.1088/1475-7516/2007/01/006}{JCAP \textbf{0701},006(2007)};
	[\href{\arxiv/hep-th/0608178}{arXiv:hep-th/0608178}]; Ekaterina O. Pozdeeva, M. Sami, Alexey V. Toporensky, Sergey Yu. Vernov, 
	\href{\doi/10.1088/1475-7516/2005/06/007}{Phys. Rev. D 100, 083527 (2019)}[ 	\href{\arxiv/hep-th/0502191}{arXiv:hep-th/0502191}].
	
\bibitem{Gumjudpai:2005ry}
B.~Gumjudpai, T.~Naskar, M.~Sami and S.~Tsujikawa,
\href{\doi/10.1088/1475-7516/2005/06/007}{JCAP \textbf{06}, 007 (2005)}
[\href{\arxiv/hep-th/0502191}{arXiv:hep-th/0502191} [hep-th]].
\bibitem{Adhikari:2020xcg}
R.~Adhikari, M.~R.~Gangopadhyay and Yogesh,
\href{\doi/10.1140/epjc/s10052-020-08460-3}{Eur. Phys. J. C \textbf{80}, no.9, 899 (2020)}
[\href{\arxiv/2002.07061}{arXiv:2002.07061} [astro-ph.CO]].
\bibitem{Copeland:2000hn}
E.~J.~Copeland, A.~R.~Liddle and J.~E.~Lidsey,
\href{\doi/10.1103/PhysRevD.64.023509}{Phys. Rev. D \textbf{64}, 023509 (2001)}
[\href{\arxiv/astro-ph/0006421}{arxiv:astro-ph/0006421} [astro-ph]].
\bibitem{Geng:2015fla}
C.~Q.~Geng, M.~W.~Hossain, R.~Myrzakulov, M.~Sami and E.~N.~Saridakis,
\href{\doi/10.1103/PhysRevD.92.023522}{Phys. Rev. D \textbf{92}, no.2, 023522 (2015)}
[\href{\arxiv/1502.03597}{arXiv:1502.03597} [gr-qc]].
\bibitem{Carroll:2003qq}
S.~M.~Carroll,
\href{doi:10.1063/1.1848314}{eConf \textbf{C0307282}, TTH09 (2003)}
[\href{\arxiv/astro-ph/0310342}{arXiv:astro-ph/0310342} [astro-ph]].
\bibitem{Sahni:2004ai}
V.~Sahni,
\href{\doi/10.1007/b99562}{Lect. Notes Phys. \textbf{653}, 141-180 (2004)}
[\href{\arxiv/astro-ph/0403324} {arXiv:astro-ph/0403324 }[astro-ph]].
\bibitem{Sahni:2006pa}
V.~Sahni and A.~Starobinsky,
\href{\doi/10.1142/S0218271806009704}{Int. J. Mod. Phys. D \textbf{15}, 2105-2132 (2006)}
[\href{\arxiv/astro-ph/0610026}{arXiv:astro-ph/0610026} [astro-ph]].
\bibitem{Mortonson:2013zfa}
M.~J.~Mortonson, D.~H.~Weinberg and M.~White,
[\href{arXiv:1401.0046}{arXiv:1401.0046} [astro-ph.CO]].
\bibitem{Sami:2013ssa}
M.~Sami and R.~Myrzakulov,
\href{\doi/10.1142/S0218271816300317}{Int. J. Mod. Phys. D \textbf{25}, no.12, 1630031 (2016)}
[\href{\arxiv/1309.4188}{arXiv:1309.4188} [hep-th]].
\bibitem{Copeland:2006wr}
E.~J.~Copeland, M.~Sami and S.~Tsujikawa,
\href{\doi/10.1142/S021827180600942X}{Int. J. Mod. Phys. D \textbf{15}, 1753-1936 (2006)}
\href{\arxiv/hep-th/0603057}{arXiv:hep-th/0603057} [hep-th]].
\bibitem{Li:2012dt}
M.~Li, X.~D.~Li, S.~Wang and Y.~Wang,
\href{doi:10.1007/s11467-013-0300-5}{Front. Phys. (Beijing) \textbf{8}, 828-846 (2013)}
[\href{\arxiv/1209.0922}{arXiv:1209.0922}[astro-ph.CO]].
\bibitem{Brax:2017idh}
P.~Brax,
\href{\doi/10.1088/1361-6633/aa8e64}{Rept. Prog. Phys. \textbf{81}, no.1, 016902 (2018)}
\bibitem{Zhang:2021ygh}
Z.~Zhang,
\href{\doi/10.1088/1361-6382/ac38d1}{Class. Quant. Grav. \textbf{39}, no.1, 015003 (2022)}
[\href{\arxiv/2112.04149} {arXiv:2112.04149}[gr-qc]].


\bibitem{Ratra:1987rm}
B.~Ratra and P.~J.~E.~Peebles,
\href{\doi/10.1103/PhysRevD.37.3406}{Phys. Rev. D \textbf{37}, 3406 (1988)}.

\bibitem{samrev1}M. Sami,     Lect.Notes Phys. {\bf 720} (2007) 219-256,Contribution to:3rd Aegean Summer School: The Invisible Universe: Dark Matter and Dark Energy.
\bibitem{samrev2} M. Sami,
Curr. Sci. {\bf 97},887(2009)[\href{\arxiv/0904.3445}{arXiv:0904.3445}] [hep-th].
\bibitem{Planck:2018jri}
Y.~Akrami \textit{et al.} [Planck],
\href{\doi/10.1051/0004-6361/201833887}{Astron. Astrophys. \textbf{641}, A10 (2020)}
[\href{\arxiv/1807.06211}{arXiv:1807.06211} [astro-ph.CO]].
\bibitem{Tristram:2021tvh}
M.~Tristram, A.~J.~Banday, K.~M.~G\'orski, R.~Keskitalo, C.~R.~Lawrence, K.~J.~Andersen, R.~B.~Barreiro, J.~Borrill, L.~P.~L.~Colombo and H.~K.~Eriksen, \textit{et al.}
[\href{\arxiv/2112.07961}{arXiv:2112.07961} [astro-ph.CO]].


\bibitem{Liddle:1998xm}
A.~R.~Liddle and R.~J.~Scherrer,
\href{\doi/10.1103/PhysRevD.59.023509}{Phys. Rev. D \textbf{59}, 023509 (1999)}
[\href{\arxiv/astro-ph/9809272}{arXiv:astro-ph/9809272} [astro-ph]].
\bibitem{Skugoreva:2019blk}
M.~A.~Skugoreva, M.~Sami and N.~Jaman,
\href{\doi/ 	10.1103/PhysRevD.100.043512}{Phys. Rev. D \textbf{100}, no.4, 043512 (2019)}
[\href{\arxiv/1901.06036}{arXiv:1901.06036} [gr-qc]].
\bibitem{Kolb:1990vq}
E.~W.~Kolb and M.~S.~Turner,
\href{\doi/10.1201/9780429492860}{Front. Phys. \textbf{69}, 1-547 (1990)}.
\bibitem{Husdal:2016haj}
L.~Husdal,
{Galaxies \textbf{4}, no.4, 78 (2016)}
[\href{\arxiv:1609.04979}{arXiv:1609.04979} [astro-ph.CO]].
\bibitem{Cyburt:2015mya}
R.~H.~Cyburt, B.~D.~Fields, K.~A.~Olive and T.~H.~Yeh,
\href{\doi/10.1103/RevModPhys.88.015004}{Rev. Mod. Phys. \textbf{88}, 015004 (2016)}
[\href{\arxiv/1505.01076}{arXiv:1505.01076} [astro-ph.CO]].
\bibitem{Caprini:2018mtu}
C.~Caprini and D.~G.~Figueroa,
\href{\doi/10.1088/1361-6382/aac608}{Class. Quant. Grav. \textbf{35}, no.16, 163001 (2018)}
[\href{\arxiv/1801.04268}{arXiv:1801.04268} [astro-ph.CO]].
\bibitem{Chiba:2009sj}
T. Chiba,
\href{\doi/10.1103/PhysRevD.80.109902}{Phys. Rev. D \textbf{79}, 083517 (2009)
[erratum: Phys. Rev. D \textbf{80}, 109902 (2009)]}
 \href{\arxiv/0902.4037}{arXiv:0902.4037} [astro-ph.CO]


\bibitem{Cyburt:2004yc}
R.~H.~Cyburt, B.~D.~Fields, K.~A.~Olive and E.~Skillman,
\href{\doi/10.1016/j.astropartphys.2005.01.005}{Astropart. Phys. \textbf{23}, 313-323 (2005)}
[\href{\arxiv/astro-ph/0408033}{arXiv:astro-ph/0408033} [astro-ph]].
\bibitem{gammascale1}
A. de la Macorra, G. Piccinelli
 \href{\doi/10.1103/PhysRevD.61.123503}{ Phys. Rev. D {\bf 61},
123503 (2000)}.
\bibitem{Nunes:2000yc}
A.~Nunes and J.~P.~Mimoso,
\href{\doi/10.1016/S0370-2693(00)00919-9}{Phys. Lett. B \textbf{488}, 423-427 (2000)}
[\href{\arxiv/gr-qc/0008003}{arXiv:gr-qc/0008003} [gr-qc]].
\bibitem{Ng:2001hs}
S.~C.~C.~Ng, N.~J.~Nunes and F.~Rosati,
\href{\doi/10.1103/PhysRevD.64.083510}{Phys. Rev. D \textbf{64}, 083510 (2001)}
[\href{\arxiv/astro-ph/0107321}{arXiv:astro-ph/0107321} [astro-ph]].
\bibitem{Wetterich:2013jsa}
C.~Wetterich,
\href{doi:10.1103/PhysRevD.89.024005}{Phys. Rev. D \textbf{89}, no.2, 024005 (2014)}
[\href{\arxiv/1308.1019}{arXiv:1308.1019} [astro-ph.CO]].

\bibitem{Caldwell:2005tm}
R.~R.~Caldwell and E.~V.~Linder,
\href{\doi/10.1103/PhysRevLett.95.141301}{Phys. Rev. Lett. \textbf{95}, 141301 (2005)}
[\href{\arxiv/astro-ph/0505494}{arXiv:astro-ph/0505494} [astro-ph]].
\bibitem{Linder:2006sv}
E.~V.~Linder,
\href{\doi/10.1103/PhysRevD.73.063010}{Phys. Rev. D \textbf{73}, 063010 (2006)}
[arXiv:astro-ph/0601052 [astro-ph]].
\bibitem{Scherrer:2007pu}
R.~J.~Scherrer and A.~A.~Sen,
\href{\doi/10.1103/PhysRevD.77.083515}{Phys. Rev. D \textbf{77}, 083515 (2008)}
[\href{\arxiv/0712.3450}{arXiv:0712.3450} [astro-ph]].
\bibitem{Linder:2015zxa}
E.~V.~Linder,
\href{\doi/10.1103/PhysRevD.91.063006}{Phys. Rev. D \textbf{91}, 063006 (2015)}
[\href{\arxiv/1501.01634} {arXiv:1501.01634}[astro-ph.CO]].

\bibitem{Urena-Lopez:2020npg}
L.~A.~Ure\~na-L\'opez and N.~Roy,
\href{\doi/10.1103/PhysRevD.102.063510}{Phys. Rev. D \textbf{102}, no.6, 063510 (2020)}
[\href{\arxiv/2007.08873}{arXiv:2007.08873} [astro-ph.CO]].
\bibitem{staro} Chao-Qiang Geng, Chung-Chi Lee, M. Sami, Emmanuel N. Saridakis, Alexei A. Starobinsky, 
\href{\doi/10.1088/1475-7516/2017/06/011}{JCAP \textbf{06}, 011 (2017)}
[\href{\arxiv/1705.01329}{arXiv:1705.01329} [gr-qc]].
\bibitem{Scherrer:2022umm}
R.~J.~Scherrer,
[\href{\arxiv/2202.01132}{arXiv:2202.01132} [gr-qc]].
\bibitem{Bartolo:1999sq}
N.~Bartolo and M.~Pietroni,
\href{\doi/10.1103/PhysRevD.61.023518}{Phys. Rev. D \textbf{61}, 023518 (2000)}
\href{\arxiv/hep-ph/9908521}{[arXiv:hep-ph/9908521} [hep-ph]].
\bibitem{Saridakis:2010mf}
E.~N.~Saridakis and S.~V.~Sushkov,
\href{\doi/10.1103/PhysRevD.81.083510}{
Phys. Rev. D \textbf{81}, 083510 (2010)}
[\href{\arxiv:1002.3478}{arXiv:1002.3478} [gr-qc]].

\bibitem{Wetterich:2013wza}
C.~Wetterich,
\href{\doi/10.1016/j.physletb.2013.08.023}{Phys. Lett. B \textbf{726}, 15-22 (2013)}
[\href{\arxiv/1303.4700}{arXiv:1303.4700} [astro-ph.CO]].
\bibitem{Hossain:2014xha}
M.~W.~Hossain, R.~Myrzakulov, M.~Sami and E.~N.~Saridakis,
\href{\doi/10.1103/PhysRevD.90.023512}{Phys. Rev. D \textbf{90}, no.2, 023512 (2014)}
[\href{\arxiv/1402.6661}{arXiv:1402.6661} [gr-qc]].
\bibitem{Sami:2021ufn}
M.~Sami and R.~Gannouji,
\href{\doi/10.1142/S0218271821300056}{Int. J. Mod. Phys. D \textbf{30}, no.13, 2130005 (2021)}
[\href{\arxiv/2106.00843}{arXiv:2106.00843} [gr-qc]].
\bibitem{Sahni:1990tx}
V.~Sahni,
\href{\doi/10.1103/PhysRevD.42.453}{Phys. Rev. D \textbf{42}, 453-463 (1990)}

\bibitem{Boyle:2005se}
L.~A.~Boyle and P.~J.~Steinhardt,
\href{\doi/10.1103/PhysRevD.77.063504}{Phys. Rev. D \textbf{77}, 063504 (2008)}
[\href{\arxiv/astro-ph/0512014}{arXiv:astro-ph/0512014} [astro-ph]].
\bibitem{Watanabe:2006qe}
Y.~Watanabe and E.~Komatsu,
\href{\doi/10.1103/PhysRevD.73.123515}{Phys. Rev. D \textbf{73}, 123515 (2006)}
[\href{\arxiv/astro-ph/0604176}{arXiv:astro-ph/0604176} [astro-ph]].
\bibitem{Kuroyanagi:2008ye}
S.~Kuroyanagi, T.~Chiba and N.~Sugiyama,
\href{\doi/10.1103/PhysRevD.79.103501}{Phys. Rev. D \textbf{79}, 103501 (2009)}
[\href{\arxiv/0804.3249}{arXiv:0804.3249} [astro-ph]].
\bibitem{Figueroa:2018twl}
D.~G.~Figueroa and E.~H.~Tanin,
\href{\doi/10.1088/1475-7516/2019/10/050}{JCAP \textbf{10}, 050 (2019)}
[\href{\arxiv/1811.04093}{arXiv:1811.04093} [astro-ph.CO]].

\bibitem{Ahmad:2019jbm}
S.~Ahmad, A.~De Felice, N.~Jaman, S.~Kuroyanagi and M.~Sami,
\href{\doi/10.1103/PhysRevD.100.103525}{Phys. Rev. D \textbf{100},no.10, 103525 (2019)}
[\href{\arxiv/1908.03742}{arXiv:1908.03742} [gr-qc]].

\end{thebibliography}
\end{document}